\documentclass[]{aastex631}
\usepackage{wrapfig,lipsum,booktabs}

\usepackage[nointegrals]{wasysym}
\usepackage{ amssymb }
\usepackage[table]{xcolor}
\usepackage{amsmath}

\begin{document}

\newcommand{\ysopy}{{\color{black}{\ttfamily YSOpy }}}
\newcommand*\lah[1]{{\color{brown} #1}}
\newcommand*\lahcomm[1]{{\color{brown} [COMMENT: #1]}}
\newcommand*\gautam[1]{{\color{magenta} #1}}
\newcommand*\asc[1]{{\color{blue} #1}}
\newcommand*\asccomm[1]{{\color{blue} [COMMENT: #1]}}

\submitjournal{ApJ}
\accepted{15 May 2026}

\title{A Parameterized YSO Accretion Disk Model with Increasing Accretion Rate: \\Predicted Outburst Lightcurves}

\author[0009-0001-0200-2574]{Gautam Das}
\affiliation{Department of Physical Sciences;
Indian Institute of Science Education and Research Kolkata;
Mohanpur 741246, Nadia, West Bengal, India}
\affiliation{Visiting Scholar, Department of Astronomy;  California Institute of Technology; Pasadena, CA 91125, USA}
\email{gau7amdas@gmail.com}

\author{Lynne A. Hillenbrand}
\affiliation{Department of Astronomy;  California Institute of Technology; Pasadena, CA 91125, USA}
\email{lah@astro.caltech.edu}

\author[0000-0002-9540-853X]{Adolfo S. Carvalho}
\affiliation{Department of Astronomy;  California Institute of Technology; Pasadena, CA 91125, USA}
\affiliation{Current Address: Center for Astrophysics, Harvard University, 60 Garden St., Cambridge, MA, 02138, USA}
\email{adolfo.carvalho@cfa.harvard.edu}

\begin{abstract}   
A sub-class among Young Stellar Objects (YSOs), known as FU Ori type stars, undergo sudden rises in luminosity by several orders of magnitude on timescales of a few months to a few years, and decay back to quiescence on timescales of a few decades. Modelling the light curves of these objects is crucial to understanding how different components of these accretion disk systems evolve during outburst. For this purpose, we use a parametric model that couples the stellar photospheric emission, magnetospheric accretion shocks, an irradiated dust disk, and a viscously heated gas disk.  We adopt time-dependent accretion rate profiles that mimic the observed morphologies of FU Ori outburst light curves, and we use the accretion model infrastructure to simulate multi-band light curves, as well as color curves. The model enables us to study how different components dominate the flux in each band over the course of an outburst, providing insight into star-magnetosphere-disk interactions throughout the outburst cycle. 
{We find that throughout an accretion outburst, red optical and near-infrared lightcurves 
generally follow the same or very similar form as the input accretion profile,
being sensitive to heating in the accretion shocks and inner gas disk, 
while mid-infrared lightcurves are more responsive to the location and heating of the innermost dust disk.}  
\end{abstract}

\keywords{FU Orionis stars (553), Young stellar objects (1834), Stellar accretion disks (1579), Light curves (918)}


\section{Introduction}

Accretion of material, first from protostellar infall, and then from the circumstellar environment, is the way in which stars accumulate their mass. 
Low-mass \citep[$<2\ M_{\astrosun}$; ][]{belt_de_wit_2016_review} Young Stellar Objects (YSOs) are embedded in a disk 
that brings material towards the stellar surface and deposits it onto the star, thus converting the gravitational potential energy to stellar radiation and kinetic energy through winds and jets. The accretion process is believed to be magnetically controlled. The stellar magnetic field stops the radial flow of matter at roughly the radius where the magnetic pressure equals the gas ram pressure, and inflowing materials follow the stellar magnetic fields towards either pole, accreting onto the stellar surface through funnel flows \citep{uchida1984,koenigl_1991_blackbody_shock_assumption}. 
The magnetospheric accretion activity creates accretion shocks, often buried under the stellar surface, 
that heat the adjacent photosphere and the base of the accretion columns with X-rays. As the radiation propagates outwards, photons are reprocessed and produce hydrogen recombination lines and hydrogen continuum. The continuum is typically approximated by a ``slab" model \citep{hartigan1991,manara2014physics}. 

For rapidly rotating accretion disks, viscous processes heat the disk to temperatures much higher than the stellar photospheric temperature. Thus, if the accretion rate is high enough, the viscous disk overwhelms in optical and infrared wavelengths.
Episodic and bursty accretion is now recognized as an important phenomenon in the evolution of young stellar objects, with accretion outbursts occurring in protostars as well as young stars that are much further along in the dissipation of their circumstellar disks \citep{fisher_hillenbrand_2023}. 

In this paper, we study the transition from low-state magnetospheric accretion to high-state outburst accretion, applying a simple parametric model for the star, the disk, and the magnetospheric accretion shocks, while varying the accretion rate over time.
We discuss the theory and formulations infrastructure of the modelling pipeline in Section \ref{sec:modeling_infra}. In Section \ref{sec:linear} we present a test case of an outburst that follows a linearly increasing accretion rate, and illustrate the predictions in order to orient the reader to the format our results including diagnostic plots. Section \ref{sec:diff_systems} presents the results for four different types of FU Ori accretion profiles, based on characteristic familes of outburst lighturves. 
Finally we conclude in Section \ref{sec:conclusion}.

\section{Modeling Infrastructure}\label{sec:modeling_infra}

\begin{figure}[t!]
\plotone{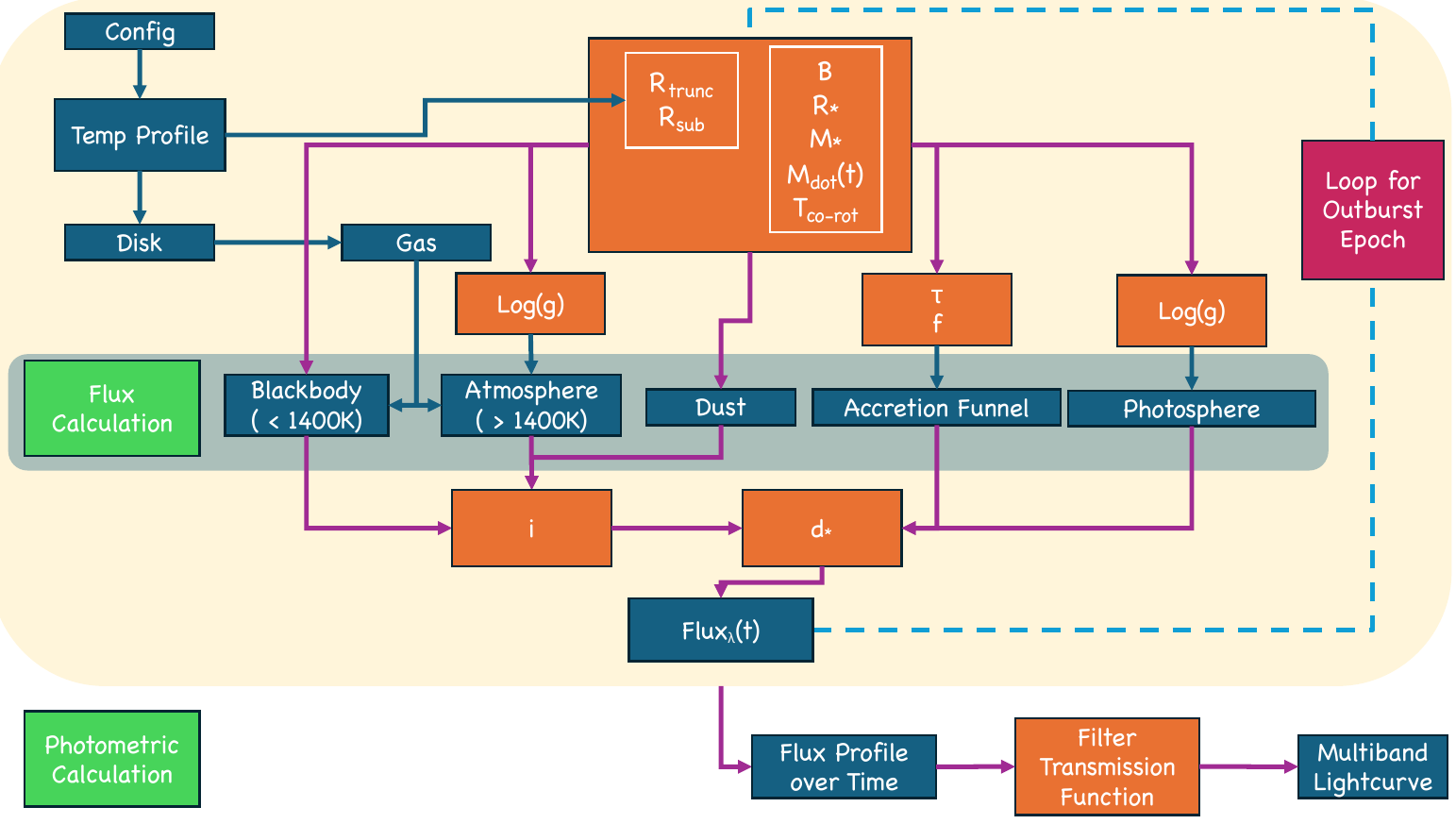}
\caption{
Flowchart showing the modelling pipeline implementation.  See text and Table~\ref{tab:variables} 
for definitions of all variables.
Key parameters (orange boxes) are either entered as config parameters (upper left aqua box), or in the case of $R_{\textrm{trunc}}$ and $R_{\textrm{sub}}$, result from application of standard formulae for YSO accretion disks {(\S\ref{subsec:formula})}.  The pipeline separately calculates the spectra corresponding to each system component (magnetospheric accretion hotspot, gas disk, dust disk). The component-wise spectra are then scaled based on the viewing angle (inclination) and distance to the system. This process is carried out serially over the whole outburst cycle by changing the disk accretion rate. At the end of the simulation, we are left with the spectral variation as a function of time. We then convolve this temporal flux profile with telescopic filter transmission functions to generate the multi-band lightcurve profiles.
}
\label{flowchart}
\end{figure}

Our lightcurve modelling infrastructure is built on top of the YSOpy code \citep[][in-preparation]{Das2025YSOs,Das2026YSOs} 
for predicting optical and near-infrared spectra of accreting young stars.  
The present model differs from the original YSOpy infrastructure in many ways. Some upgrades improve the self-consistency of the calculations, and some simplifications improve computational efficiency. We also developed tools to simulate time series photometry resulting from a time-dependent model. These modifications are described in Appendix~\ref{sec:ysopy}. 
The basic elements and parameters of our model,  and a flowchart showing the steps and the order of the calculations  is provided in Figure~\ref{flowchart}, described in detail below. 

The pipeline separately calculates spectra corresponding to each of the system components:  
star, accretion shock on the stellar surface, and disk.
Thereafter the components of the spectrum are rescaled based on the viewing inclination and the distance to the stellar system.
The process of modelling an evolving accretion rate is carried out serially by changing the disk accretion rate, step-by-step, through an outburst cycle, as the defined by the assumed accretion rate profile. At the end of the cycle, we are left with a high-resolution spectrum varying as a function of time. We then convolve this temporal flux profile with telescopic filter transmission functions in order to generate the multi-band lightcurve profiles. 

We emphasize that we are using a parameterized model, rather than a radiative transfer calculation,
or the self-consistent physics of a numerical simulation \cite[e.g.][]{nayakshin2024,pavlyuchenkov2025, roberts2025, roberts2026}. 

\subsection{Parameters}\label{subsec:params}

Model inputs{ to \ysopy are shown as \textit{config} in Figure \ref{flowchart}. 
They} include stellar parameters like the mass of the star ($M_{*}$), radius of the star ($R_{*}$), 
dipolar magnetic field strength ($B$), and stellar rotation period ($P_{\textrm{rotation}}$).   
Additional parameters to model the accretion shock component include the hotspot area filling factor ($f$), 
and {gas} optical depth ($\tau$). 

The circumstellar disk has a temperature gradient with radius as described {in Section \ref{subsec:formula}} 
and an assumed surface gravity (log\hspace{1mm}$g$). 
Disk parameters include an assumed accretion rate ($\dot{M}$), 
inclination to our line-of-sight ($i$), 
and an inner disk radius ($R_{in}$). {The accretion rate varies in time but is constant with radius. }
The inner gas disk radius ($R_{in}$) is set by the balance of accretion pressure inward and magnetic pressure outward, and evolves over time as the accretion rate changes. 
The inner dust disk radius ($R_{\textrm{dust,inner}}$) is the same as the inner gas radius $R_\textrm{{in}}$, except when $R_{in}$ is close enough
to the star such that the temperature exceeds the {dust sublimation temperature, at which point the dust radius is set by} $R_{sub}$. $R_{\textrm{dust,inner}}$ also evolves in time as the accretion rate changes.

The assumed outer dust disk radius ($R_{outer}$) is much larger than the viscously dominated inner region.
The outer boundary for the viscous gas disk is set to 1 AU based on the fact
that, for viable model parameters, the gas disk generates little viscous heating {beyond} 1 AU, 
{and its flux contribution is} negligible compared to that from the passive disk at these same annuli. 
Finally, we assume a source distance ($d$). 

For convenience to the reader, we collect all model parameters in Table \ref{tab:variables}.

\subsection{Formulations}\label{subsec:formula}

For the disk, we {separately track}
the active (gas accreting) component
and the  non-active (passive) disk component that is predominantly dust.
The former governs when the accretion rate is high and in the innermost disk regions,
while the latter controls when the accretion rate is low and at larger radii.
The contributions to disk heating and cooling, and therefore the overall T(r) in the disk also depend on  
several critical radii in the disk.

The  modified \cite{shakurasunyaev1973} temperature profile for gas accretion is given by 

\begin{equation}\label{Temp_viscous}
    T_{visc}^{4} = \dfrac{3GM_{\ast}\dot{M}}{8\pi\sigma R^3} \left[1-\left(\dfrac{R_{in}}{R}\right)^{1/2}\right]
\end{equation}
while the temperature profile from \cite{chiang_goldreich_97} for the dust-dominated disk is given by 

\begin{equation}\label{eq:Temp_dust}
    T_{\textrm{eff,dust}} = \left(\dfrac{\alpha_{0}}{2}\right)^{1/4} \cdot \left(\dfrac{R_{\ast}}{R}\right)^{1/2} T_{0}
\end{equation}
where 
$T_{0}$ is the effective temperature of radiation reaching the flaring dust disk. Since the radiation field is complicated, we follow the prescription in \cite{liu2022diagnosing} and make an approximation by combining the gaseous disk, stellar photosphere and magnetospheric accretion components into a single parameter $T_{0}$; at low accretion rates $T_0=T_{\textrm{photo}}$.

For $\alpha_{0}$, the angle between the incident radiation and the surface of the disk,  
we follow \cite{chiang_goldreich_97} and
consider $H$ as the disk height above the mid-plane, with

\begin{equation}\label{eq:a_0tempprof1}
        \alpha_{0} = \dfrac{0.4 R_{\ast}}{R} + R \dfrac{d}{dR} \left(\dfrac{H}{R}\right).
\end{equation}
Adopting the \cite{liu2022diagnosing} parameterization, we get
\begin{equation}\label{eq:a_0tempprof2}
        \alpha_{0}= 0.003\dfrac{R_{\ast,1.6}}{R_{au}} + 0.05\dfrac{T_{\ast,3400}^{4/7}\, R_{\ast,1.6}^{2/7}}{M_{\ast,0.3}^{4/7}} R_{au}^{2/7},
\end{equation}
where typical values of $T_{\ast}, R_{\ast}$ and $M_{\ast}$ are given in the subscript. $T_{\ast}$ is in Kelvins, $R_{*}$ is in units of $R_{\astrosun}$ and $M_{*}$ is in units of $M_{\astrosun}${, all scaled to the values noted in subscripts}. 

{We use the \citet{chiang_goldreich_97} temperature profile for the dust at low accretion rates, but not high}. When the accretion rate is low, since the gas disk temperature is low, the {heating of the dust} disk is dominated by {irradiation from} the stellar photosphere. Thus fixing $T_{0} = T_{\textrm{photo}}$ makes sense at small accretion rates. For high accretion cases, the {accretion heating of dust can dominate over irradiation in the inner disk}.  {We therefore compute the temperature of the dust using both Equation \ref{Temp_viscous} and \ref{eq:Temp_dust} and select the greater of the two:} 
 \begin{equation}\label{eq:t_dust_cases}
        T_{\textrm{dust}}(r) = 
        \begin{cases}
            T_{\textrm{eff,dust}} \hspace{10pt}\text{(Equation \ref{eq:Temp_dust})} & \text{if $T_{\textrm{eff,dust}}(r) > T_{\textrm{visc}}(r)$}\\
            T_{visc} \hspace{14pt}\text{(Equation \ref{Temp_viscous})}
            & \text{if $T_{\textrm{eff,dust}}(r) < T_{\textrm{visc}}(r)$}.\\
        \end{cases}
    \end{equation}
{We do not include emission from dust located along the inner disk ``wall" 
\citep{isella2005,mcclure2013} that directly faces the star.  
This may mean that near-infrared fluxes are underestimated for low accretion rates.
However, the number of timesteps for which emission from such a component would be relevant, are few.
}    

The magnetospheric truncation radius $R_{\textrm{trunc}}$ is defined as the boundary where the accretion pressure from the disk equals the magnetic pressure from the stellar magnetic field. {At typical accretion rates, disk matter is channeled along the magnetic field lines to the star, setting an inner boundary for the thin, viscous accretion disk geometry}, i.e., $R_{\textrm{in}}$ = $R_{\textrm{trunc}}$. The expression for {$R_{\textrm{trunc}}$} is given by{ \citet{hartmann2016accretion} as,}

\begin{equation}\label{eq:r_trunc}
    R_{\textrm{trunc}} = 18 \eta \xi \dfrac{B_{3}^{4/7} R_{2}^{12/7}}{M_{0.5}^{1/7} \dot{M}_{-8}^{2/7}} \ R_{\astrosun},
\end{equation}
where the typical values of $B$ (in units of G), 
$R_{\ast}$ (in units of $R_{\astrosun}$), 
$M_{\ast}$ (in units of $M_{\astrosun}$), 
$\dot{M}$ (in units of $M_{\astrosun}$/yr) 
are given in the subscript. 
Here, $\eta$ is the factor accounting for equatorial accretion disk and taken to be $0.5$ according to \cite{long2005locking}, and $\xi$ is the factor relating to the disk-magnetosphere interaction and taken to be $0.7$ as in \cite{hartmann2016accretion}. 

The inner boundary of disk {is set to be}
$R_{\textrm{in}} = max\left\{R_{*}, min\left[R_{\textrm{co-rot}}, R_{\textrm{trunc}} \right]\right\}$, which we discuss in more detail in Section \ref{subsubsec:calc_evo_radii_temp}.
Here $R_{\textrm{co-rot}}$ denotes the co-rotation radius of the star derived from the stellar rotation period,

\begin{equation}\label{eq:corotation_radius}
    R_{\textrm{co-rot}} = \left( \dfrac{\sqrt{GM_{*}}T_{*\textrm{,rot}}}{2\pi}\right)^{2/3},
\end{equation}
where we fix the stellar rotation period to be $7$ days, in agreement with low mass stellar rotation periods observed by \citep{bouvier_review_rotation_period}.

Much of the following exposition is based on the calculation presented in \cite{hartmann2016accretion}, which is itself based on \cite{calvet_and_gullbring_1998}. 
For a given accretion rate, $\dot{M}$, the accretion luminosity $L$,  for the infall of matter from the inner truncation radius $R_{\textrm{in}}$ given by 

\begin{equation}\label{eq:lumi_acc}
    \begin{split}
        L_{\textrm{acc}} &= \dfrac{1}{2} \Dot{M} v_{\textrm{ff}}^{2}\\
        &= \dfrac{G M \Dot{M}}{R_{*}} \left(1 - \dfrac{R_{\ast}}{R_{\textrm{in}}} \right).\\
    \end{split}
\end{equation}

This means, upon rearranging the above expressions, that the free-fall velocity $v_{\textrm{ff}}$ through the accretion column, near the stellar surface, is given by:

\begin{equation}\label{eq:v_freefall}
    v_{\textrm{ff}} = \left[\left(\dfrac{2 G M_{\ast}}{R_{\ast}} \right) \left( 1 - \dfrac{R_{\ast}}{R_{\textrm{in}}}\right)\right]^{1/2}
\end{equation} 

The flux from magnetospheric accretion {was initially} assumed to be radiated as blackbody radiation with an effective temperature $T_{\textrm{hotspot}}$ \citep{koenigl_1991_blackbody_shock_assumption, hartmann_1994_radiative_transfer_acc_column} 
{and this assumption is still sometimes used \citep{liu2022diagnosing}. 
It enables an easy calculation of $T_\mathrm{hotspot}$ from $L_\mathrm{acc}$ via the Stefan-Boltzmann equation,} 
with the shock luminosity $L_{\textrm{hotspot}}$ equal to $L_\mathrm{acc}$ ({or more realistically some fraction of it}, given that processes like winds use some of the accretion energy).  Thus the fractional area of the stellar surface covered by accretion hotspots can be estimated.

When {an optically thin} plane parallel slab is assumed in place of a blackbody, {as we do here}, setting $T_{\textrm{slab}}$ = $T_{\textrm{blackbody}}$ {reduces the flux and hence $L_{\textrm{slab}}$ relative} to the blackbody model.
{To enable the direct connection of $T_\mathrm{hotspot}$ to $\dot{M}$ in the disk while conserving the accretion energy, one must scale the $T_\mathrm{hotspot}$ and $n_e$ inputs to the slab model such that $L_{\textrm{slab}}= L_{\textrm{blackbody}}$.}

We thus introduce a parameter, {$\chi$, which is the scale factor in the plane-parallel slab model 
approximation necessary to maintain $L_{\textrm{slab}}= L_{\textrm{blackbody}}$.
When the hotspot is approximated as a blackbody, $\chi= 1$. 
However, in the slab model, to ensure luminosity conservation with $\tau_{3000\AA}=1$, 
we find that $\chi$ is a function of $\dot{M}$ and we calculate the scaling at each time step. 
Using, as an example, the parameters in the case of the linear accretion rate increase discussed below,
we find values of $\chi \approx$5, 1.9, 1.35, and 1.3 for accretion rates of log $\dot{M} = -8, -7, -6, -5$ 
dex $M_{\astrosun} yr^{-1}$. For convenience, we also define $\zeta = \chi \times f$, where $f$ is the
filling factor of the magnetospheric funnel on the stellar surface.
Effectively we are thus changing the hotspot area from the fiducial 10\% to the value needed to conserve energy.}

We thus calculate the mass density within the {free falling} accretion column, $\rho_{\textrm{ff}}$, as 

\begin{equation}
    \begin{split}
        \rho_{\textrm{ff}} &= \dfrac{\dot{M}}{A_{\textrm{shock}} \cdot v_{\textrm{ff}}}\\
        &= \dfrac{\dot{M}}{\zeta \cdot 4 \pi R_{\ast}^{2} \cdot v_{\textrm{ff}}}.\\
    \end{split}
    \label{eq:density}
\end{equation}
Since we are concerned with the 
{pre-shock region, we can use} 
the mass density $\rho_{ff}$ and the velocity of mass flow $v_{ff}$  
{together to calculate}
the flux $F$ coming from shock region as 

\begin{equation}\label{eq:flux_acc}
    \begin{split}
        F &= \dfrac{1}{2}\rho_{\textrm{ff}}v_{\textrm{ff}}^{3}\\
        &= \dfrac{1}{4} \dfrac{G M \Dot{M}}{\pi \zeta R_{\ast}^{3}} \left( 1 - \dfrac{R_{\ast}}{R_{\textrm{in}}}\right).
    \end{split}
\end{equation}

As matter falls onto the shock region, $1/2$ of the energy propagates inwards towards the post-shock region and $1/2$ of the energy {propagates} outwards to the pre-shock region, which reprocesses the energy {and re-radiates} $\sim 1/4$ {of it} into 
the post-shock region. This region in turn heats the photosphere locally, producing a hotspot, which is what we see as the blue/uv excess the spectra of accreting young stars.
We can estimate the temperature of the hotspot region assuming that $3/4$ of the shock energy is generating the relevant temperature, or

\begin{equation}
    \begin{split}
        \sigma T_{\textrm{hotspot}}^{4} &= \dfrac{3}{4} F,
    \end{split}
\end{equation}
which leads to 

\begin{equation}\label{eq:t_hotspot}
        T_{\textrm{hotspot}} = \left[\dfrac{3}{32} \dfrac{\dot{M}}{\pi \sigma \zeta R_{\ast}^{2}} v_{ff}^2 \right]^{1/4} = \left[\dfrac{3}{16} \dfrac{G M \Dot{M}}{\pi \sigma \zeta R_{\ast}^{3}} \left( 1 - \dfrac{R_{\ast}}{R_{\textrm{in}}}\right)\right]^{1/4}.
\end{equation}

We can similarly calculate the electron density $n_{\textrm{e}}$ as given by:

\begin{equation}\label{eq:numberdensity_total}
    \begin{split}
        \rho_\textrm{ff} &= (\mu m_{\textrm{H}}) n_{\textrm{tot}}\\
        &= \dfrac{1}{2} m_{\textrm{H}} (n_\textrm{H} + n_e) \\
        &= n_{\textrm{e}} m_{\textrm{H}}\\
    \end{split}
\end{equation}
where we are assuming the material to be mostly composed of fully ionized hydrogen, so $\mu=0.5$ and $n_e = n_\mathrm{H}$.
From the expression for $\rho_\mathrm{ff}$ in equation \ref{eq:density}, and substitution, the electron density at post shock region is:

\begin{equation}
    \begin{split}
        n_{\textrm{e}} &= \dfrac{\Dot{M}}{m_{\textrm{H}} \zeta \pi R_{\ast}^{2}} \cdot \left[\left(\dfrac{2 G M_{\ast}}{R_{\ast}} \right) \left( 1 - \dfrac{R_{\ast}}{R_{\textrm{in}}}\right)\right]^{1/2}.
    \end{split}
    \label{n_e}
\end{equation}

{{The three parameters $\tau(\lambda)$, $T_{\textrm{hotspot}}$, and $n_e$ are inputs to the accretion shock model. The hydrogen slab model consists of H and $\textrm{H}^{-}$ continuum. Emissivity is calculated from radiative recombination and free-free radiation when free-electrons encounter protons, thus giving H emission. Similarly the continuous absorption from $\textrm{H}^{-}$ is obtained by calculating the absorption coefficients corresponding to bound-free transition in photo-detachment mechanism and in free-free transitions, 
leading to the emissivity corresponding to $\textrm{H}^{-}$ emission. 
The total emissivity from these two processes, H and $\textrm{H}^{-}$ emissions, gives the flux from the slab model.}

\begin{figure}[hb!]
    \centering
        \includegraphics[width=0.32\linewidth]{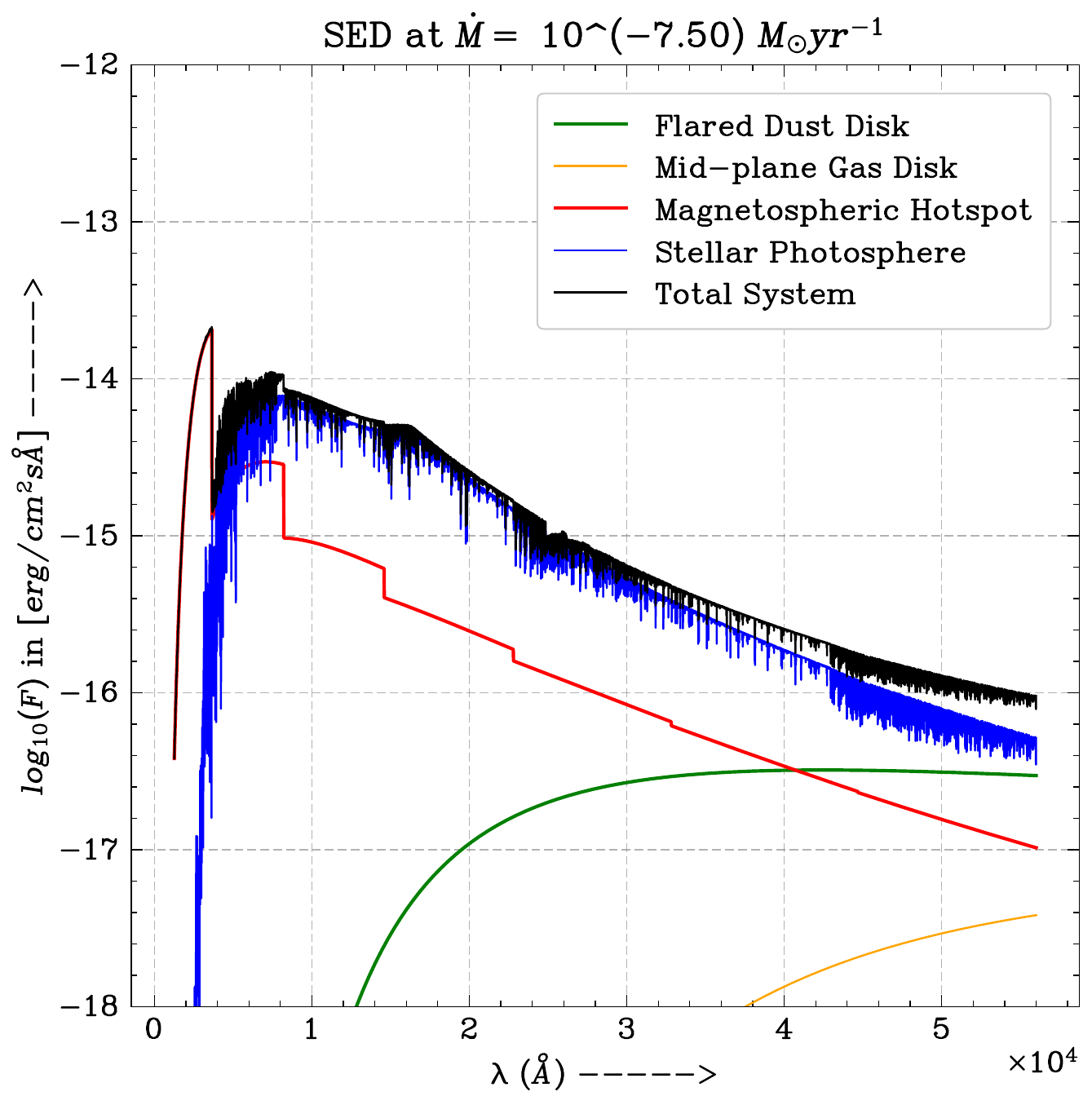}   
        \includegraphics[width=0.32\linewidth]{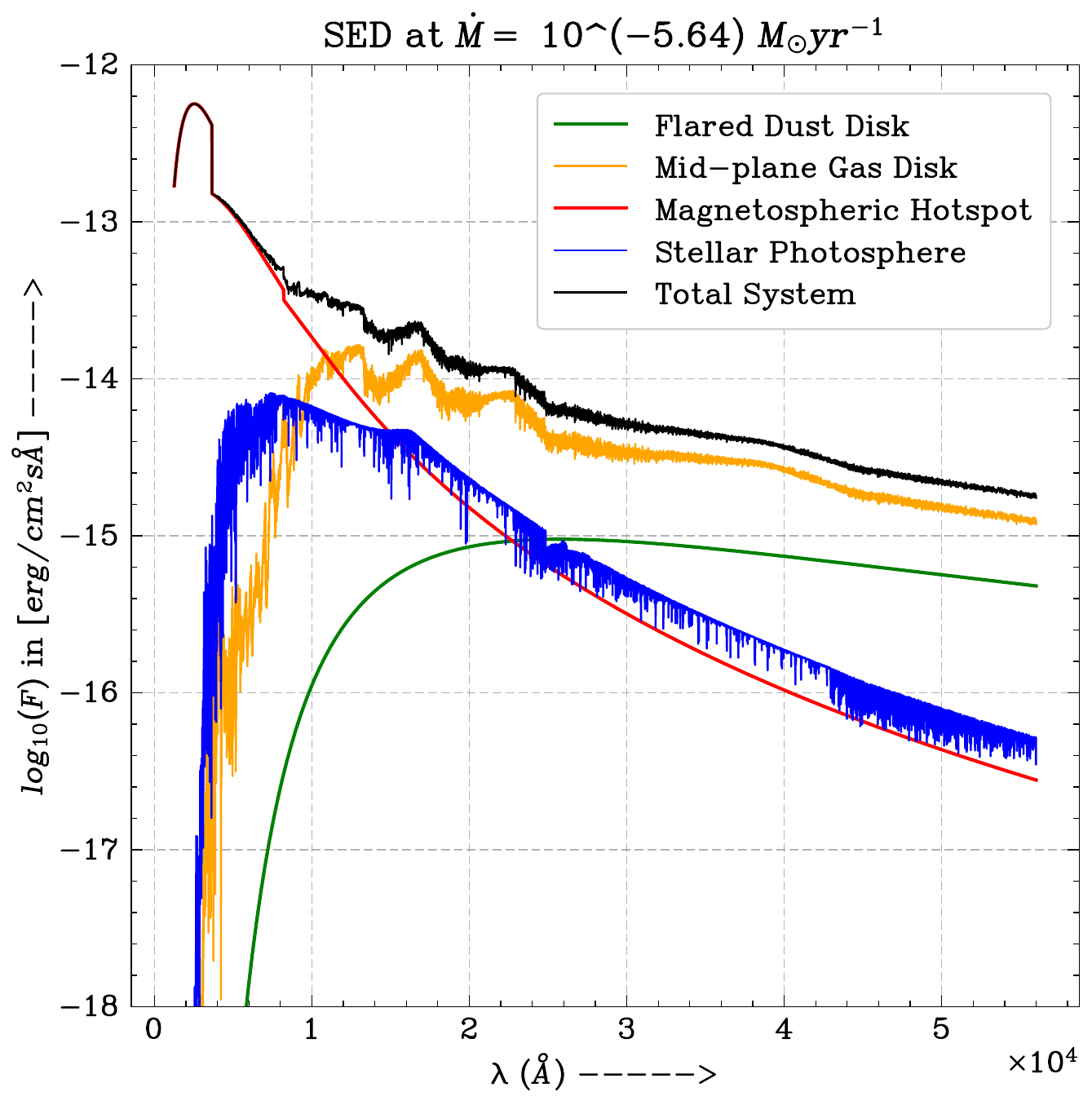}
        \includegraphics[width=0.32\linewidth]{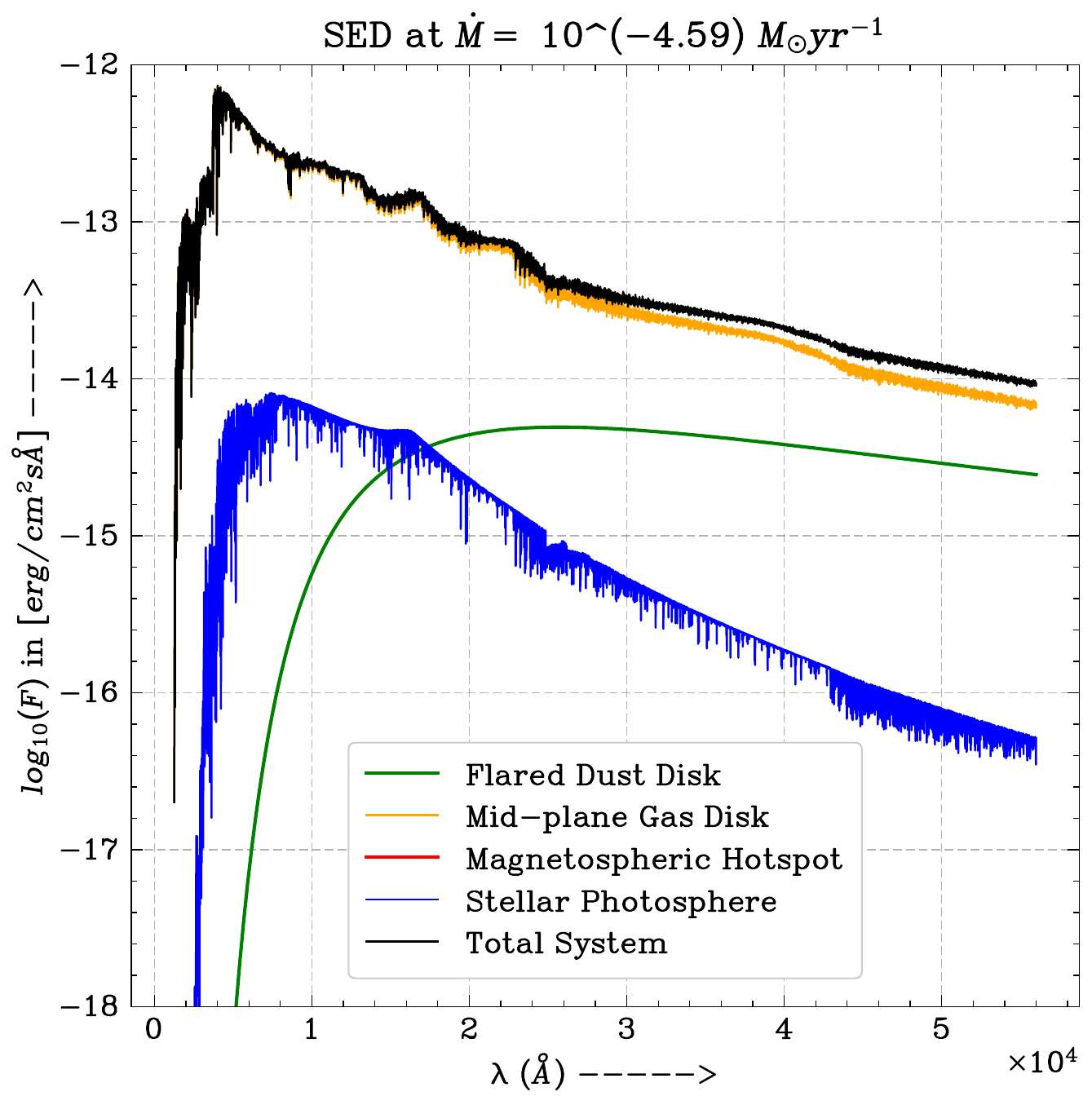}
    \caption{ {Component-wise spectral energy distributions for low, intermediate, and high accretion scenarios. {Left:} At low accretion rates, the flux from the viscous gas disk component (yellow) and from the hotspots on the stellar surface due to magnetospheric accretion (red) are essentially negligible compared to the stellar photosphere (blue), 
    with the dust component (green) contributing only at the longest wavelengths shown.
    {Middle:} At intermediate accretion rates, all four components are present and contribute to the total flux. The accretion shock gives rise to the Balmer jump that is prominent in the u band.
    The viscous accretion produces the spectral features in the near-infrared JHK bands, 
    and the dust continuum contributes to the mid-infrared KLM bands. 
    {Right:} At high accretion rates, the poloidal accretion activity has switched-off and the viscous disk component dominates the flux.} 
    }
    \label{fig:seds}
\end{figure}

\subsection{Overview of Spectral Flux Calculation and Filter Profile Convolution}\label{sec:filter}

We compute the spectral flux from four different emission components: (i) the viscously heated gas disk, (ii) the passively heated dust disk, (iii) the accretion shock, and (iv) the photosphere of the central star. Each of these components is computed separately and then summed to produce the total flux of the system. The disk components are scaled according to the inclination of the disk prior to summation and then the entire system flux is scaled by the source distance {(squared).  The individual component spectral energy distributions and the composite are shown in Figure~\ref{fig:seds} for three different values of the accretion rate corresponding to: a low state, an intermediate state, and a high state.  The stellar and other parameters are the same as
the linear rise accretion case described in \S~\ref{sec:linear}.}

For computing the viscous disk emission, we use the pre-calculated BT-Settl grid {of stellar spectra} \citep{bt_settl_paper}.  While our radial mesh is uniform, the atmospheres are available only at specific temperature intervals, with maximum resolution of 100 K.  We thus bin the discrete $r_i$ in the model to the temperature resolution of the grid. 
For example, if a ring has a computed temperature of 2375 K, the ring becomes a part of the isothermal annulus at 2400 K. 
However, the radial thickness of each isothermal annulus is determined by the temperature profile {and thus follows a log scale}. 
{Where the temperature gradient is sharp}, the isothermal annuli are radially narrow and the number of rings in each annulus is small, whereas in shallower parts of the temperature profile the annuli are wide and thus have larger number of rings. To account for the varying number of bins per annulus,
we draw the theoretical spectrum corresponding to the temperature of each annulus, multiply by the number of rings in that annulus, and divide by the total number of rings constituting the viscous disk.  

Finally, we convolve the full-resolution spectrum with various filters to produce simulated lightcurves. We retrieved filter functions and zero-point calibrations for $u, BP, RP, G, J, H, K, W1 \textrm{ and } W2$ bands from the SVO Filter Service \citep{svo_profile}.  

\begin{table}[!h]
    \centering
    \begin{tabular}{l|c|l}
       Symbol  & Definition & Reference \\
       \hline
       $M_{*}$&Stellar Mass& Table \ref{tab:fid_parameters}\\
       $R_{*}$&Stellar Radius& Table \ref{tab:fid_parameters}\\
       $T_{\mathrm{photo}}$&Stellar Temperature & Table \ref{tab:fid_parameters}\\
        $B$&Stellar Magnetic Field Strength& Table \ref{tab:fid_parameters}\\
        $P_{\textrm{rotation}}$&Stellar Rotation Period& fixed at 7 days\\
        $f$&Accretion Hotspot Filling Factor& fixed at 1\% \\
        $\tau$&Optical Depth of Hydrogen Slab& fixed at unity \\
        $\Dot{M}$&Disk Accretion Rate& Figures \ref{stline_mdot_rin} (left) and \ref{mdot}\\
        log $g$&Disk Surface Gravity& fixed at 3.5 or 1.5 dex cm$^2$/g  (see text) \\ 
        $i$&Disk Inclination Angle& Table \ref{tab:fid_parameters}\\
        $R_{outer}$&Outer Boundary of Disk& fixed at 1 AU\\
        $R_{in}$&Inner Boundary of Gas Disk&Equation \ref{eq:in_disk_boundary}\\
        $R_{\textrm{dust,inner}}$&Inner Boundary of Dust Disk&Equation \ref{eq:dust_inner_boundary}\\
        $R_{sub}$&Dust Sublimation Radius&  Equation \ref{eq:Temp_dust} exceeding {$T_{sub}$}\\
        $R_{\textrm{trunc}}$&Magnetospheric Truncation Boundary&Equation \ref{eq:r_trunc}\\
        $R_{\textrm{co-rot}}$&Co-rotation Radius&  Equation \ref{eq:corotation_radius} \\
        $d$&Source Distance &Table \ref{tab:fid_parameters}\\
        $T_{visc}$&Temperature of Viscous Disk&Equation \ref{Temp_viscous}\\
        $T_{\textrm{eff,dust}}$& Effective Temperature of Irradiated Dust Disk &Equation \ref{eq:Temp_dust}\\
        $T_{0}$&Effective Temperature of Radiation Heating the Dust Disk& \cite{liu2022diagnosing} \\
        {$T_{\textrm{dust}}$ }&{Dust Temperature}&{Equation \ref{eq:t_dust_cases} }\\
        $T_{sub}$&Temperature of Dust Sublimation & assumed 1400 K; {\cite{Posch_2007_dust_2400K}}\\
        $\alpha_{0}$&Angle between Disk Surface and Incident Radiation&Equations \ref{eq:a_0tempprof1}, \ref{eq:a_0tempprof2}\\
        $\eta$&Correction Factor for Equatorial Accretion&{assumed 0.5; \cite{long2005locking} }\\
        $\xi$&Correction Factor for Disk-Magnetosphere Interaction& {assumed 0.7; \cite{hartmann2016accretion} }\\
        $L_{acc}$&Accretion Luminosity&Equation \ref{eq:lumi_acc}\\
        $F$&Flux from Accretion Shock & Equation \ref{eq:flux_acc} \\
        $\chi$&Scale Factor Maintaining $L_{\textrm{slab}}$ = $L_{\textrm{Blackbody}}$& {calculated (see text)}\\
        $v_{\textrm{ff}}$&Free-fall Velocity& Equation \ref{eq:v_freefall}\\
        $\rho_{\textrm{ff}}$& Mass Density during Free-fall& Equation \ref{eq:density}\\
        $T_{\textrm{hotspot}}$&Temperature of Hotspot&Equation \ref{eq:t_hotspot}\\
        $n_{\textrm{tot}}$&Total Number Density&Equation \ref{eq:numberdensity_total}\\
        $n_{\textrm{e}}$&Electron Number Density&Equation \ref{n_e}\\
        $\mu$&Mean Molecular Weight& fixed at 0.5\\ 
    \end{tabular}
    \caption{Table of variables used in this work.}
    \label{tab:variables}
\end{table}

\subsection{Procedure}
\subsubsection{Assuming an Accretion Rate Profile}

We create synthetic accretion rate profiles in a log-linear form, 
varying $\dot{M}$ in the model at each step. 
As presented below, we begin with a simple assumption of a log-linearly rising value of $\dot{M}$
{(\S\ref{sec:linear})}
and then we consider accretion profiles that are based on the observed lightcurves of example real FU Ori objects
{(\S\ref{sec:diff_systems})}. 
The accretion outburst profiles are run through the pipeline illustrated in Figure~\ref{flowchart}.

\subsubsection{Calculating the Evolution of Radii, Temperatures, and Velocities}\label{subsubsec:calc_evo_radii_temp}

The parameterized increase in $\dot{M}$ affects many other parameters of the accretion system. A rising mass accretion rate first increases the magnetospheric shock temperature, and viscously heats the inner disk{, which consequently increases} the sublimation radius. At higher values of $\dot{M}$, the inner magnetospheric truncation radius of the disk also decreases\footnote{{For extreme $\dot{M}$, in principle the system would transition from magnetospheric to more equatorial accretion. However, our accretion shock model does not include the emission from such an equatorial boundary layer between the disk and the star. We note that the boundary layer in FU Ori stars, if present, appears to have a very small filling factor and contributes primarily in the FUV/NUV \citep{adolfo_uv}.}}. 
We use the equations in \S\ref{subsec:formula} to track the following parameters.

\begin{itemize}

    \item {$R_{\textrm{in}}$:} In general, $R_{\textrm{in}} = R_{\textrm{trunc}}$. The magnetospheric truncation radius $R_{\textrm{trunc}}$ is given by Equation \ref{eq:r_trunc}. At very low accretion rates, $R_{\textrm{trunc}} $ diverges and becomes arbitrarily large. However
    at large $r$ the coupling between the disk and the stellar field is expected to be weak as the disk ionization fraction and temperature decreases. Thus, for very low accretion rates, instead of the magnetic field controlling $R_{\textrm{in}}$ we have $R_{\textrm{in}} = R_{\textrm{co-rot}}$ as given by Equation \ref{eq:corotation_radius}, with the rotation period of the star controlling $R_{\textrm{in}}$. 
    As the accretion rate increases, $R_{\textrm{trunc}}$ decreases, {eventually reaching $R_\mathrm{corot}$ so that $R_{\textrm{in}}$ then becomes set by $R_{\textrm{trunc}}$. At very high accretion rates} the inner disk boundary {continues moving} inwards and $R_\mathrm{in} = R_{\textrm{trunc}}$ determined from Equation \ref{eq:r_trunc} becomes equal to or smaller than $R_{*}$. In this accretion regime, $R_{\textrm{in}} = R_{*}$. {In general, then,} $R_{\textrm{in}}${ can be parameterized by} the value of $T_{\textrm{hotspot}}(r)$ and $R_{\textrm{trunc}}(r)$ as shown below 
    
    \begin{equation}\label{eq:in_disk_boundary}
        R_{\textrm{in}}(r) = 
        \begin{cases}
            R_{\textrm{co-rot}} & \text{if $T_{\textrm{hotspot}} < T_{\textrm{photo}}$ and $R_{\textrm{trunc}}(r) \geq R_{*}$}\\
            R_{\textrm{trunc}}(r) & \text{if $T_{\textrm{hotspot}} \geq T_{\textrm{photo}}$ and $R_{\textrm{trunc}}(r) \geq R_{*}$}\\
            R_{*} & \text{if $R_{\textrm{trunc}}(r) < R_{*}$}
        \end{cases}
    \end{equation}
    
    \item {$R_{\textrm{sub}}$:} The dust sublimation radius is defined as the boundary interior {to which dust particles sublimate} and the disk is present as gas only. 
    $R_{\textrm{sub}}$ is thus denoted by where the dust temperature computed from 
    Equation \ref{eq:Temp_dust} exceeds 1400 K \citep{Posch_2007_dust_2400K}.

    \item {$R_{\textrm{dust,inner}}$:} The inner boundary of the passively heated dust disk {need not equal} the inner boundary of the gas disk. When $\Dot{M}$ is low, accretion heating is minimal or nonexistent, so the maximum temperature of the {dust} disk is set by $T_{*}$ alone.  
    For low-mass stars, $R_{\textrm{co-rot}} > R_{\textrm{sub}}$, so in the low accretion rate case, $R_{\textrm{dust,inner}} = R_{\textrm{in}}$. 
    As the accretion rate rises, the temperature of the disk at $R_{\textrm{in}}$ (Equation \ref{Temp_viscous}) rises above the dust sublimation temperature and a region composed only of gas, free from dust, can be sustained. Once, $R_{\textrm{sub}} > R_{\textrm{in}}$, we set $R_{\textrm{dust,inner}} = R_{\textrm{sub}}$. The sublimation radius moves outwards as the accretion rate increases. The inner boundary of the dust disk {is thus given by}
    \begin{equation}\label{eq:dust_inner_boundary}
        R_{\textrm{dust,inner}} = 
        \begin{cases}
            R_{\textrm{in}} & \text{if $T_{\textrm{eff,dust}}(R_\textrm{in}) \leq$ 1400 K}\\
            R_{\textrm{sub}} & \text{otherwise.}
        \end{cases},
    \end{equation}
    
    \item {$T_{\textrm{hotspot}}$:} The accretion hotspot temperature is assumed to be that of the base of the accretion funnel on the stellar surface, computed using Equation \ref{eq:t_hotspot}. At very low accretion rates, $T_{\textrm{hotspot}}$ turns out to be lower than the $T_{\textrm{photo}}$, the photospheric temperature of the star. In this regime, even if poloidal accretion is present, it cannot be studied properly in our current framework and we simply set $T_{\textrm{hotspot}}= T_{\textrm{photo}}$.  This has the effect of conserving the total flux from the photosphere. As the accretion rate increases, $T_{\textrm{hotspot}}$ also increases. But once the accretion rate becomes so high that $R_{\textrm{in}}$ comes close to $R_{*}$, {the $v_\mathrm{ff}$ term in} equation \ref{eq:t_hotspot} {rapidly decreases, causing} $T_{\textrm{hotspot}}$ to decrease {as well}, until $R_{\textrm{in}} = R_{\textrm{*}}$ where $T_{\textrm{hotspot}}$ then becomes $0$.  As the disk is touching the stellar surface at this time, we interpret this situation as the shutdown of poloidal accretion activity. We would not see any hotspot emission with further increase in $\Dot{M}$ since it is completely turned off. Such high accretion rate implies equatorial accretion; although 
    as noted earlier, our current framework does not include this emission source.
    
    \item {$v_{\textrm{freefall}}$:} We calculate the freefall velocity for matter falling on the stellar surface from the $R_{\textrm{in}}$ boundary using equation \ref{eq:v_freefall}. 
    As $R_{\textrm{in}}$ decreases, the $v_{\textrm{freefall}}$ also decreases. 
    {$R_\mathrm{in}$ varies with the accretion rate}, and $v_{\textrm{freefall}}$ thus acts as a coupling parameter between the magnetospheric accretion and the evolution of the inner disk. In equations \ref{eq:density} and \ref{eq:numberdensity_total}, $v_{\textrm{freefall}}$ is used to determine the $n_e$ for the hydrogen slab model. 
    \end{itemize}

\subsubsection{Predicting the Evolution of Observable Radiative Properties}   

As $\dot{M}$ increases in our model, the corresponding physical changes in radii, temperatures, and velocities result in increasing radiative flux.  
We follow the total luminosity evolution of the system.  For more direct comparison to observations,
we also present monochromatic (filter-integrated) flux variation at different wavelengths as well as color evolution 
using combinations of optical and infrared filters among $gri$, $JHK$, and $W1, W2$.
In particular, we are interested in understanding how magnitudes and colors change as outbursting YSOs brighten,
and how to interpret outburst behavior that is bluer-when-brighter versus redder-when-brighter 
\citep[e.g.][]{neha2025}.

In the accretion rate profiles, we {increment $\dot{M}$ over 120 ``time steps''.} 
{For all but the linear case, which by construction does not have a peak, the} lightcurve peak occurs at time step $60$. 
The model time steps are not connected to any real unit of time,  but one can keep in mind 
that the {rise time} from quiescence to peak brightness 
of an FU Ori outburst is generally on the order of months to years. 
For each of our assumed five accretion rate profiles, subsections for different time intervals 
contain detailed descriptions of how the increasing accretion rate affects the various radii, temperatures, and velocities.

\begin{figure}[ht]
\plottwo{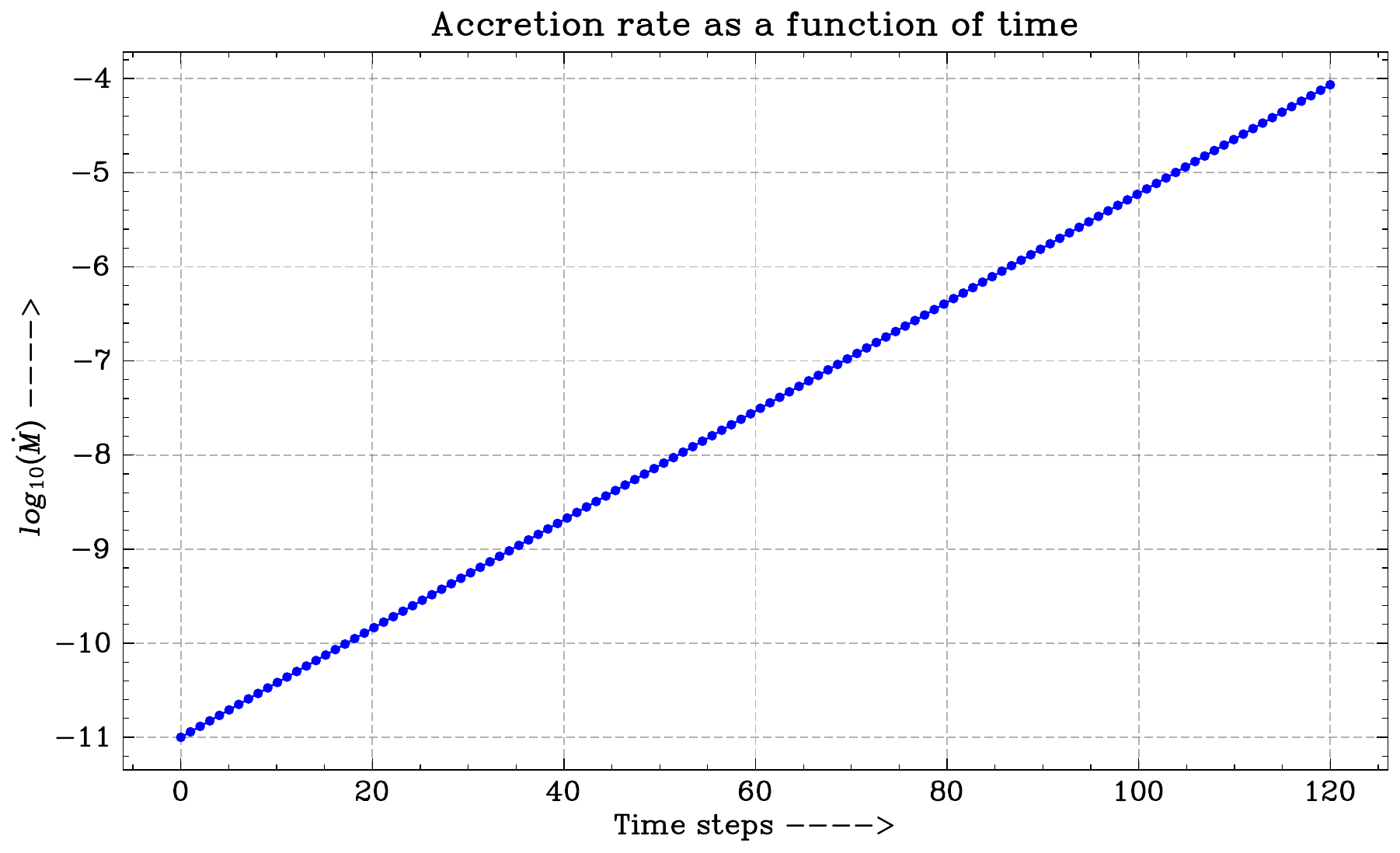}{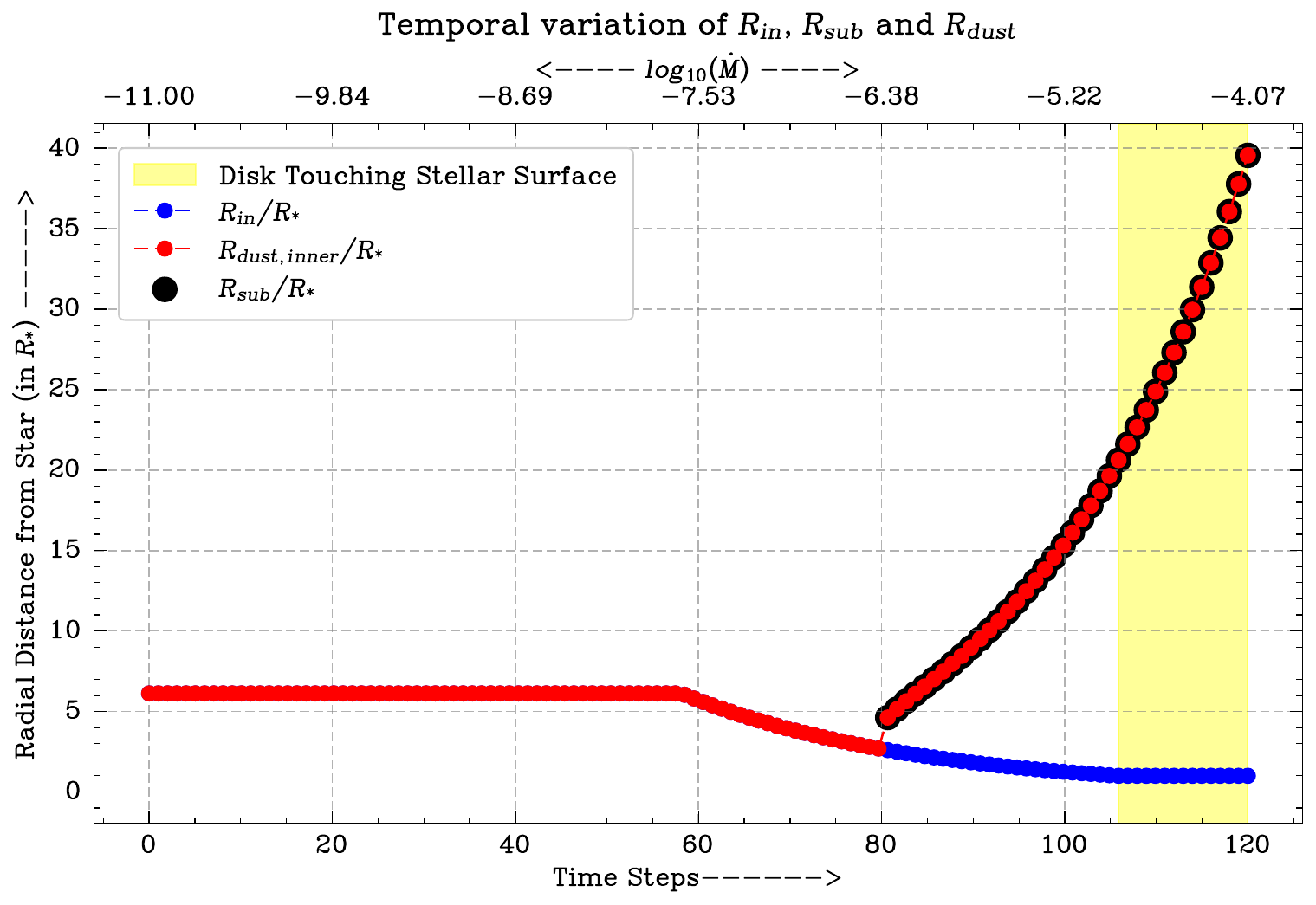}
\caption{{Left:} Assumed variation in disk accretion rate, for the simple linearly increasing case. {Right:} Variation in location of the radii $R_{\textrm{trunc}}$, $R_{\textrm{sub}}$ and $R_{\textrm{dust, inner}}$, as accretion increases. 
Until time step $\sim$ 60, the inner disk boundary ($R_{\textrm{in}}$) remains at the co-rotation radius. 
The inner boundary of the dust disk ($R_{\textrm{dust,inner}}$) is equal to the inner (gas) disk boundary, 
since the increasing viscous temperature  remains $\leq T_{\textrm{sub}}$. 
During time steps [60,80], $R_{\textrm{in}}$ starts moving inwards, thus $R_{\textrm{dust, inner}}$ also moves inwards. In time steps [80,105], the viscous temperature reaches $1400$ K, and only a gaseous disk can be sustained at the innermost disk radii. As the accretion rate continues to increase, the 1400 K isotherm moves outwards, thereby moving the $R_{\textrm{dust, inner}}$ outwards. The $R_{\textrm{in}}$ continues to move towards the stellar surface until it touched the stellar surface at time step $\sim$ 105.
}
  \label{stline_mdot_rin}
\end{figure}

\section{Case Study of a Linearly Increasing Accretion Rate}\label{sec:linear}

In this section, we illustrate our modeling results for the simplest possible case of an accretion outburst.
We suppose a log-linearly increasing accretion rate that rises from a minimum value of $10^{-11}\ M_{\astrosun} \textrm{yr}^{-1}$ to a maximum value of $10^{-4}\ M_{\astrosun} \textrm{yr}^{-1}$, 
as illustrated in  Figure~\ref{stline_mdot_rin} (left).
This basic profile can be used to develop intuition regarding how contributions from each system component (photosphere, magnetospheric accretion, gas disk, dust disk) change in response. It is also useful for appreciating the rate at which certain diagnostic parameters like $R_{\textrm{in}}$, $R_{\textrm{sub}}$, $T_{\textrm{hotspot}}$ evolve as accretion strength increases. 
{In our simple model, the increase of $\dot{M}$ takes place simultaneously at all radii,
rather than originating at a certain point in the disk and propagating through the disk and magnetosphere.}

{For our linear accretion case, the remaining model parameters are those of} the V960 Mon disk, as 
provided in Table \ref{tab:fid_parameters}.
The evolution of $R_{\textrm{in}}$, $R_{\textrm{sub}}$, and $R_{\textrm{dust,inner}}$
is shown in  Figure~\ref{stline_mdot_rin} (right).   
Disk temperature profiles $T_{visc}(r)$ and $T_{\textrm{eff, dust}}(r)$,
and their evolution {as accretion increases,} appear in Figure~\ref{st_line_temp_heatmap}.
The evolution of $T_{\textrm{hotspot}}$ and $v_{\textrm{freefall}}$  is shown in  Figure \ref{stline_hotspot_lumi}.
Finally, we display the light curves and color curves produced by the model in Figure \ref{stline_lightcurve_color} .

For this first case of the log-linear rise in accretion, there are four time periods worthy of discussion,
detailed in the subsections below. 

\begin{figure}[ht!]
\includegraphics[width=0.5\linewidth]{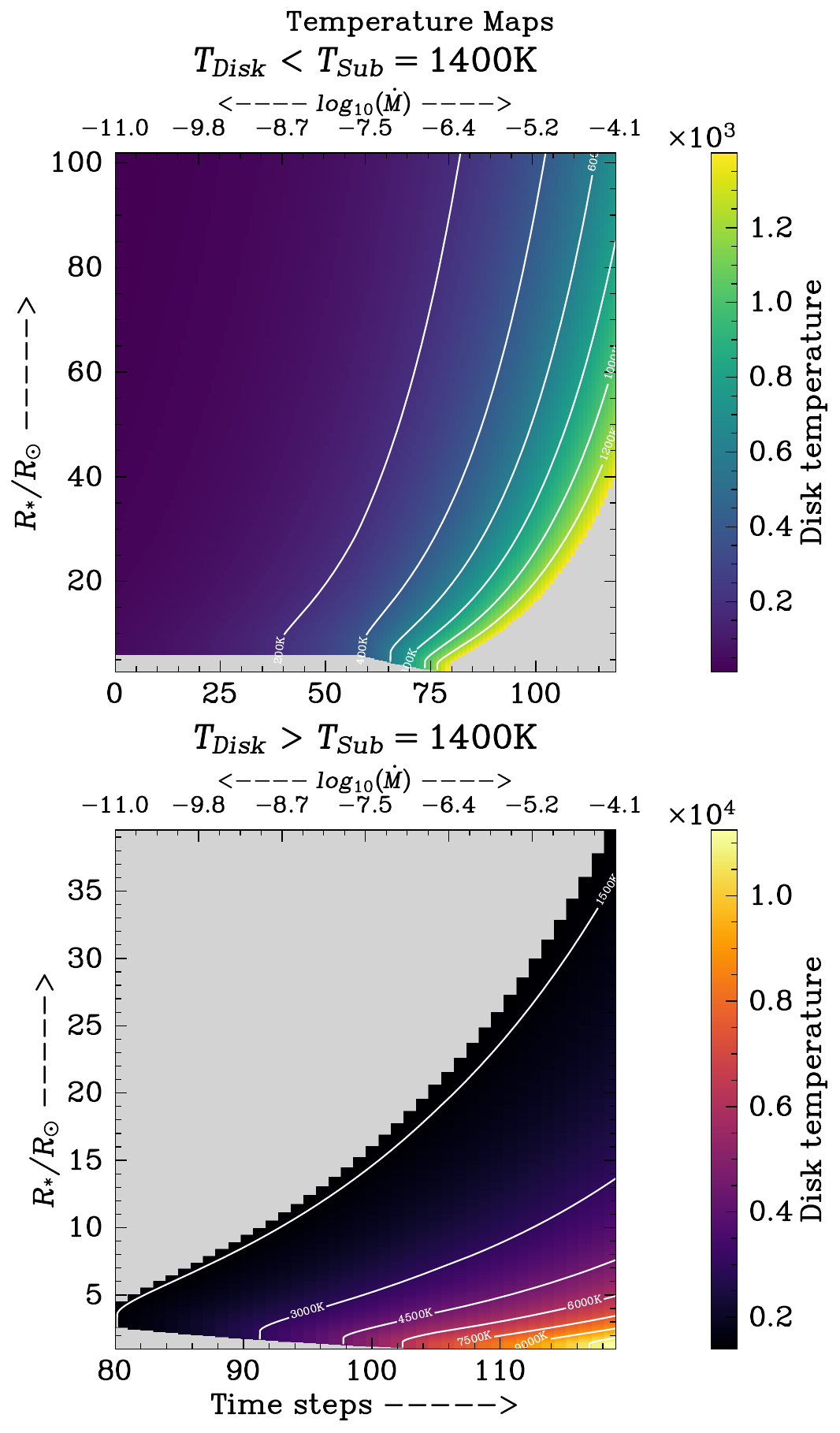}
\caption{Temperature in disk midplane as a function of radius (vertical axis) and accretion rate (horizontal axis), in the simple case of a linearly increasing accretion rate. 
The gray regions represent where either the {dust disk (top panel) or gas disk (bottom panel)} is not present. Contours show how the isothermal surfaces evolve as the accretion rate rises with time.
}
  \label{st_line_temp_heatmap}
\end{figure}

\begin{figure}[ht!]
\plottwo{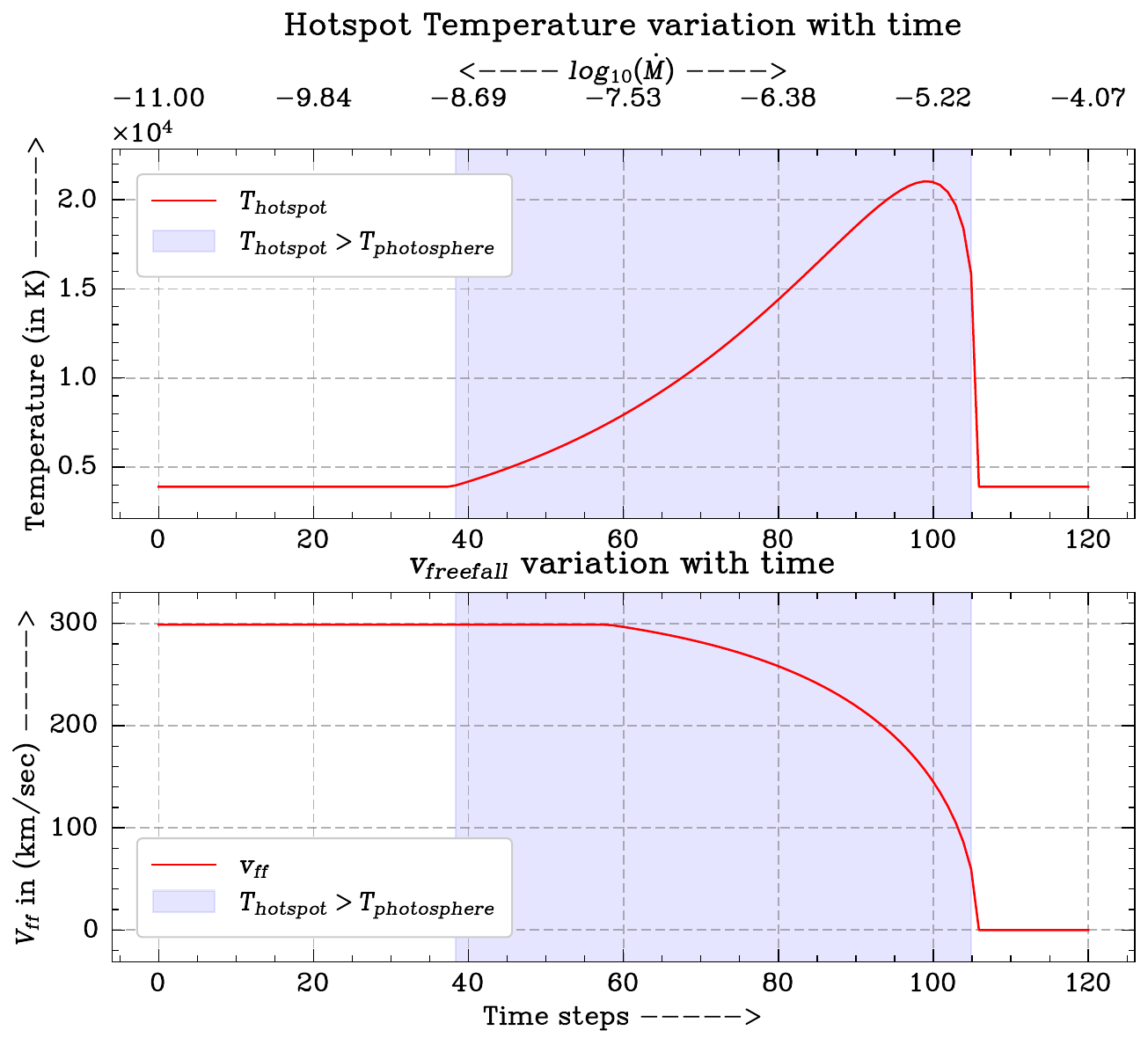}{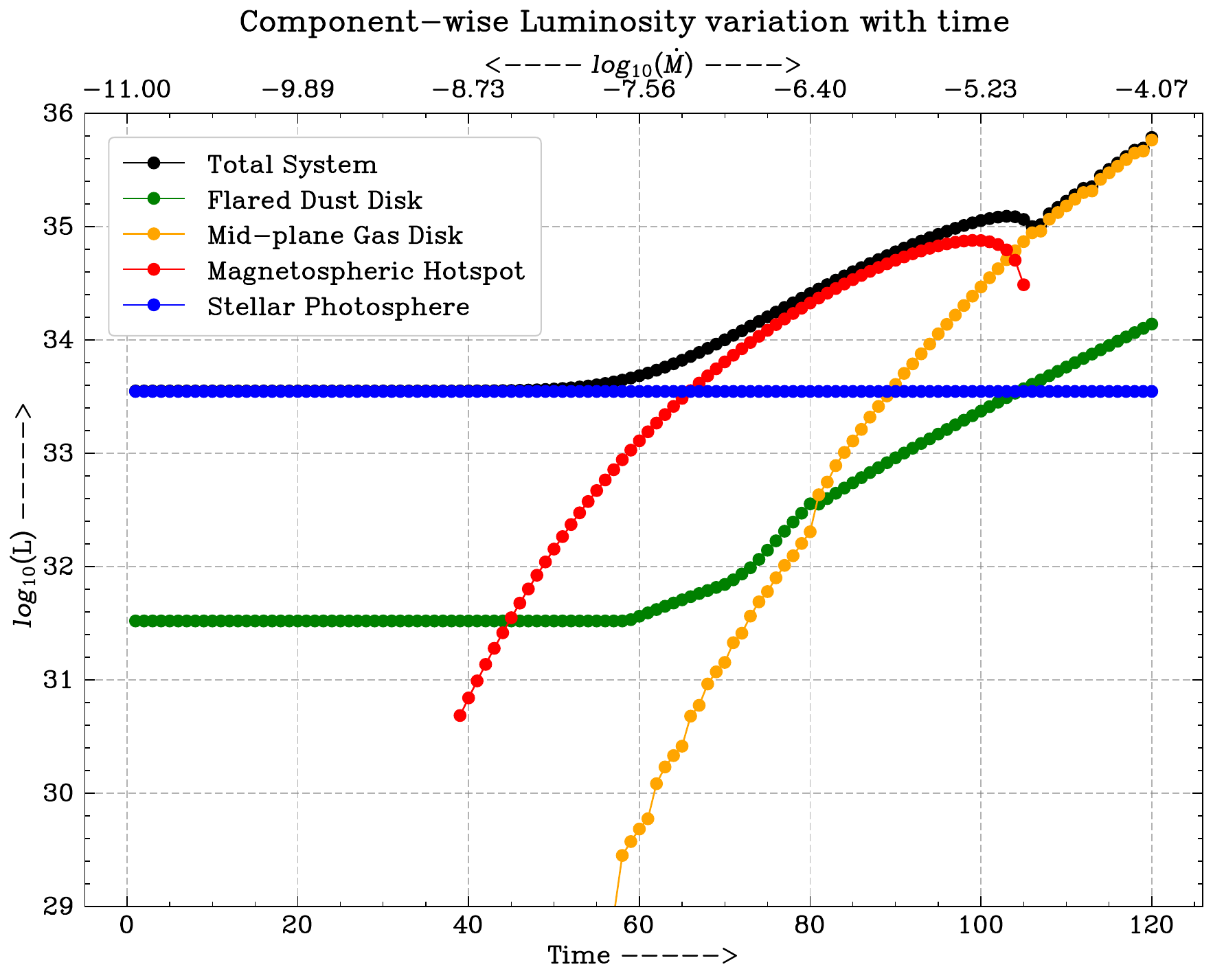}
\caption{
For the simple case of a linearly increasing accretion rate, 
{Left}: Temperature of the accretion hotspot on the stellar surface, and freefall velocity of matter from the inner truncation boundary, $R_{in}$, onto stellar surface. {Right}: Luminosity {evolution for each emitting} component over the course of outburst.  
}
  \label{stline_hotspot_lumi}
\end{figure}

\subsection{Prolonged Low-State Behavior: Time Steps 
{[0, 40]}}
During the first {third}
of the modeled time period, the accretion rate increases from $10^{-11}\ M_{\astrosun} \textrm{yr}^{-1}$ to 
$1.6 \times 10^{-9}\ M_{\astrosun} \textrm{yr}^{-1}$. (Figure \ref{stline_mdot_rin}, left). Because the accretion pressure is very low, $R_\mathrm{trunc} > R_\mathrm{co}$, so the inner disk boundary $R_{\textrm{in}}$ stays fixed at the star's co-rotation radius (see Equation \ref{eq:in_disk_boundary}). Additionally for lower accretion rates, $T_{\textrm{hotspot}} \leq T_{\textrm{photo}}$, so we treat the poloidal accretion activity is switched \textit{off}. During this period, the highest temperature in the disk is lower than the dust sublimation temperature, so $R_{\textrm{dust,inner}} = R_{\textrm{in}} = R_{\textrm{co-rot}}$ and dust and gas components share an inner boundary, as shown in Figure \ref{stline_mdot_rin} (right). 

The viscous midplane heating during this period is weaker than the passive heating of the dust disk by stellar irradiation, resulting in a higher maximum dust disk temperature than the gas disk (see Figure \ref{st_line_temp_heatmap}). Because of this, the dust disk has the second highest contribution to the total luminosity after the stellar photosphere, and more than the gas disk (see Figure \ref{stline_hotspot_lumi}, right). Since there is little magnetospheric accretion, there is not enough heating to observe any flux from the magnetospheric accretion component of the model.

The photosphere {still} dominates the total luminosity during these early time steps
(Figure \ref{stline_hotspot_lumi}, right), and thus 
the lightcurves and color curves (Figure \ref{stline_lightcurve_color}) remain constant. 

\subsection{Magnetospheric Brightening: Time Steps [40, 60]}
At these stages with accretion continuing to rise, from $1.6\times10^{-9}\ M_{\astrosun} \textrm{yr}^{-1}$ to $\sim 2.7 \times 10^{-8}\ M_{\astrosun} \textrm{yr}^{-1}$ (Figure \ref{stline_mdot_rin}, left), {the $R_{in}$, $R_{sub}$ and $R_{\textrm{Gas, Dust}}$ values remain constant (Figure \ref{stline_mdot_rin}, right), indicating that the accretion rate is still not high enough to affect the inner boundary. } 
The disk mid-plane is also still cool, and at these low temperatures the gas and dust are both represented by blackbodies.

We observe in Figure \ref{stline_hotspot_lumi} (left) that the temperature of the hotspot increases rapidly during this period, even though the free-fall velocity is remains constant at $\sim 300 \mathrm{km} \mathrm{s^{-1}}$. This rise in the temperature is solely due to increase in the accretion rate.
The hotspot luminosity rises very rapidly in this epoch (Figure \ref{stline_hotspot_lumi}, right) and almost equals in the flux contribution from the photosphere by the end of this epoch.

Finally among the lightcurves (Figure \ref{stline_lightcurve_color}, left), the shortest wavelength u band is seen to show a rapid rise by almost $\sim 4$ mag, followed by optical bands showing small change. The higher wavelength bands almost are inert to the hotspot activity at these accretion rates. Interestingly in the color curves (Figure \ref{stline_lightcurve_color}, right), except the mid-infrared color, optical and infrared colors seem to respond towards the later part of the epoch.

\subsection{Inner Gas Disk Brightening: Time Steps [60, 100]}
In this epoch, the accretion rate rises from $\sim 2.7 \times 10^{-8} M_{\astrosun} \textrm{yr}^{-1}$ to $\sim 6 \times 10^{-6} M_{\astrosun} \textrm{yr}^{-1}$ (Figure \ref{stline_mdot_rin}, left). The evolution of $R_{in}$, $R_{sub}$ and $R_{\textrm{Gas,Dust}}$ (Figure \ref{stline_mdot_rin}, right) indicates that the accretion pressure has increased so much that it is pushing the truncation boundary to a smaller radius until time step of 80. The disk mid-plane is still cool, and at these low temperatures the gas and dust are both represented by blackbodies. However, as seen in Figure \ref{st_line_temp_heatmap} (left), as the accretion rate increases, the disk's inner boundary temperature rises much faster and reaches $1400$ K by time step of 80, which affects the dust disk's boundaries. This is represented by formation of a gap in Figure \ref{st_line_temp_heatmap} at time step of 80. However from time step of 80-100, the accretion heating warms the disk mid-plane above $1400$ K such that stellar atmosphere models can reasonably represent the disk spectrum. 
During this time interval, $R_{\textrm{in}}$ decreases as the accretion rate pushes the truncation boundary inwards towards the star, while $R_{\textrm{dust,inner}}$ moves outwards because of enhanced accretion heating which moves the $1400$ K isotherm to larger radii (Figure \ref{stline_mdot_rin}, right). 
This is also seen in Figure \ref{st_line_temp_heatmap} (right), where in the time step window [80,100], the boundaries of the viscous disk move inwards as well as outwards.

The temperature of the hotspot and the freefall velocity plots show a monotonic behavior in this epoch (Figure \ref{stline_hotspot_lumi}, left). Notably, the freefall velocity is monotonically falling because the inner boundary of {viscous} gas disk ($R_{in}$) is moving inwards.  

Figure \ref{st_line_temp_heatmap} illustrates that in these epochs, the gaseous {viscous} disk temperature 
exceeds the temperature of the flaring dust disk, and the disk dynamics at this point become dominantly driven by viscous heating. 
As the accretion rate increases, the disk maximum temperature keeps increasing. The viscous disk is modelled by annuli of temperature $T_{\textrm{visc}}$ with the spectrum given by the BT Settl atmosphere models when $T_{\textrm{visc}} \geq$ 1400 K and by blackbodies when $T_{\textrm{visc}} < $ 1400 K, mainly at the larger annuli. We highlight where the switchover happens since it can affect the predicted spectrum. 

There is a transition from photosphere-dominated luminosity at the start of the period, to magnetospheric accretion dominated luminosity and finally to gas disk dominated luminosity. Figure \ref{stline_hotspot_lumi} (right) shows that the maximum flux from the system comes, for a brief period, from the magnetospheric accretion. The total flux is not dominated over this period by any single component, and it is the only period where there is a significant contribution coming from magnetospheric accretion. 
At other times, the hotspots flux is overwhelmed by other components of the system.
By the end of this epoch, the total luminosity is contributed mostly by the {viscous} gas disk (yellow curve).

The luminosity changes also manifest in Figure \ref{stline_lightcurve_color} (left), where all lightcurves show a rise in brightness. As the rate of rise in $T_{\textrm{hotspot}}$ slows (Figure \ref{stline_hotspot_lumi}, left), this affects primarily the $u$-band lightcurve. The colors, show a rapid changes across all bands during the initial periods due to transition from photosphere dominated flux to magnetosphere dominated flux. However in the later part of the epoch where the transition occurs between magnetospheric accretion dominated flux to gas disk dominated flux, the color changes are relatively flatter (Figure \ref{stline_lightcurve_color}, right).

\begin{figure}[ht!]
\plottwo{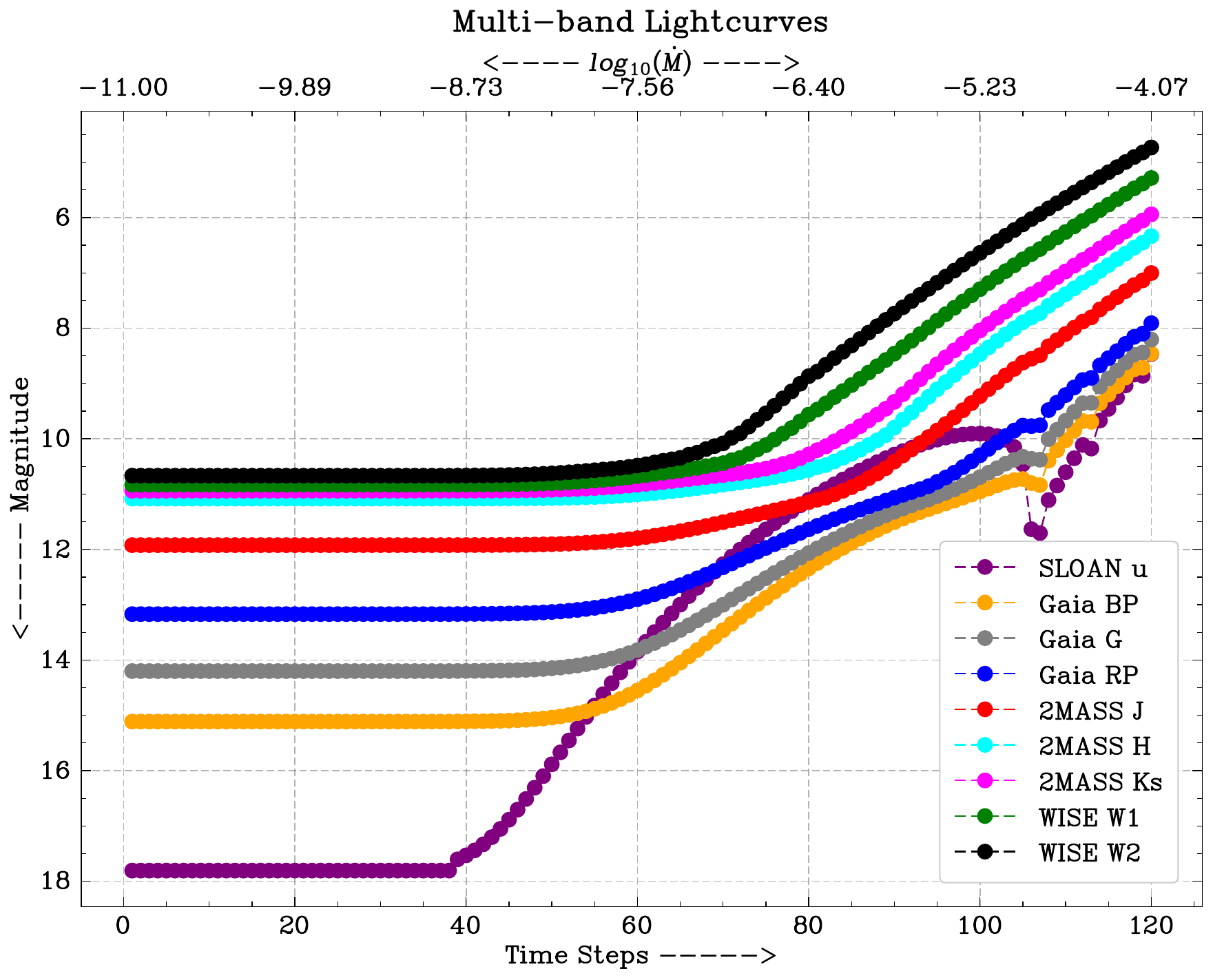}{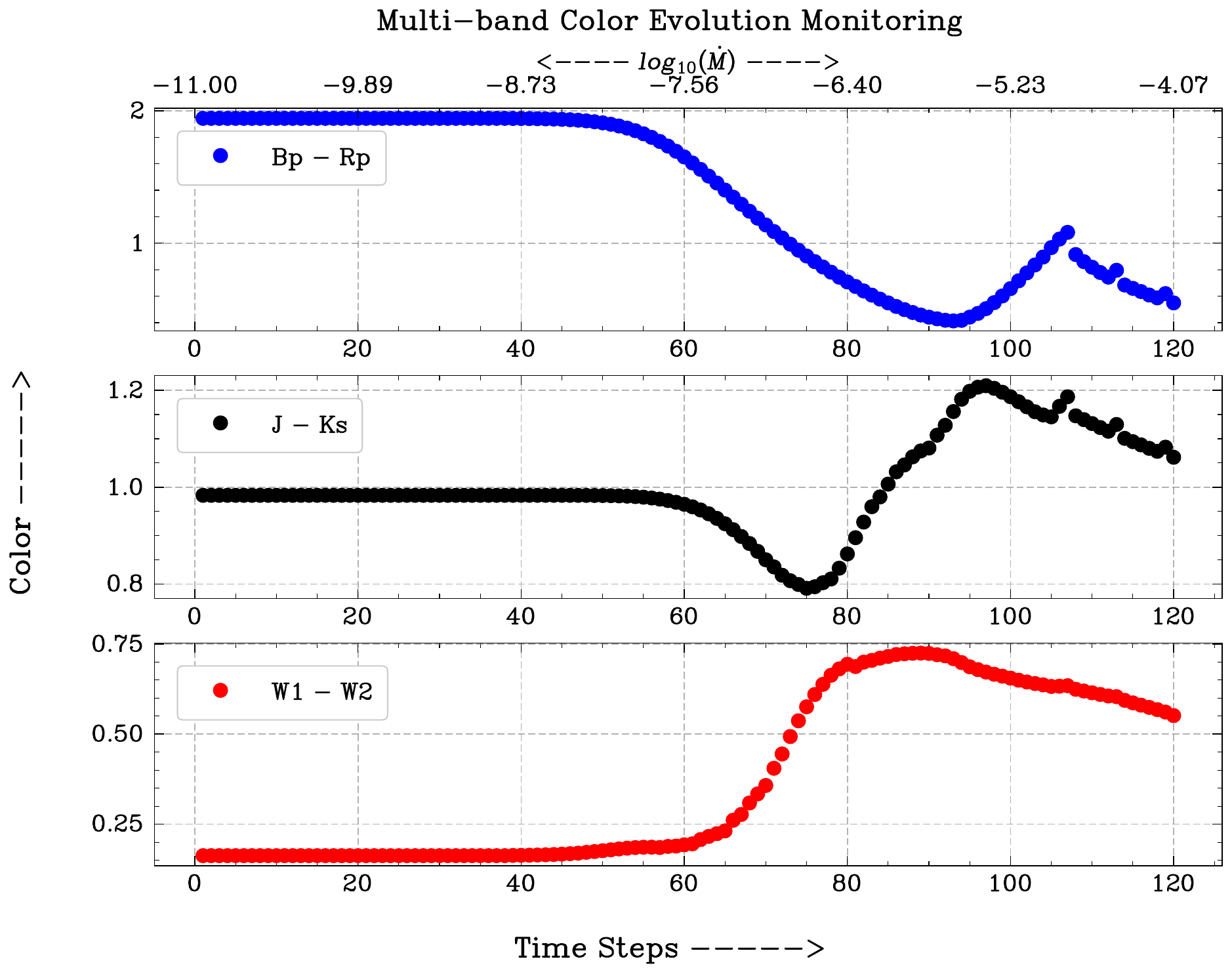}
\caption{{Left}: Multi-band lightcurves constructed by integrating the model high-resolution spectra over various filter bandpasses. {Right}: Color curves constructed from the lightcurves. Both plots are for the simple case of a linearly increasing accretion rate.}
  \label{stline_lightcurve_color}
\end{figure}

\subsection{Continued High-State Evolution: Time Steps [100, 120]}

In this final time period, with accretion continuing from
$6\times10^{-6}\ M_{\astrosun} \textrm{yr}^{-1}$ to the maximum of $\sim 8.6\times10^{-5}\ M_{\astrosun} \textrm{yr}^{-1}$
(Figure \ref{stline_mdot_rin}, left),  we see  that $R_{\textrm{trunc}}$ remains fixed at $R_{*}$ 
(Figure \ref{stline_mdot_rin}, right).  
This is  because the accretion pressure makes the disk reach the stellar surface. 
In Figure \ref{st_line_temp_heatmap} (right), the maximum temperature of the disk can be seen to increase continuously. 
The minimum temperature of the gas disk is also seen to rise but this is because we have assumed
an outer boundary for the viscous disk of 1 AU. 

In Figure \ref{stline_hotspot_lumi} (left) shows the temperature of the shock region starting to turn towards cooler temperatures. 
This behavior illustrates the competing nature of $\Dot{M}$ and $v_{\textrm{ff}}$ in the expression for $T_{\textrm{hotspot}}$ (see Equation \ref{eq:t_hotspot}). If the hotspot is present at all (i.e., $R_{\textrm{in}}(r) = R_{\textrm{trunc}}(r)$ in Equation \ref{eq:in_disk_boundary}), as the accretion rate increases, there is simultaneous increase in $T_{\textrm{hotspot}}$. However when the disk is very close to the stellar surface, $v_{\textrm{ff}}$ starts to dominate over $\Dot{M}$ in Equation \ref{eq:t_hotspot}. This leads to a decreasing $T_{\textrm{hotspot}}$ as the $v_{\textrm{ff}}$ is decreasing, even if the $\Dot{M}$ keeps on increasing.
At even higher accretion rate, the inner disk boundary drops to the stellar radius, and the poloidal accretion activity ends, leading to the shut-down of magnetospheric accretion. 

The luminosity plot in Figure \ref{stline_hotspot_lumi} (right) shows the shut-down of magnetospheric accretion. This is also evident in the multi-band lightcurve plot in Figure \ref{stline_lightcurve_color} (left), where the $u$-band lightcurve is seen to evolve from a smooth curve characterizing the increasing shock emission over time, to a differently shaped curve that increases in brightness similar to the Gaia bands. The change in behavior is due to increasing contribution from very hot gaseous emissions in the inner disk.
The colors (Figure \ref{stline_lightcurve_color}, right)  all become bluer during this period.
We note that the occasional spikes in the color curves are a result of the changing temperature grid towards higher temperatures
in the spectral templates. The incorporated BT-Settl grid \citep{bt_settl_paper} is available at intervals of only 500 K for the higher temperature templates, versus 100 K at the lowest temperatures.

\begin{figure}[ht!]
\plottwo{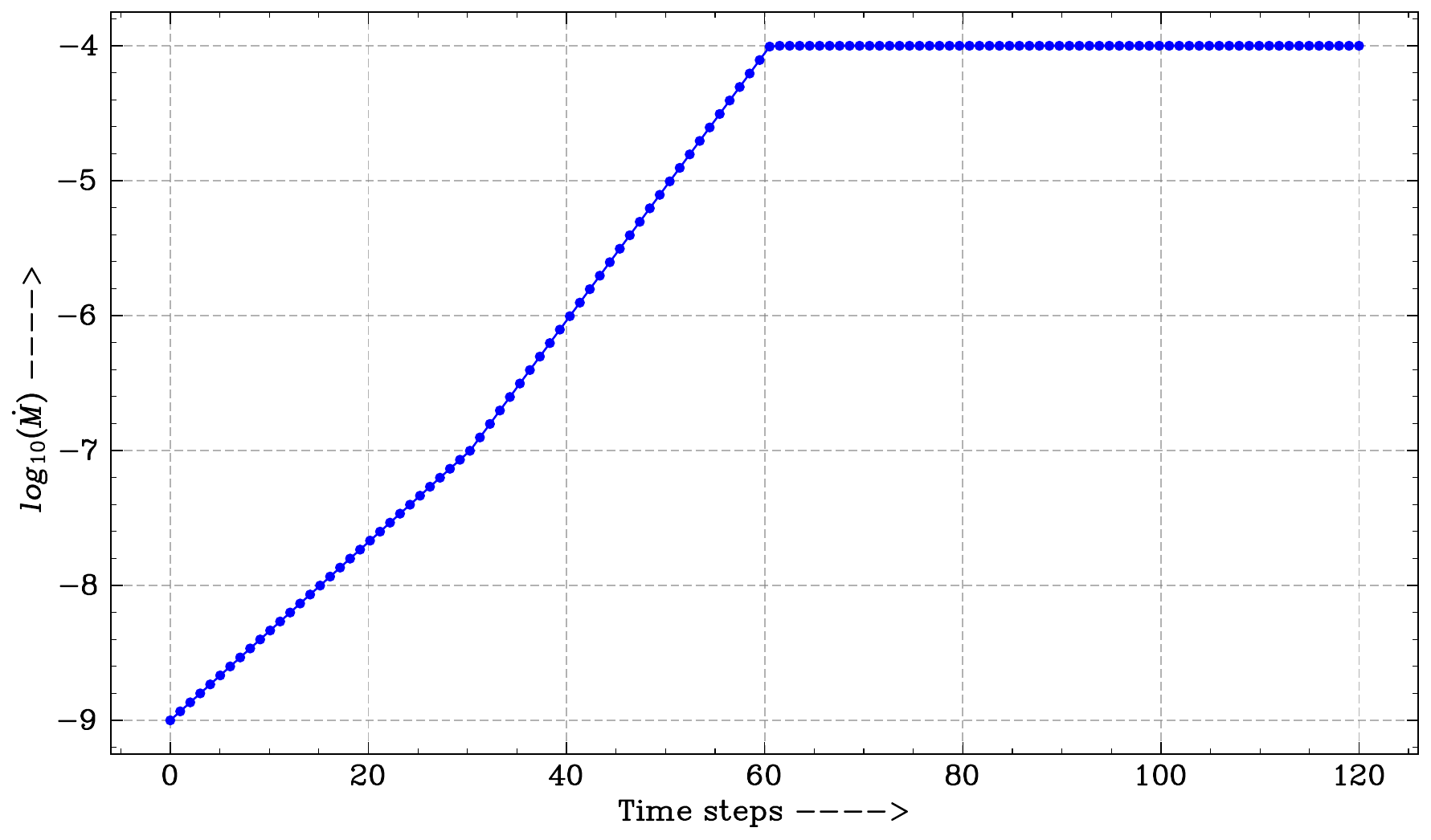}{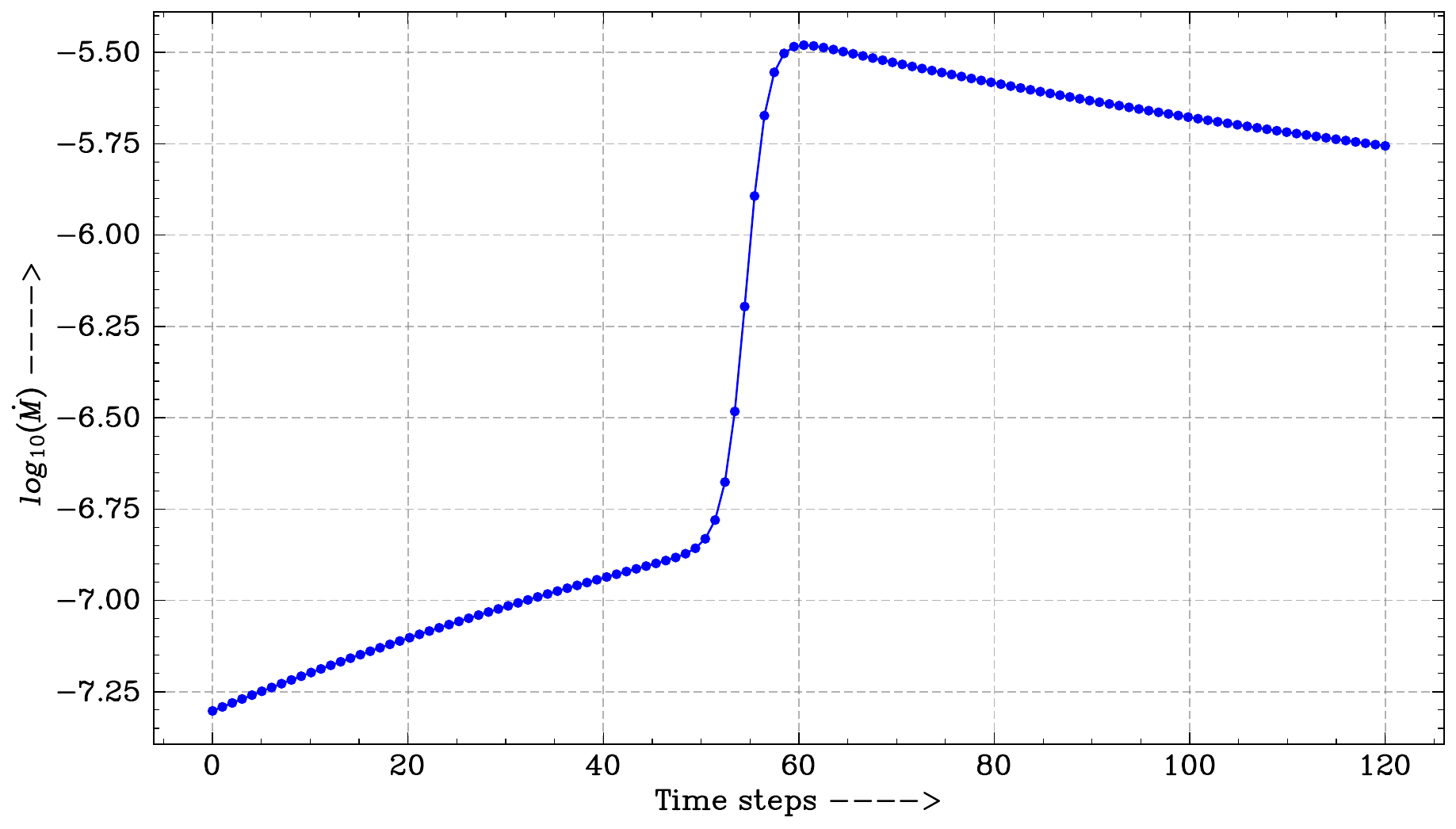}
\plottwo{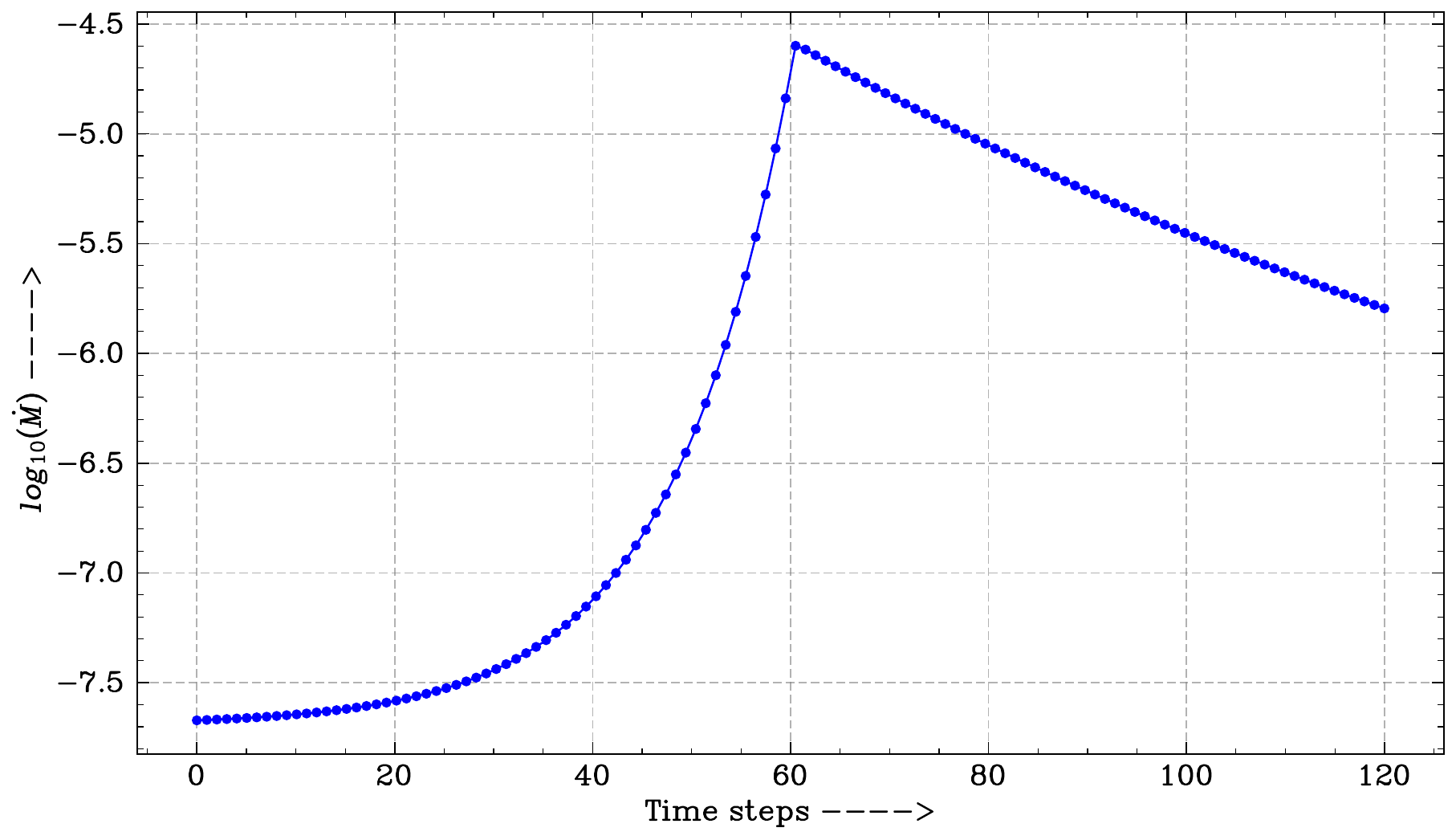}{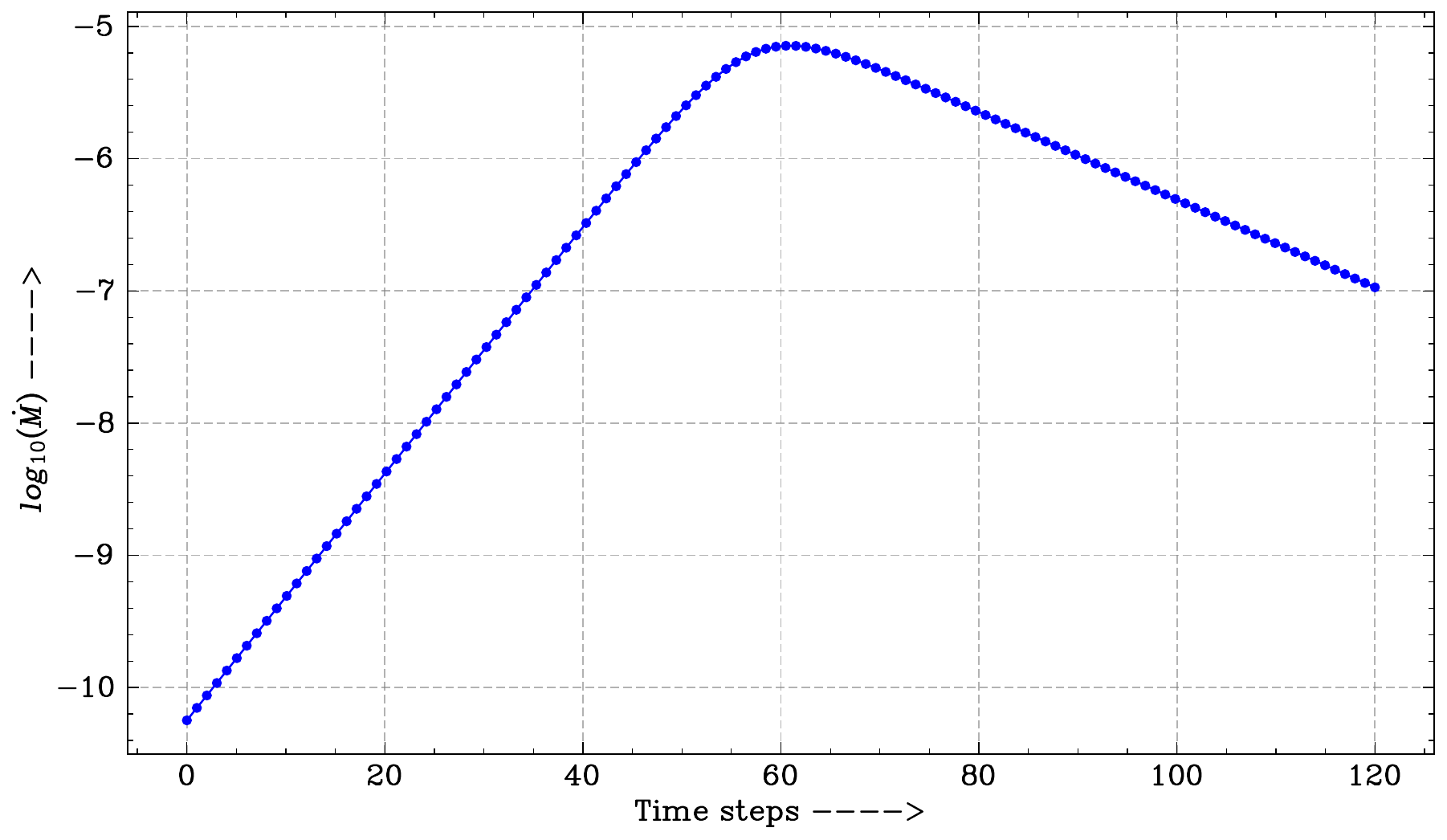}
\caption{Assumed variation in disk accretion rate for {four outburst cases:}
HBC 722-Like {represented as a three-piece linear function (top left), 
V890 Aur-Like represented as a hyperbolic tangent (top right),
V960 Mon-Like represented by a two-sided exponential (bottom left), 
and Gaia 17bpi-Like represented by a logistic gaussian (bottom right)}.
}
\label{mdot}
\end{figure}

\section{Application to Different Accretion Rate Profiles}\label{sec:diff_systems}

In this section, we apply the infrastructure illustrated above for the assumed linearly increasing accretion rate, to explore other forms for the accretion rate profile.  
The adopted mathematical functions are based on observed lightcurves for several recently outbursting YSOs.
Four examples are chosen, each mimicking an actual source that has brightened by a factor of 10-100, and exhibited FU Ori-Like spectral features in its bright state.

\begin{table}[ht]
\centering
\begin{tabular}{l|l|ccccccc}
Accretion Profile & Example& $B$ & $M_{*}$ & $R_{*}$ & $T_{\textrm{photo}}$ & $log(\Dot{M})$ &  $i$ & $d$\\
 & & [kG] & $[M_{\odot}]$ & $[R_{\astrosun}]$ & [K]& [dex $M_{\astrosun} yr^{-1}$] & [deg] & [kpc]\\
\hline
Linear Rise            & \nodata &1.4 & 0.59 & 2.11 & 3900 & {-4}& 15 & 1120\\
Three-Piece Linear     &HBC 722     & 1.4   & 0.20  & 2.80 & 3200& {-4} & 55 & 745 \\
Hyperbolic Tangent     &V890 Aur   &  0.8  & 0.17 &  1.46 & 3100 & {-5.48} & 6 & 1500   \\
Two-Sided Exponential  &V960 Mon   & 1.4 & 0.59 & 2.11 & 3900 & {-4.59} & 15 & 1120\\
Logistic $\times$ Gaussian    &Gaia 17bpi & 2.0   & 0.23  & 0.80 & 3300 & {-5.14} & 86 & 1150\\
\end{tabular}
\caption{Tabulation of the shapes of the accretion rate profiles, example objects having this lightcurve shape, and assumed values for the model input parameters. Magnetic field stengths $B$ are arbitrary, but within the range $1-3$ kG \citep{john_krull_2007_mag_field}, though in the case of V890 Aur, we had to assume a lower $0.8$ kG so that the disk actually reaches the stellar surface during peak outburst. 
Stellar masses $M_{*}$ and $R_{*}$ are from the literature cited in the text.  The slab optical depth parameter $\tau$ is assumed to be 1 in all cases, and the surface area filling factor for hotspots $f$ is taken to be $1\%$ of the total photospheric surface area. 
Rotation period of the star $P_{\textrm{rotation}}$ is assumed to be $7$ days in all cases, in agreement with observations \citep{bouvier_review_rotation_period}}
\label{tab:fid_parameters}
\end{table}

Table \ref{tab:fid_parameters} provides the sources and the set of fiducial parameters for the different systems. 
{For the Gaia 17bpi-Like system, in Appendix~\ref{sec:parametric} we perform parameter studies around our fiducial values of several of the stellar parameters.}

Our motivation for adopting these particular sources is that they illustrate the
range of lightcurves exhibited by different FU Ori accretion outburst systems.
HBC 722 has had limited variation in brightness since reaching its lightcurve plateau \citep[e.g.][]{adolfo_hbc722_paper}. 
It is relatively straightforward to approximate as a three-piece linear accretion profile. 
V890 Aur is one of the most recent additions to the catalog of FU Ori stars, and was fit in \cite{AUR_0544_Hillenbrand_2025} with a hyperbolic tangent function. 
V960 Mon is one of the most well-studied FU Ori objects, with multi-band lightcurves available. 
Its lightcurve rose to a sharp peak, with a subsequent decline to a long-lived high plateau \citep[e.g.][]{adolfo_v960_2},
and can be modelled by a two-sided exponential profile.
Gaia 17pi is the first FU Ori object to show different degrees of brightness change across wavelength \citep{hillenbrand2018},
and experienced a relatively short-lived peak phase, with evidence for a lightcurve decline.

By construction, in our simulations the accretion peak occurs exactly halfway throughout the simulation, 
at time step $60$ within the set [0,120] interval.
We need to make an assumption regarding the value of the peak accretion rate. 
We choose these based on detailed studies in the literature of the particular sources,   
adopting the best fit values from the references cited above.
The minimum or low-state accretion rate is unknown, but 
an initial estimate can be made from the amplitude of the lightcurve rise.  
From this basis, we then conduct iterative trials, feeding the rescaled accretion rate profile 
through the \ysopy pipeline to generate a preliminary lightcurve. 
Although the generated lightcurve peak may be close to the actual lightcurve peak, the low-state plateau can differ from observational constraints. This is rectified by changing the low-state accretion rate, then running the pipeline again to compare the resulting lightcurve. This process is repeated until we are satisfied with the approximate match,
including to any available pre-outburst photometry.  

{We discuss each of the four objects in its own} sub-section below. 
The {accompanying} figures show the HBC 722-Like and V890 Aur-Like accretion profiles side-by-side, 
since they are similar in shape but different in amplitude and peak accretion rate,  
and the V960 Mon-Like and Gaia 17bpi-Like figures together, 
since both sources have evolved since their peaks and had similar peak accretion rates, 
though different morphology both pre- and post-peak.
We re-iterate that we are not claiming to model these particular outburst sources in detail.
We are instead illustrating the consequences of assuming an accretion profile like that of the observed outburst lightcurve. {One aim of this work is to determine which range of wavelengths
mostly closely traces the underlying accretion program in observed lightcurves.}

\subsection{HBC 722-Like Lightcurve}
HBC 722 had a photometric outburst and spectroscopic change consistent with suddenly elevated accretion rates. 
Its lightcurve has the following phases: (i) initial slow rise, (ii) steeper rise in outburst, (iii) dipping brightness immediately post-peak, (iv) gradual re-brightening, and (v) long plateau phase \citep{Miller_PTF10qpf, Kospal_HBC_722}. 

For our model of the abrupt increase in accretion rate in the inner disk region, we  
take fiducial values for the $B$, $M_{*}$, $R_{*}$ and $i$ parameters from \cite{adolfo_hbc722_paper}. 
The distance of $745$ pc to this source is taken from \cite{mike_and_hillenbrand_2020}.
{The assumed accretion profile is shown in Figure~\ref{mdot},}
where we are ignoring the finer structure in the observed lightcurve.
The initial gradual rise, and then the steeper rise to peak occupies the first half of the time steps. 
The plateau phase occurs for the rest of the model duration. 

The left panels of Figures  \ref{r_in_hbc_722_v_890}, \ref{heatmap_hbc_722_v_890}, \ref{hotspot_hbc_722_v_890}, and \ref{lum_lightcurve_hbc_722_v_890}
represent the behavior of the various parameters for this accretion profile.

\subsubsection{Time [0, 40]}

The early accretion rate is very low at the beginning of the outburst epoch (Figure~\ref{mdot}).
Although $\Dot{M}$ is increasing, there is only a small amount of heating and no observable change in the inner disk truncation radius or the sublimation radius (Figure \ref{r_in_hbc_722_v_890}). We should remind the readers that the rise in accretion rate becomes steeper at time step of $\sim 30$. The inner boundary of the disk is the stellar co-rotation radius.
The {viscous gas} disk temperature is below $1000$ K (Figure \ref{heatmap_hbc_722_v_890}). Also, the temperature of the dust disk is below the sublimation temperature. 
At this stage, the dynamics of the disk are driven by passive heating due to photospheric radiation. 

As matter falls from $R_{\textrm{co-rot}}$ onto the stellar surface by magnetospheric accretion, the free-fall velocity is $\approx$ 140 km $s^{-1}$ (Figure \ref{hotspot_hbc_722_v_890}). However due to rise in the accretion rate, the temperature of the hotspot rises above the $T_{\textrm{photo}}$ at time step $\sim 10$ and reaches to about $10000K$ by the end of this period, due to elevated accretion levels through the poloidal accretion funnels.
Figure \ref{lum_lightcurve_hbc_722_v_890} shows that the stellar photosphere dominates the total luminosity during the initial period, but the dominant contribution quickly shifts to hotspot emission by the end of this period.
The rise in flux contribution from hotspots is seen predominantly in the $u$-band lightcurve (Figure \ref{lum_lightcurve_hbc_722_v_890}), which shows a rise happening exactly when the hotspot starts to appear on the stellar surface. The optical bands are the next to show the rise followed by the IR-bands. Interestingly the rise in brightness is direct consequence of the fact that hotspot emission affects the flux at the $u$-bands the most while the low-energy bands are relatively less affected. Also we find that the WISE bands show earlier onset of outburst compared to near-infrared bands. The earlier rise at the longer wavelengths is due to an increase in the effective area of the low-temperature dusty disk, as it is pushed farther from the star.

\begin{figure}[ht!]
\plottwo{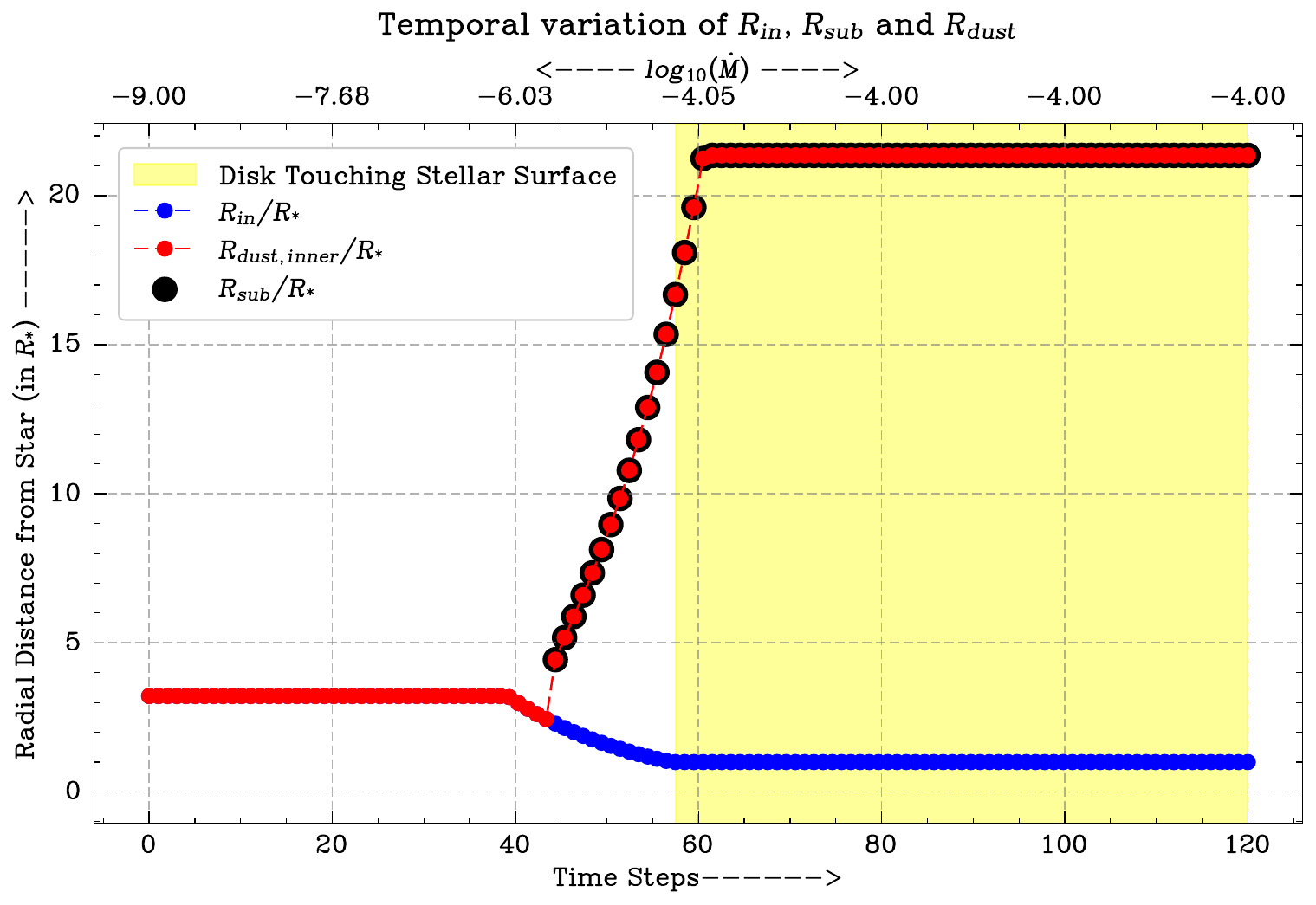}{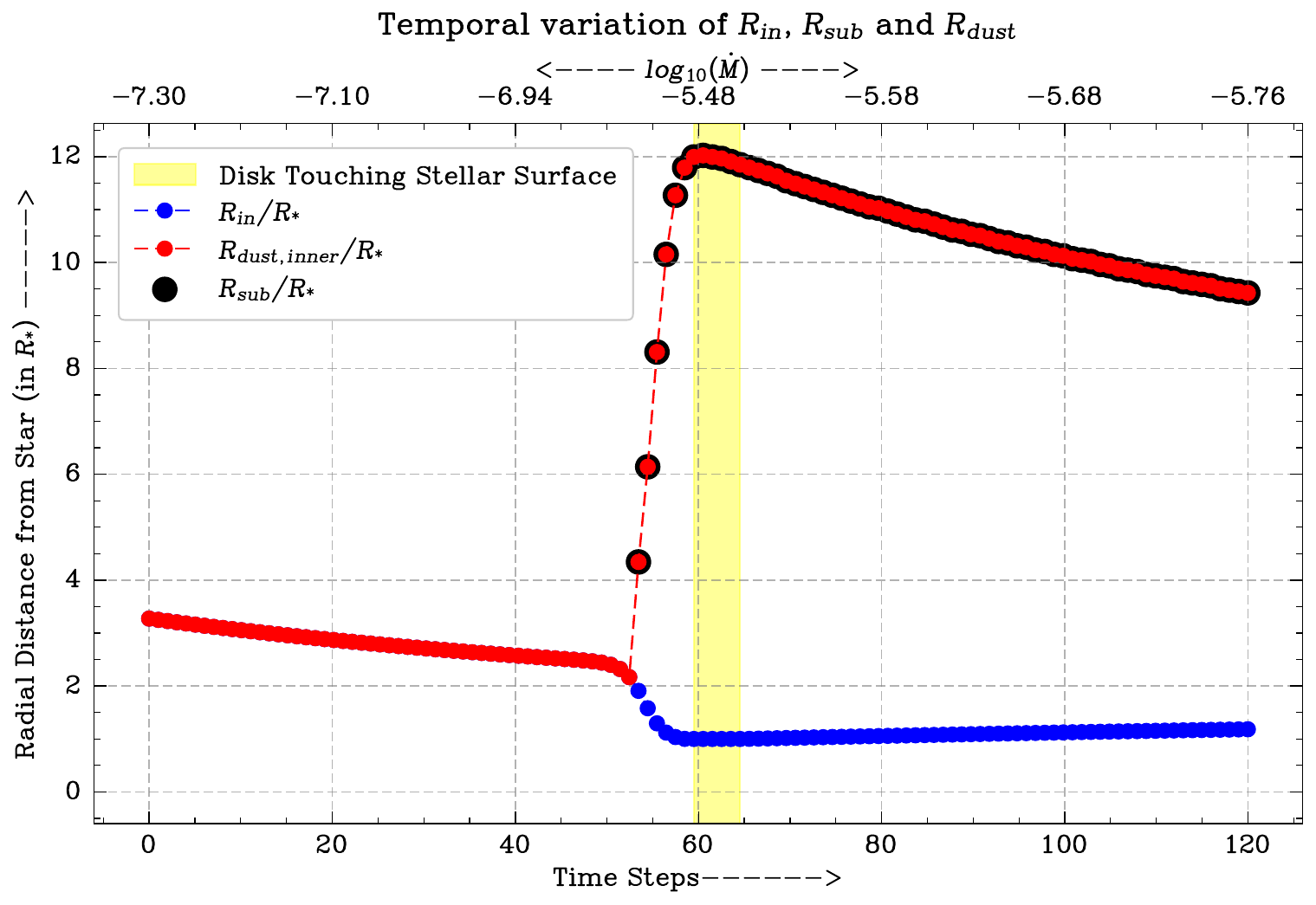}
\caption{Variation in $R_{in}$, $R_{sub}$ and $R_{\textrm{Gas,Dust}}$ for the HBC 722-Like (left) and V890 Aur-Like (right) cases.}
  \label{r_in_hbc_722_v_890}
\end{figure}
\begin{figure}
    \centering
    \plottwo{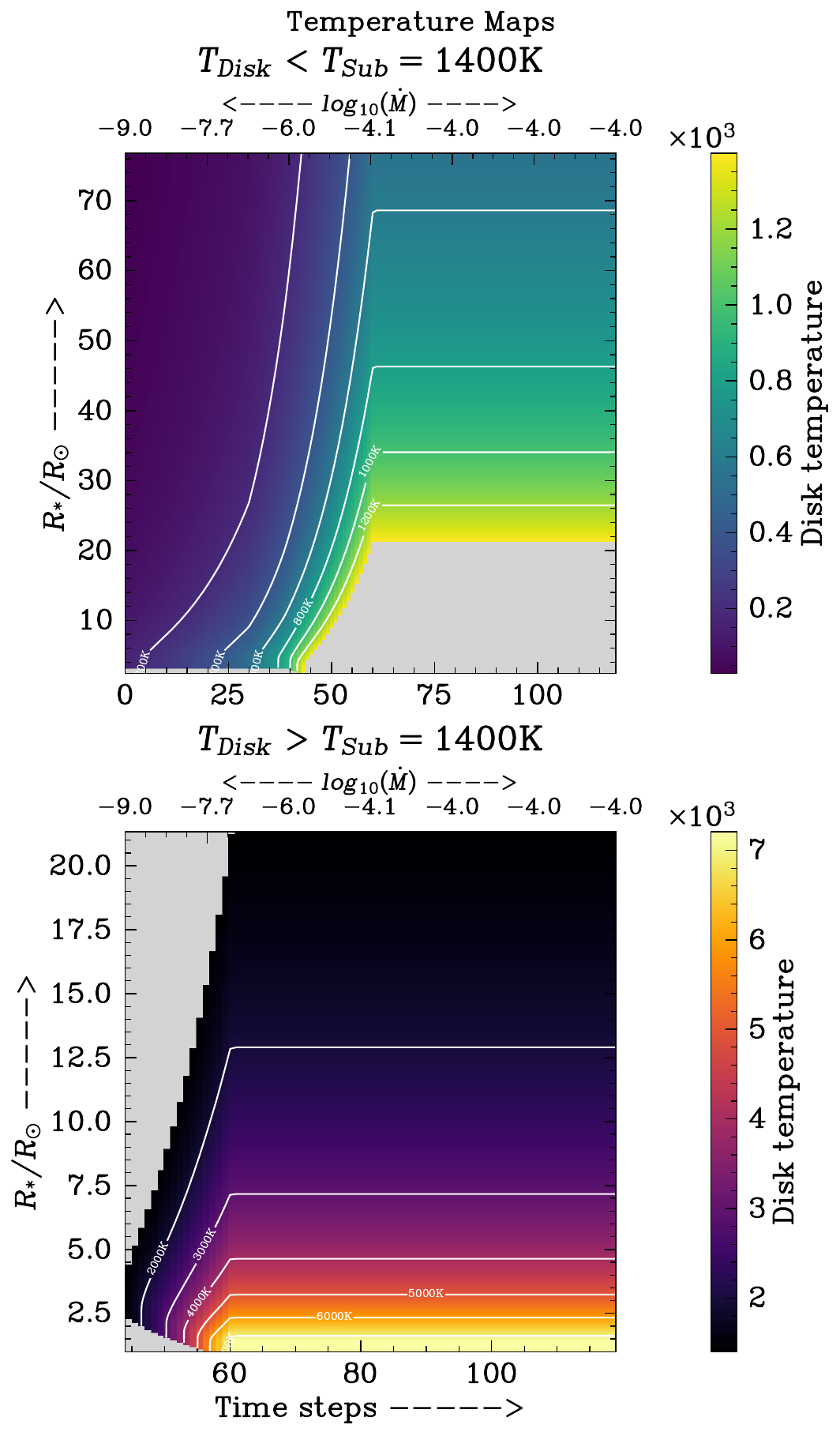}{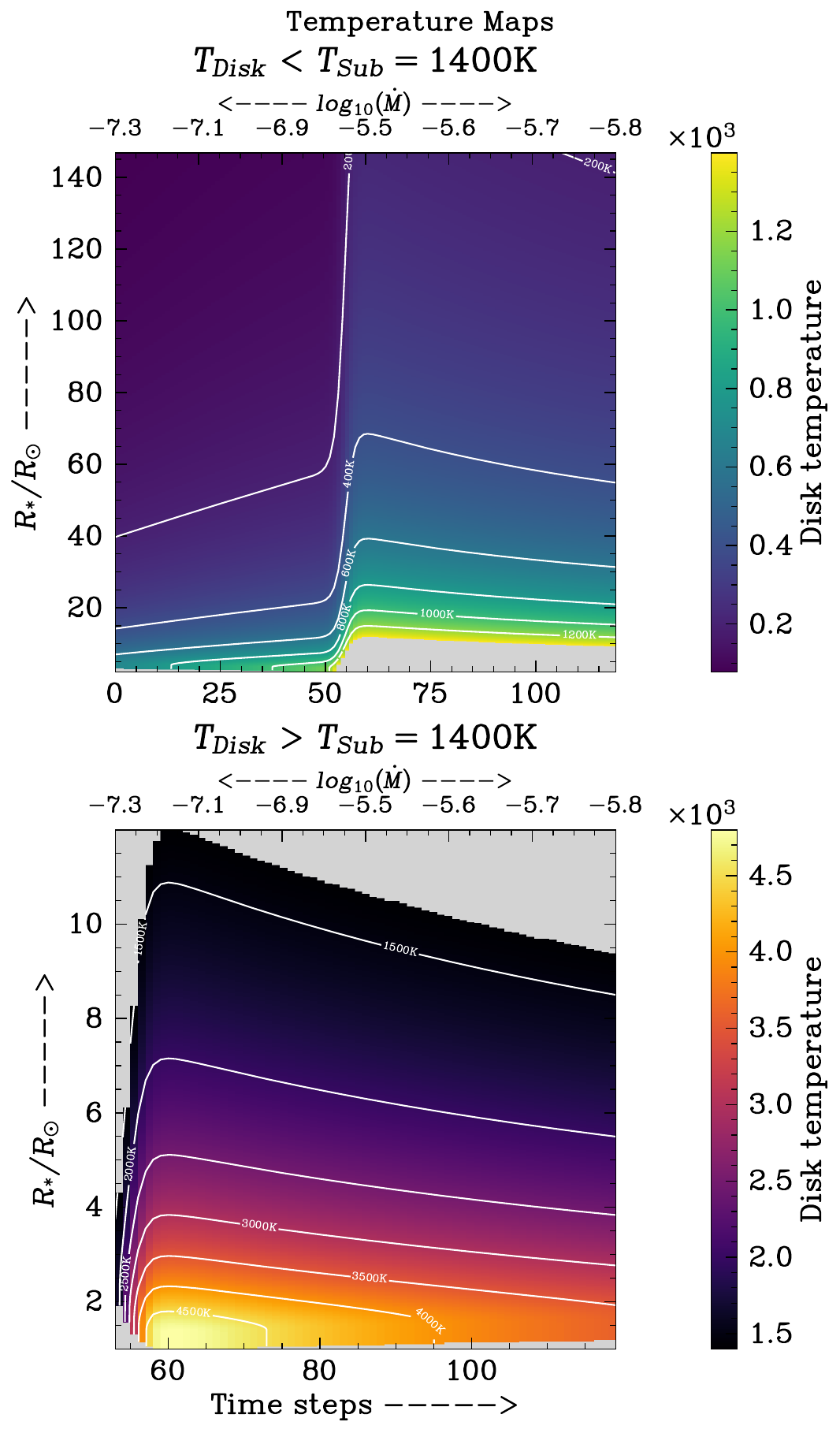}
    \caption{Temperature variation in the {viscous gas} disk {and dust disk} mid-plane for the HBC 722-Like (left) and V890 Aur-Like (right) cases.}
    \label{heatmap_hbc_722_v_890}
\end{figure}

\begin{figure}[ht!]
\plottwo{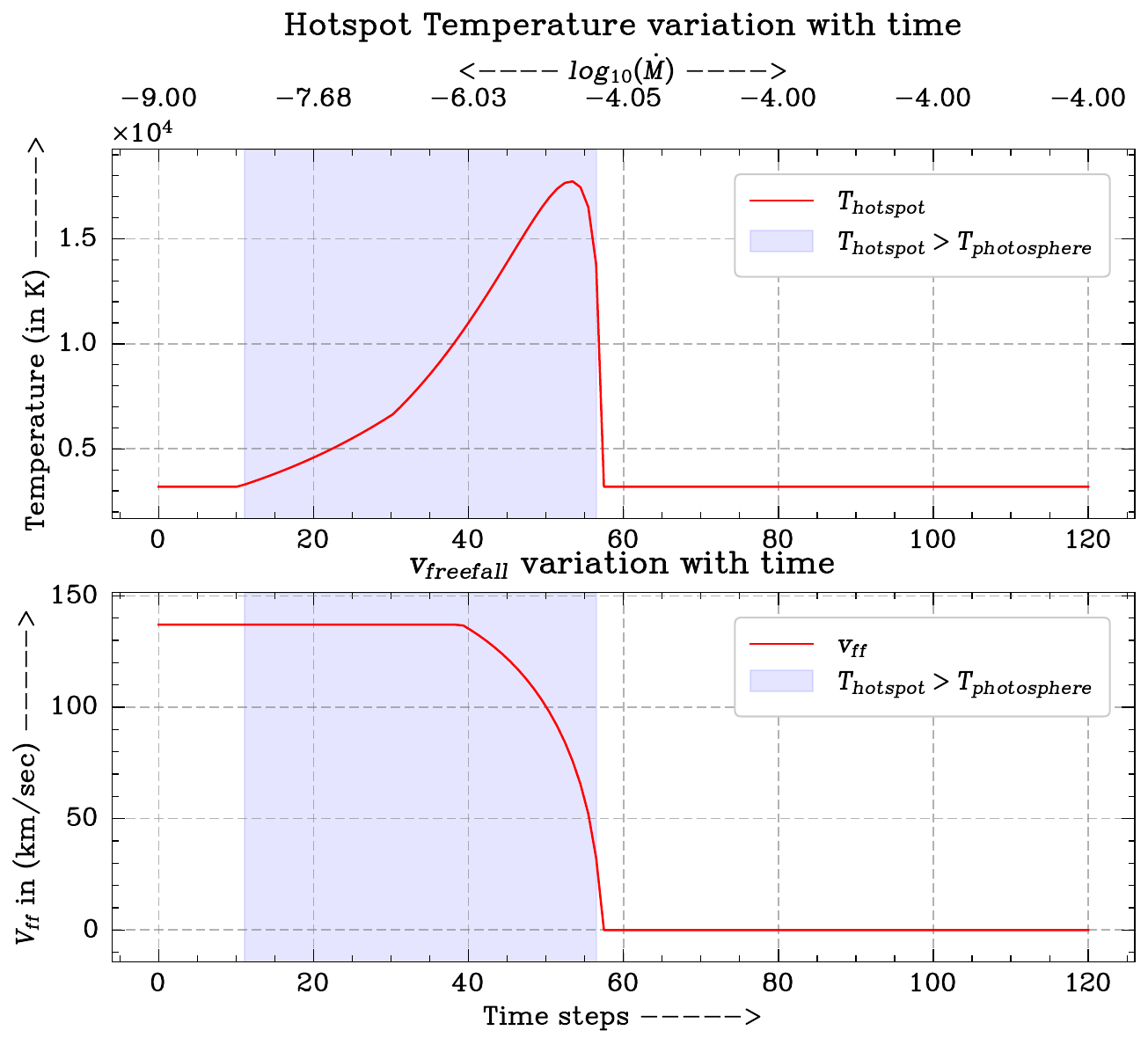}{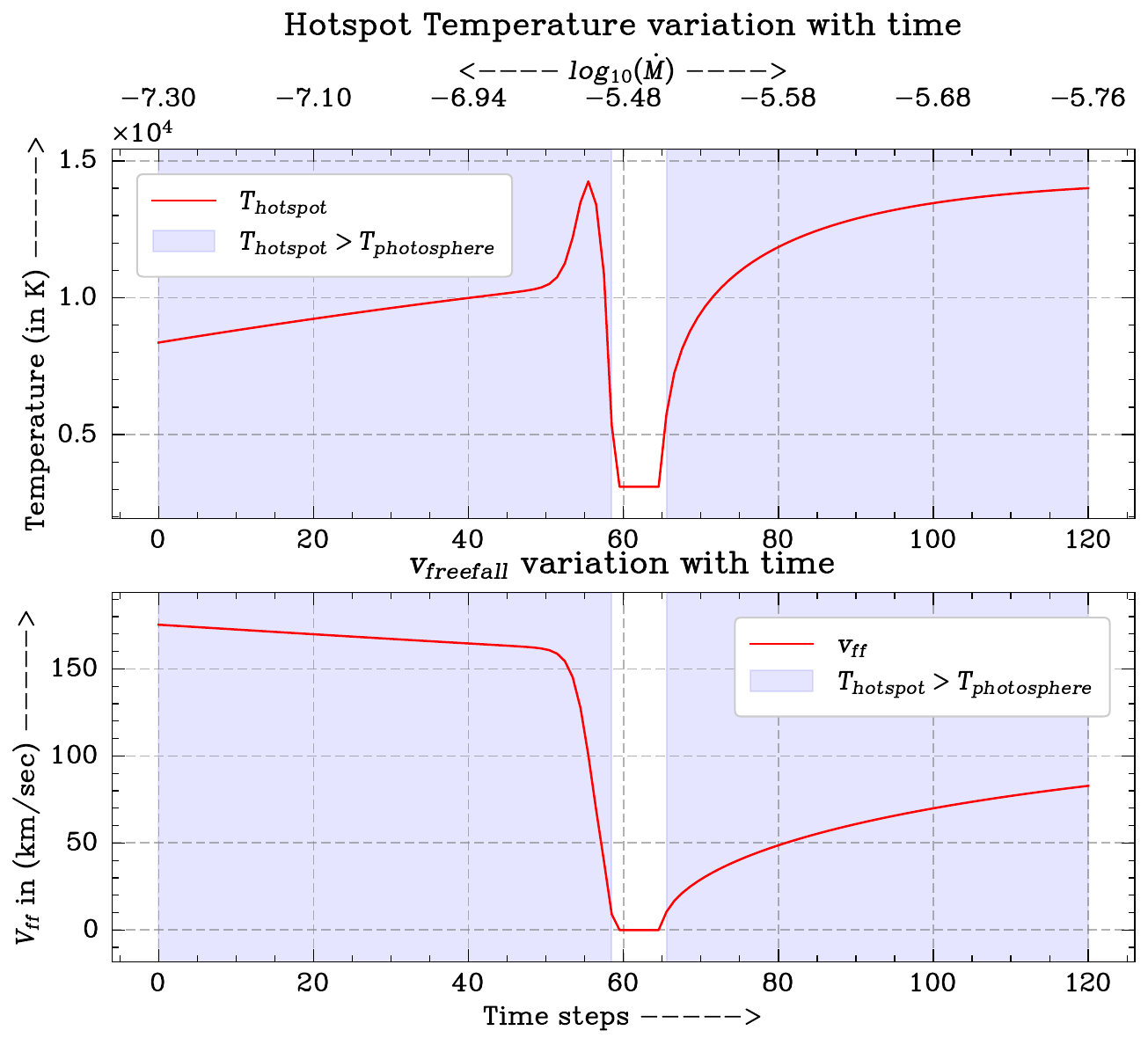}
\caption{Variation in the hotspot temperature and the free-fall velocity at the stellar surface for the HBC 722-Like (left) and V890 Aur-Like (right) cases. }
\label{hotspot_hbc_722_v_890}
\end{figure}

\begin{figure}[ht!]
\plottwo{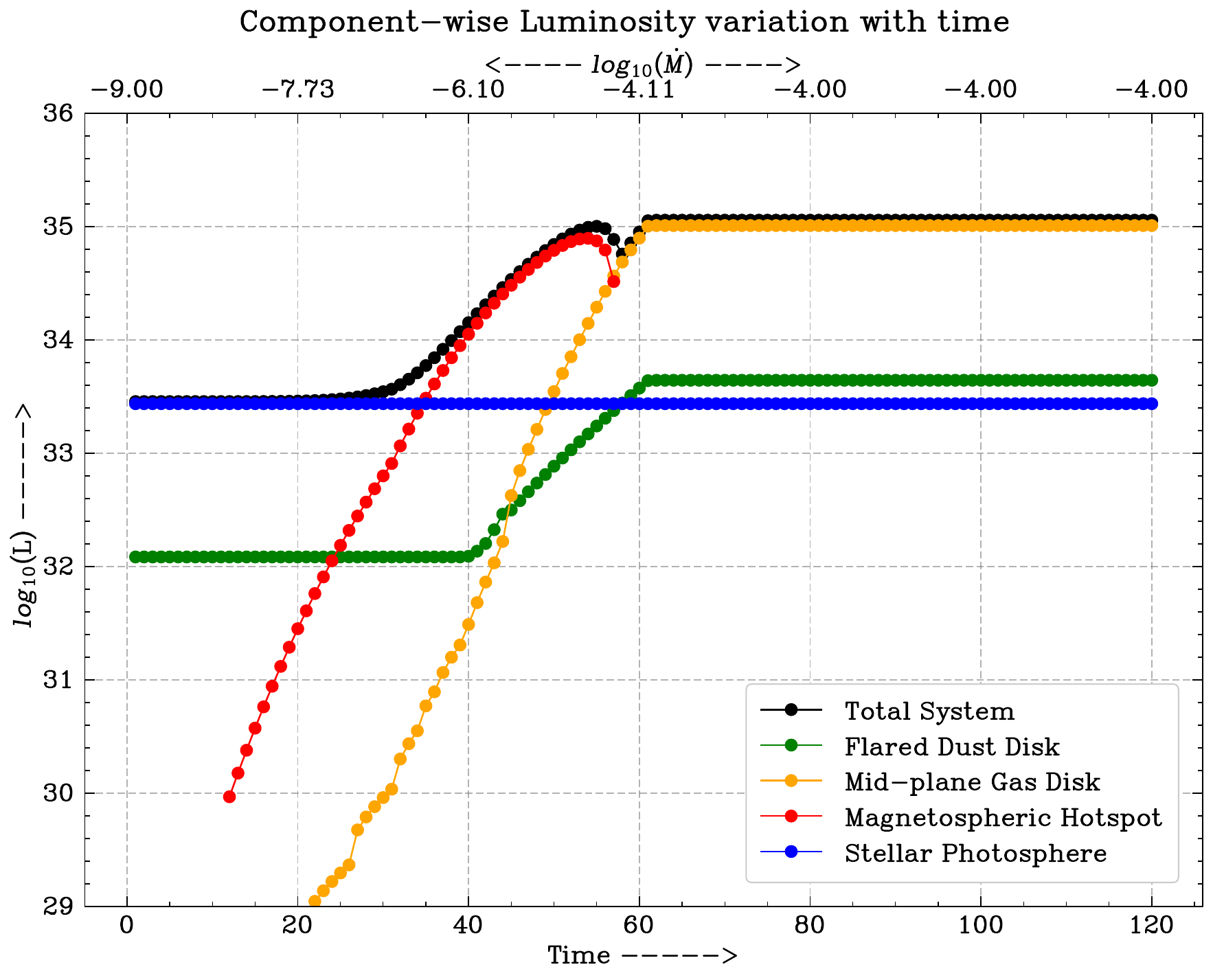}{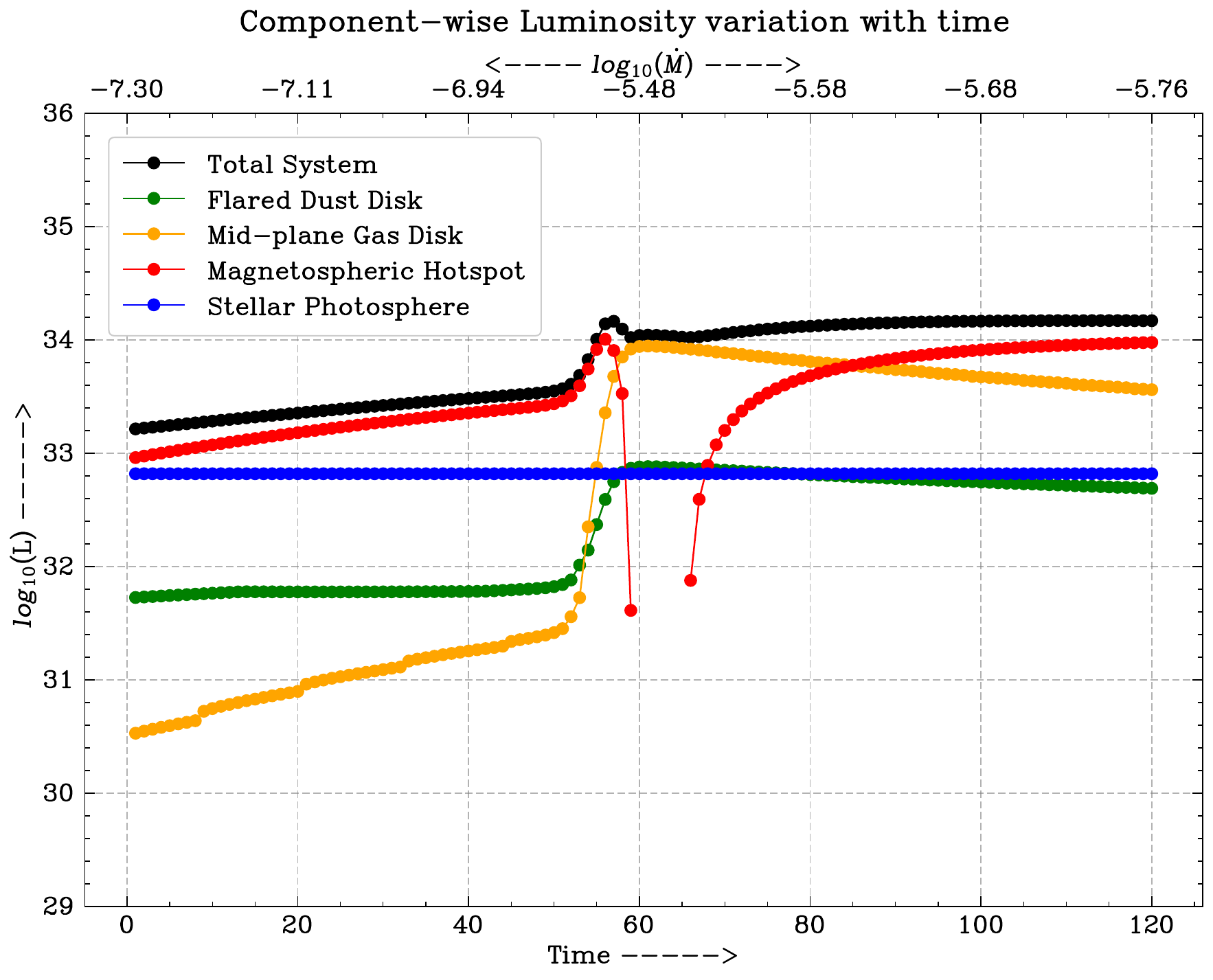}
\plottwo{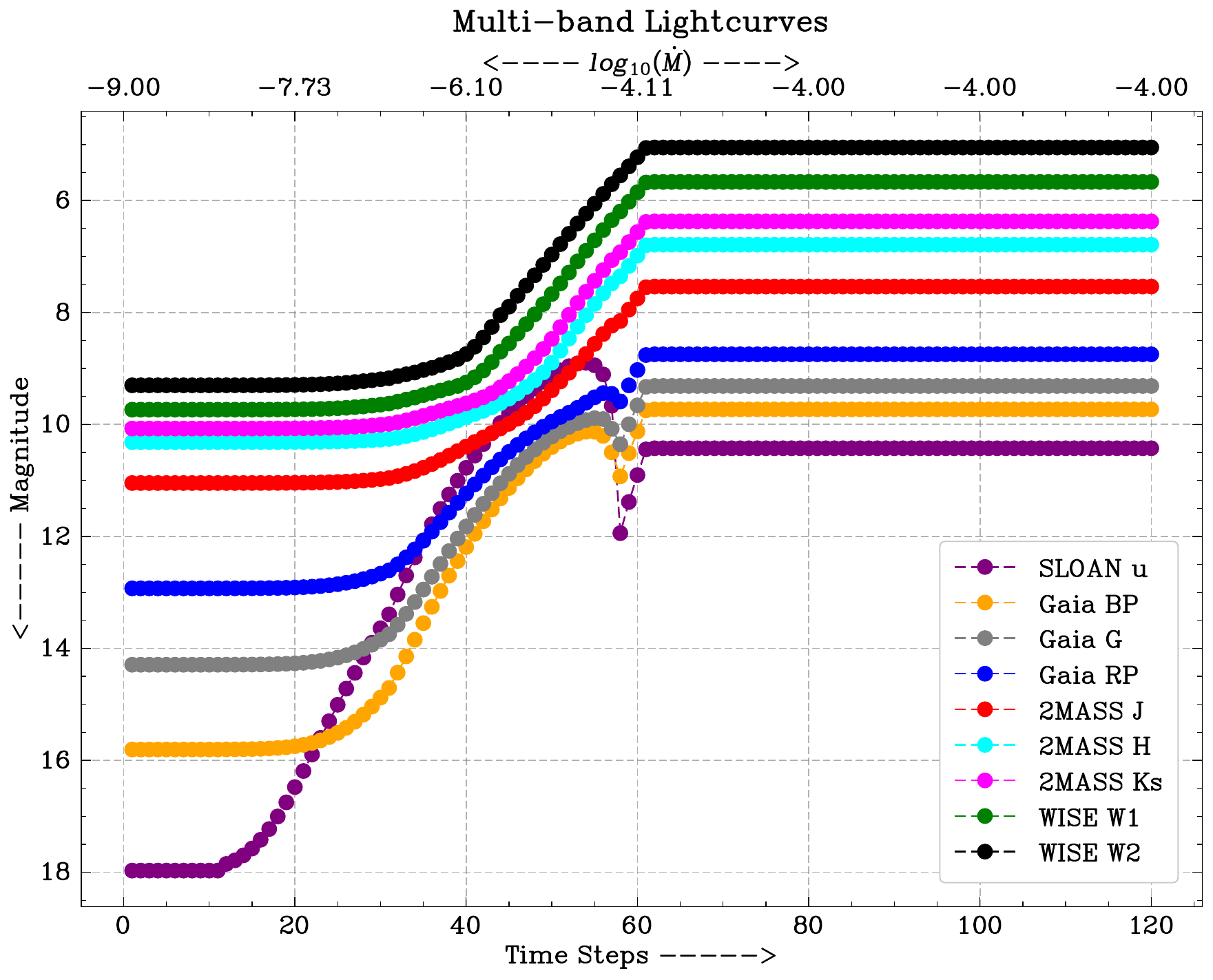}{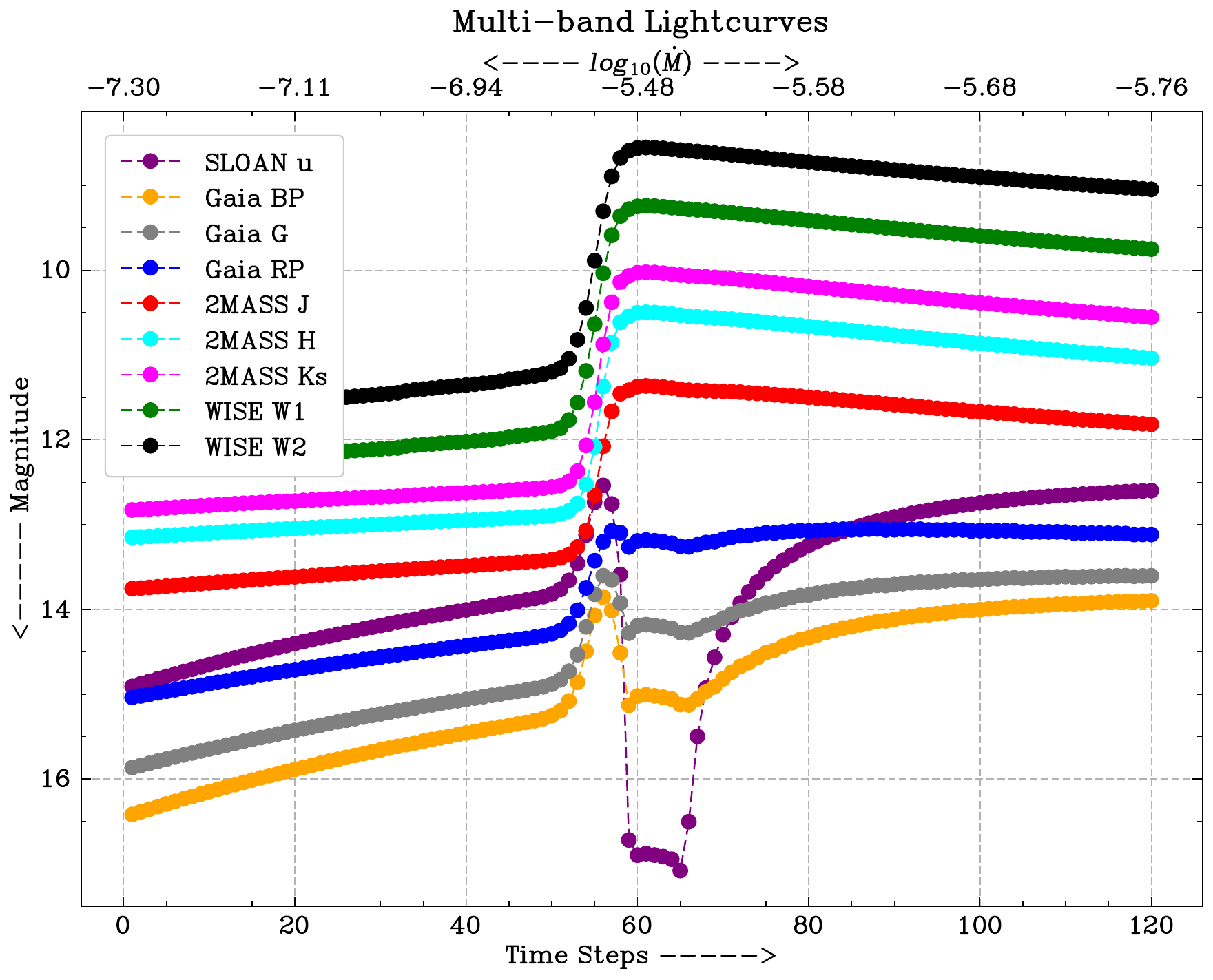}
\plottwo{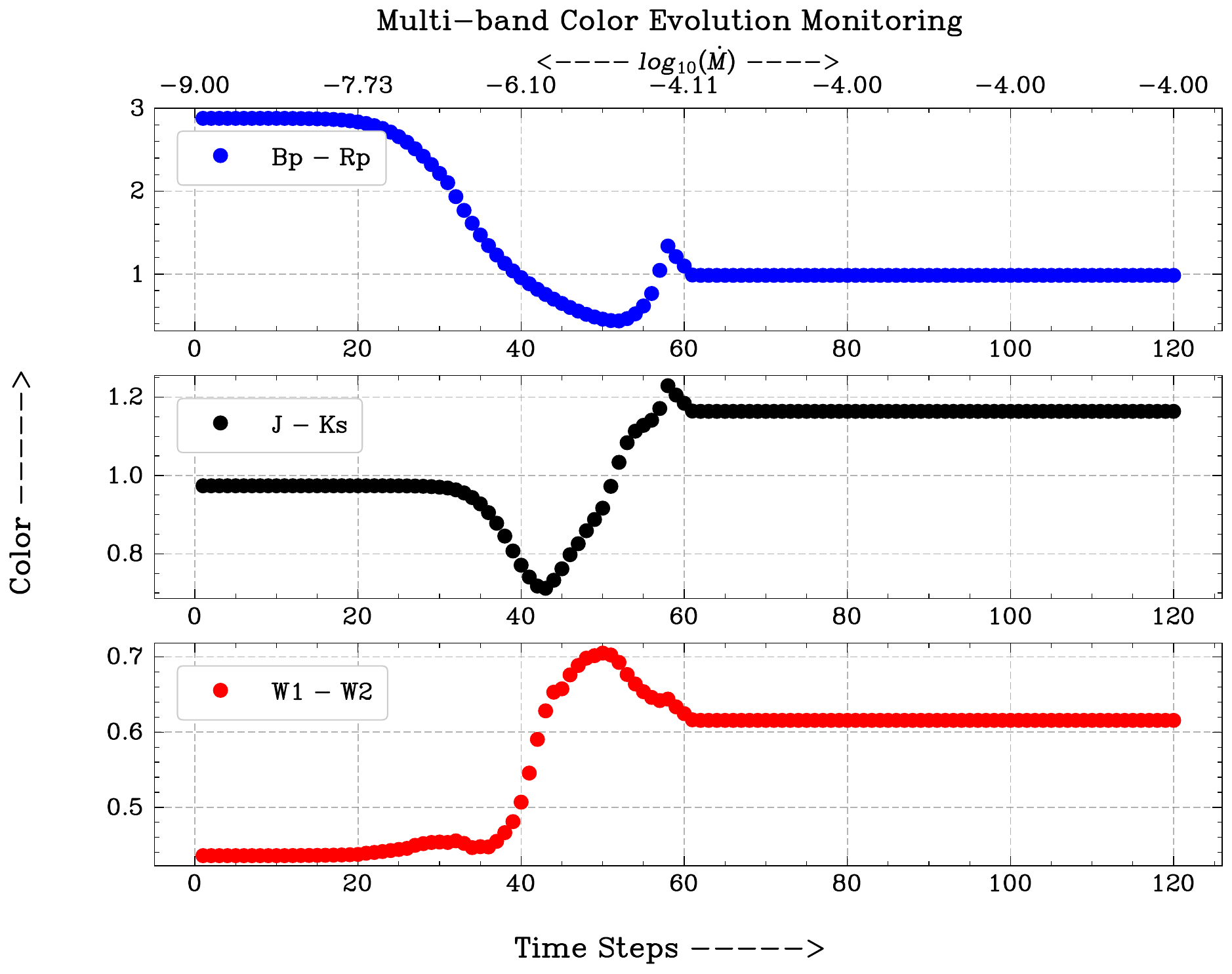}{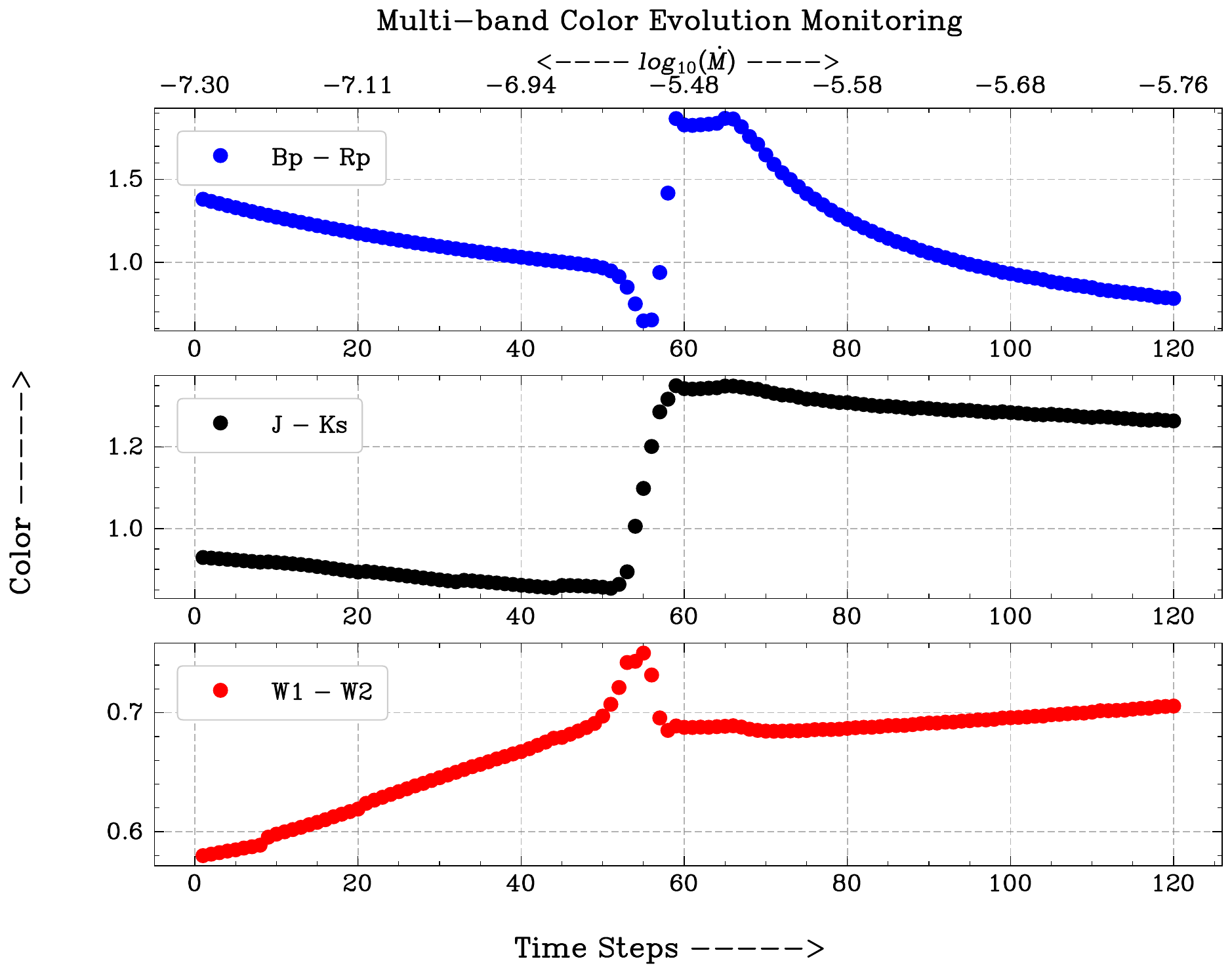}
\caption{{Variation in luminosity (top)}, model lightcurves (middle) and color curves (bottom) for HBC 722-Like (left) and V890 Aur-Like (right) accretion profiles. }
  \label{lum_lightcurve_hbc_722_v_890}
\end{figure}

\subsubsection{Time [40, 60]}

This period sees the accretion rate rise to its peak value, and is the most dynamic stage of evolution.
During this epoch the effective temperature of dust disk, $T_{\textrm{eff,dust}}$, starts to rise and saturates at $1400$K. This happens due elevated radiation from the visously heated gas disk, heating up the passively heated disk now, in addition to the heating received from stellar photospheric flux.

Figure \ref{r_in_hbc_722_v_890} shows
the magnetic truncation radius of the disk starts moving towards the star,  
due to increased gas pressure from the increasing accretion rate dominating the magnetospheric pressure.  The inner boundary of dust also moves inwards. But soon enough, the gas disk sets in due to enough viscous heating. This gas disk then pushes the $R_{dust, inner}$ outwards to a large radius $\approx 20 R_{*}$. 
By the end of this period, $R_{in}$ has decreased from $R_{\textrm{co-rot}}$ to almost $R_{*}$. 
Figure \ref{r_in_hbc_722_v_890} shows
that near the peak outburst, when the disk is about to touch the stellar surface, the temperature of the hotspot starts decreasing again rapidly after a maximum temperature of $\approx$ 18000 K. At this point, the disk completely touches the stellar surface and turns off the poloidal magnetospheric accretion flux component.

In the lightcurves (Figure~\ref{lum_lightcurve_hbc_722_v_890}),
there is a rise in brightness across all wavelength bands. 
We highlight the $BP$ behavior as it is transitional between measuring the increasingly heated inner disk, and measuring the hotspots. 
Interestingly, although the \textit{BP} band has overlap with most of the \textit{u} band, $BP$ is much less sensitive to the increasing accretion rate compared to $u$ band (3200-4000 \AA). $BP$ appears to capture less of the high temperature flux from the shock emission. 
Towards the end of this epoch, the $u$ band flux falls below $BP$ band flux because of the wide $BP$ filter.

\subsection{V890 Aur-Like Lightcurve}

The source V890 Aur (FUOr-Aur 0544+3330)
had photometric monitoring during its outburst in at least seven different filters. 
It shows a prolonged high state with gradually declining brightness \citep{AUR_0544_Hillenbrand_2025}. 
{The adopted accretion profile is shown in Figure~\ref{mdot}.}
Compared to HBC 722, the initial rise of V890 Aur as much shallower, and the approach to peak much steeper.

The right panels of Figures~\ref{r_in_hbc_722_v_890}, \ref{heatmap_hbc_722_v_890}, \ref{hotspot_hbc_722_v_890},  and \ref{lum_lightcurve_hbc_722_v_890} 
represent the behavior of various parameters throughout the accretion rate profile.

\subsubsection{Time [0, 40]}
This time period represents the initial shallow rise from the low state of YSO accretion 
(Figure \ref{mdot}). $R_{\textrm{in}}$ moves inwards at a constant velocity (Figure~\ref{r_in_hbc_722_v_890}).
The accretion hotspot is present on the stellar surface (Figure \ref{hotspot_hbc_722_v_890}), 
and {except briefly around the peak $\dot{M}$} persists for the whole outburst cycle. 
The temperature of the hotspot is increasing linearly during this period 
The free-fall velocity is slightly decreasing at these epochs.

Figure \ref{lum_lightcurve_hbc_722_v_890} shows only slightly rising luminosity during this epoch, 
driven by the increasing emission from the accretion shock and gaseous disk.
The rise in flux from the dust disk is only minor
since the change in the inner disk extent is minuscule compared to the overall size of the dust disk. 

In the multi-band lightcurves of Figure \ref{lum_lightcurve_hbc_722_v_890}, the W1 and W2 bands show elevated brightness, 
characteristic of a slow rise in dust disk luminosity. 
The $u$ band also shows an increase in brightness, characteristic of the increasing accretion shock luminosity.
The WISE and 2MASS bands show minor signs of the outburst behavior. This is because the overall flux in the optical bands is dominated by the slow rising hotspot emission. 
Changes happening in the viscously heated disk are not apparent within this epoch.

\subsubsection{Time [40, 60]}
This time period is marked by a more rapid increase in the accretion rate  (Figure \ref{mdot}) 
and features the outburst peak towards the end of this epoch.  
Various changes in the disk also happen quickly. The inner disk moves inward at an accelerated speed  (Figure \ref{r_in_hbc_722_v_890}). The disk temperature increases due to viscous heating (Figure~\ref{heatmap_hbc_722_v_890}), and once $T_{\textrm{sub}} = 1400$K is reached, {the innermost location of} dust disk and of the viscous disk separate. 
The dust disk's inner boundary moves outwards due to $T_{\textrm{sub}}$ isotherm moving outwards from the star.

The rapid rise of V890 Aur has the temperature of the accretion hotspot doubling from the low state of $\sim7000$K 
to a little over $14000$K (Figure \ref{hotspot_hbc_722_v_890}). 
Regarding the luminosity from each of the components, Figure \ref{lum_lightcurve_hbc_722_v_890} show
that the total luminosity from the system increases by about one order of magnitude, dominated by flux contributions, first from the hotspot emission and later by gas disk.

Interestingly, during the outburst,
the brightness in most filters (Figure~\ref{lum_lightcurve_hbc_722_v_890})
seems to increase around the same time,
with only the $u$ band brightening starting slightly earlier.
This is different than in the case of HBC 722 discussed above, 
which had a shallower rise to peak accretion rate, and differential behavior between the bands. 

\subsubsection{Time [60, 120]}
This epoch features the decay from the peak outburst state. However the rate of decay is small with a decrease in $\Dot{M}$ of only $10^{0.25}\ M_{\astrosun} yr^{-1}$.  
As the accretion rate decreases, the influence from the inner disk appears to subside and
the hotspot again becomes relevant. 
The lightcurves do not show much change from the peak state, but are gradually declining as well.
The $u$ flux starts to rise toward the end of the epoch.

\subsection{V960 Mon-Like Lightcurve}

V960 Mon 
exhibited a brightness increase of $\Delta W2 \approx -2.2$ mag and $\Delta B \approx -3$ mag over a few months. 
This source is also a rare FU Ori outburst with a recorded low state and its rapid rise and subsequent post-outburst fade have been continuously tracked with well-sample multi-band lightcurves. 
{The adopted accretion profile is shown in Figure~\ref{mdot}.}

The left panels of Figures~\ref{r_in_v960_17bpi}, \ref{heatmap_v960_17bpi}, \ref{hotspot_v960_17bpi}, and \ref{lum_lightcurve_v960_17bpi} 
represent the behavior of various parameters throughout the accretion rate profile.

\begin{figure}[ht!]
\plottwo{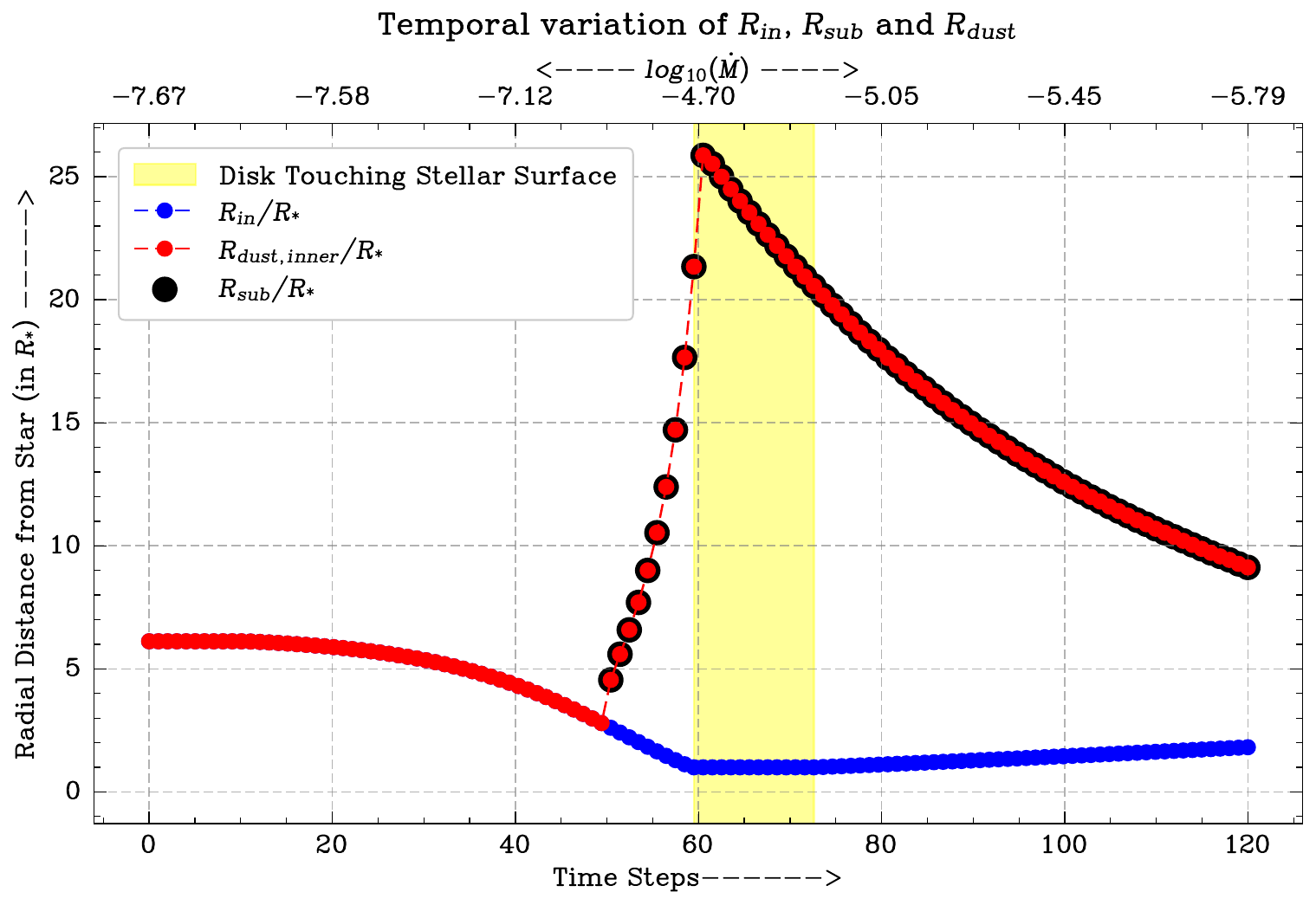}{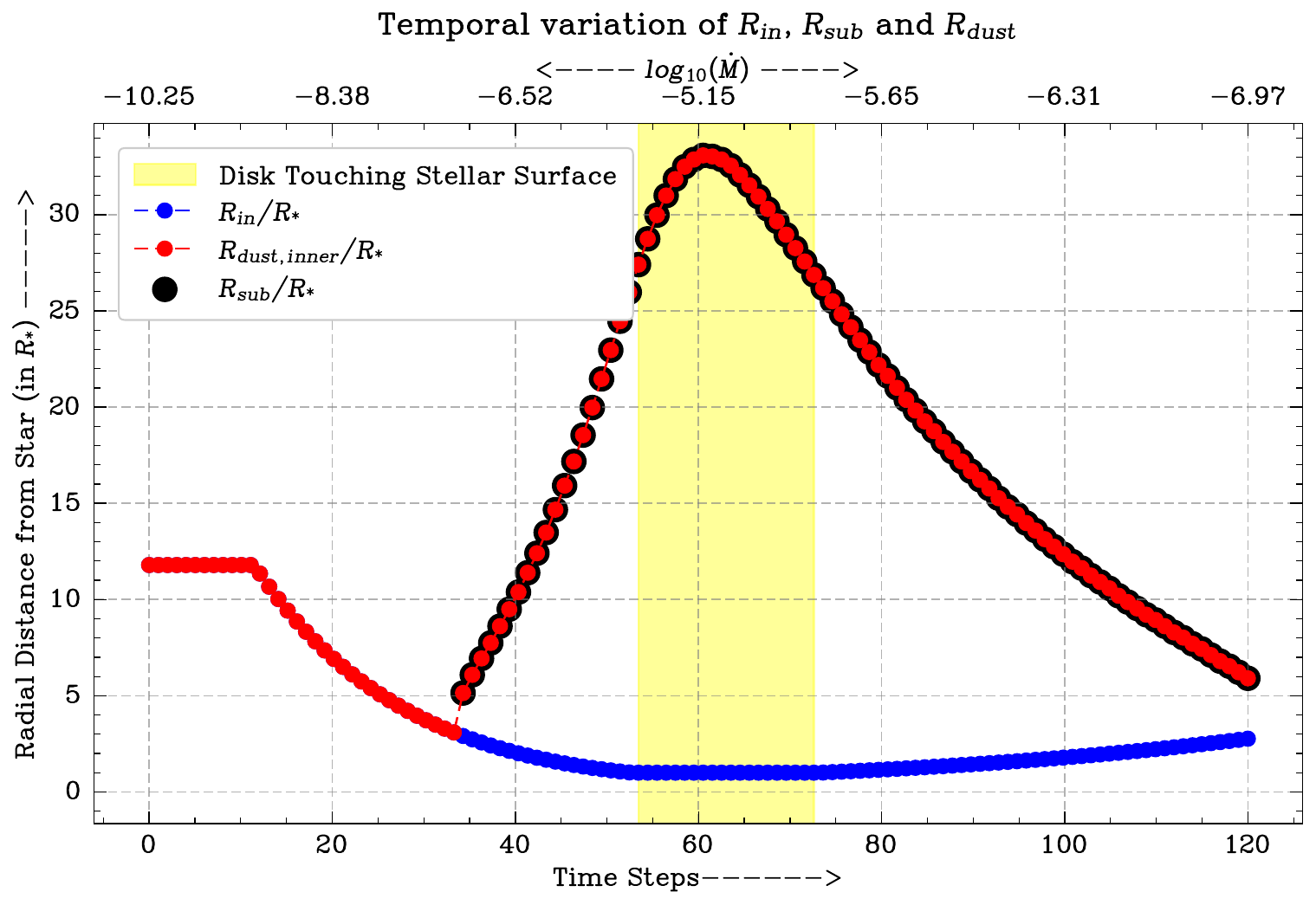}
\caption{Variation in $R_{in}$, $R_{sub}$ and $R_{\textrm{Gas,Dust}}$ for the V960 Mon-Like (left) and Gaia 17bpi-Like (right) cases.}
  \label{r_in_v960_17bpi}
\end{figure}

\begin{figure}
    \centering
    
    \plottwo{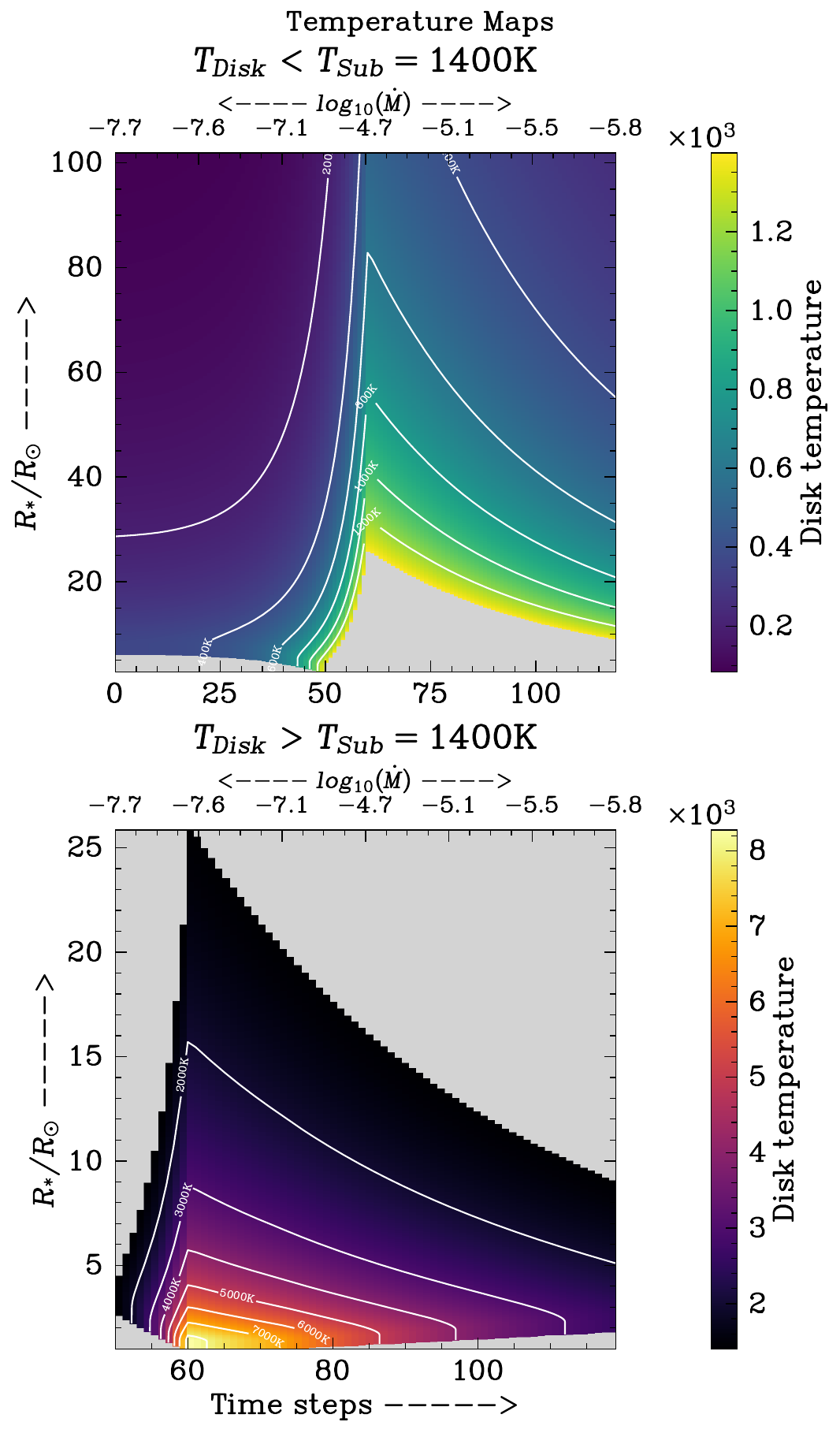}{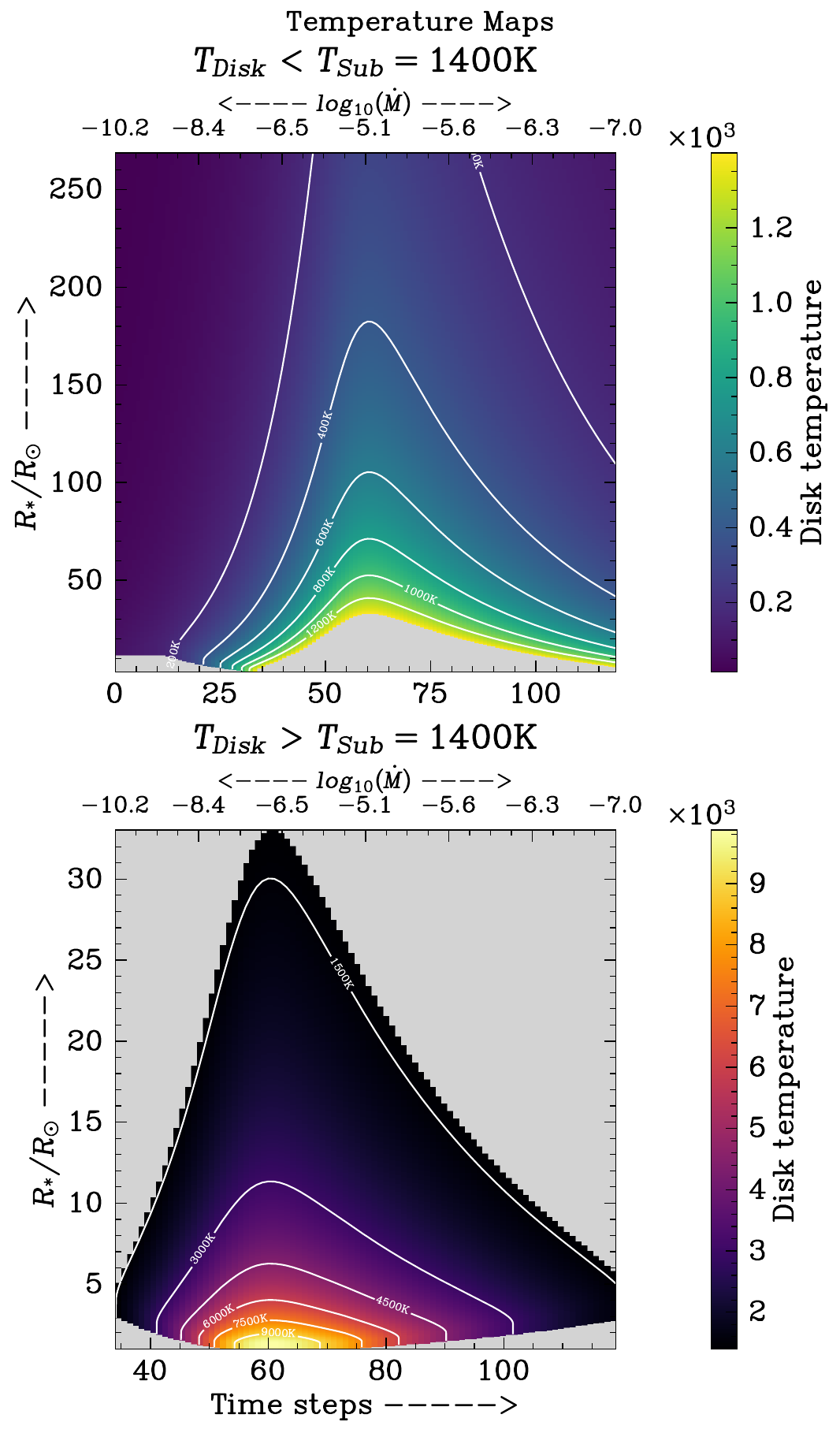}
    \caption{Temperature variation in the {viscous gas} disk {and dust disk} mid-plane for V960 Mon-Like (left) and the Gaia 17bpi-Like (right) cases.}
    \label{heatmap_v960_17bpi}
\end{figure}

\begin{figure}[ht!]
\plottwo{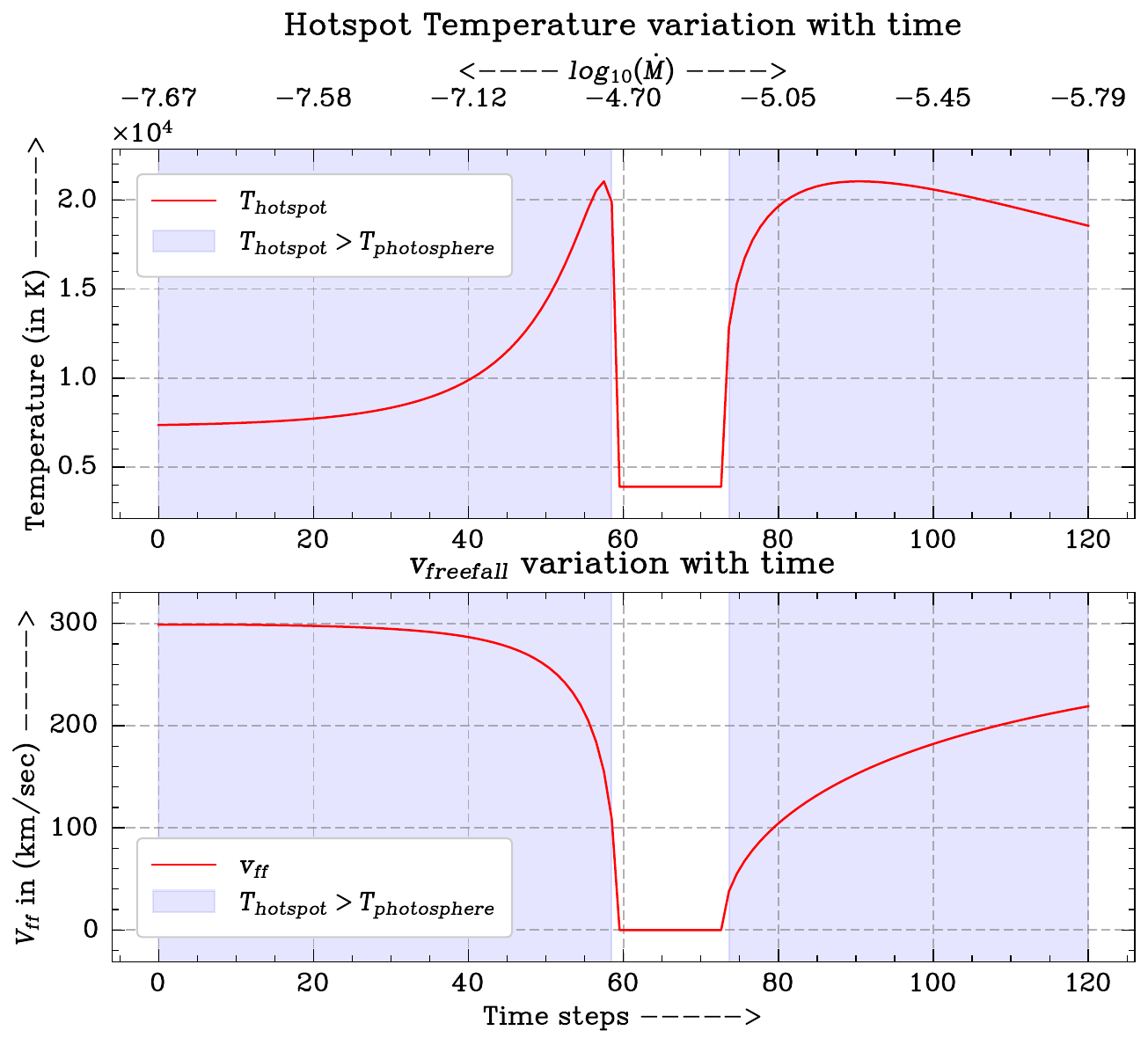}{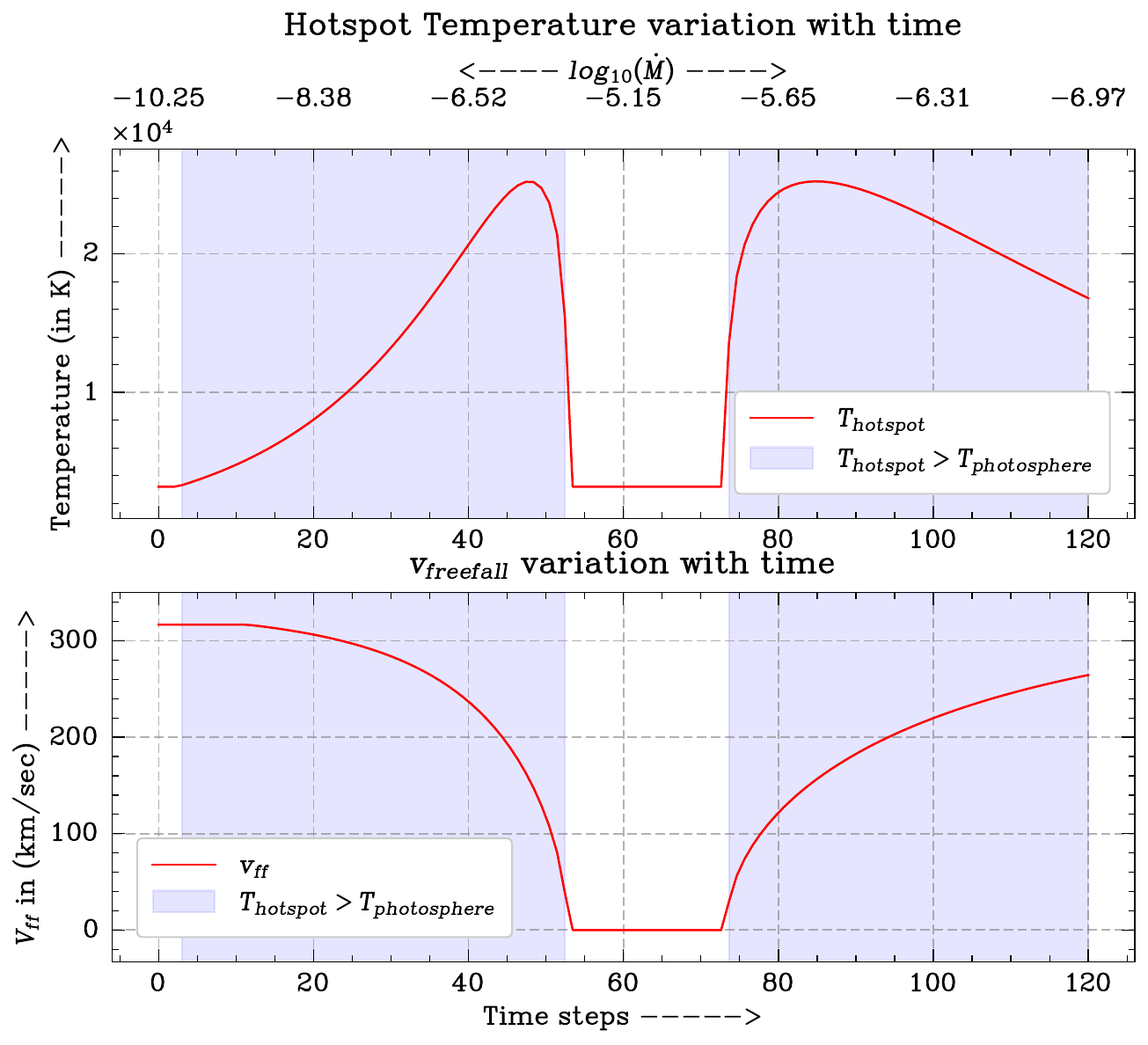}
\caption{Variation in the hotspot temperature and the free-fall velocity at the stellar surface for the V960 Mon-Like (left) and Gaia 17bpi-Like (right) cases.}
\label{hotspot_v960_17bpi}
\end{figure}

\begin{figure}[ht!]
\plottwo{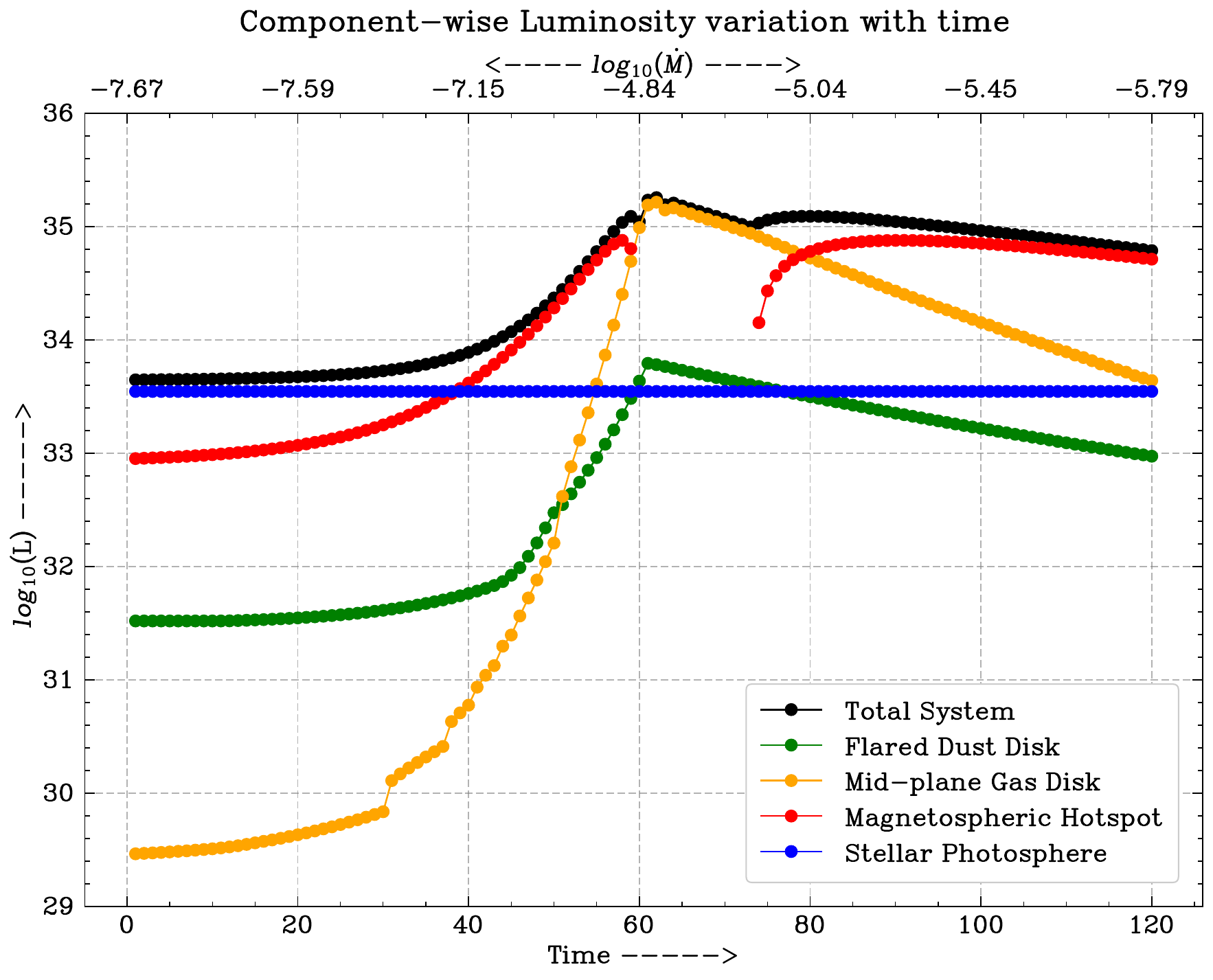}{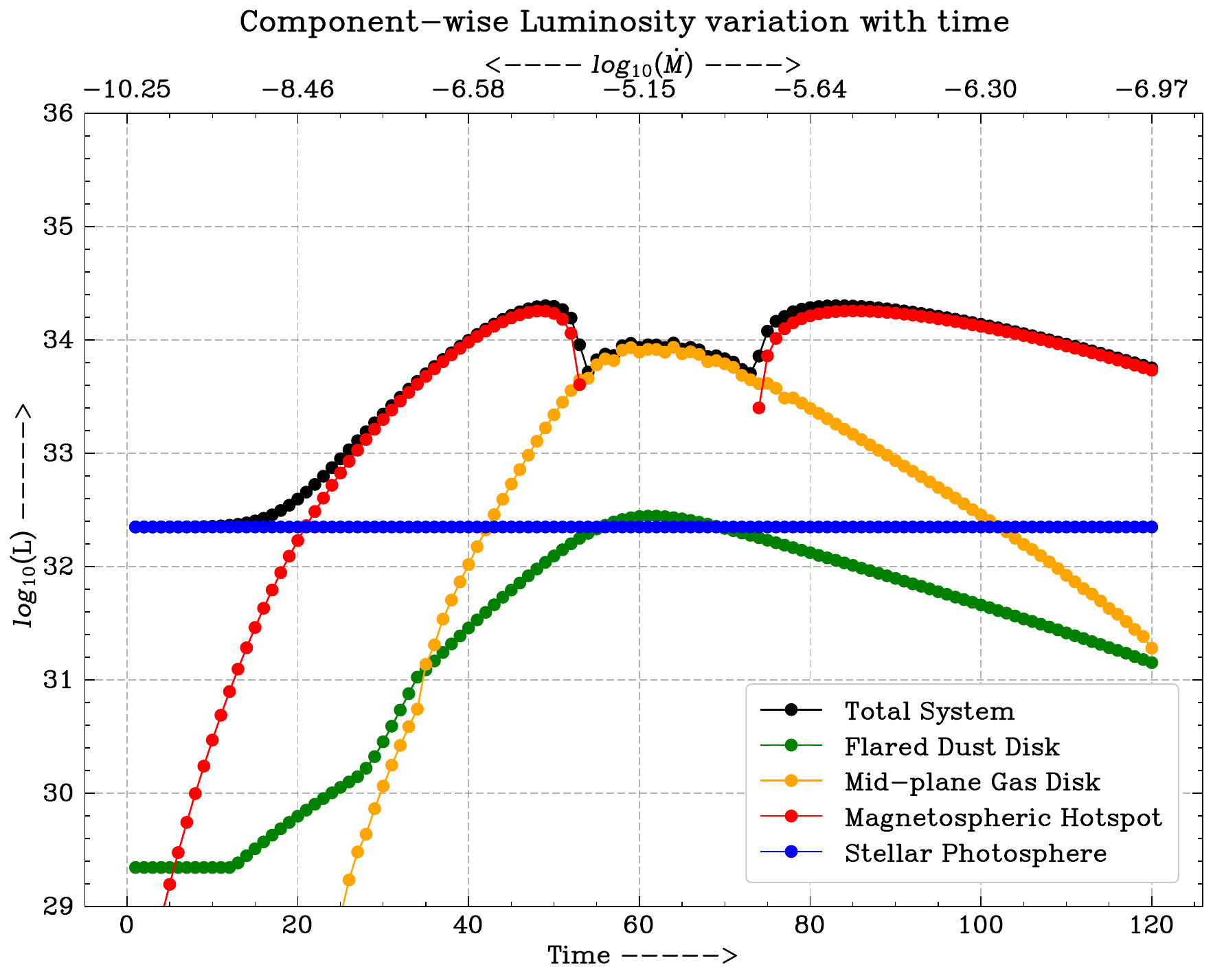}
\plottwo{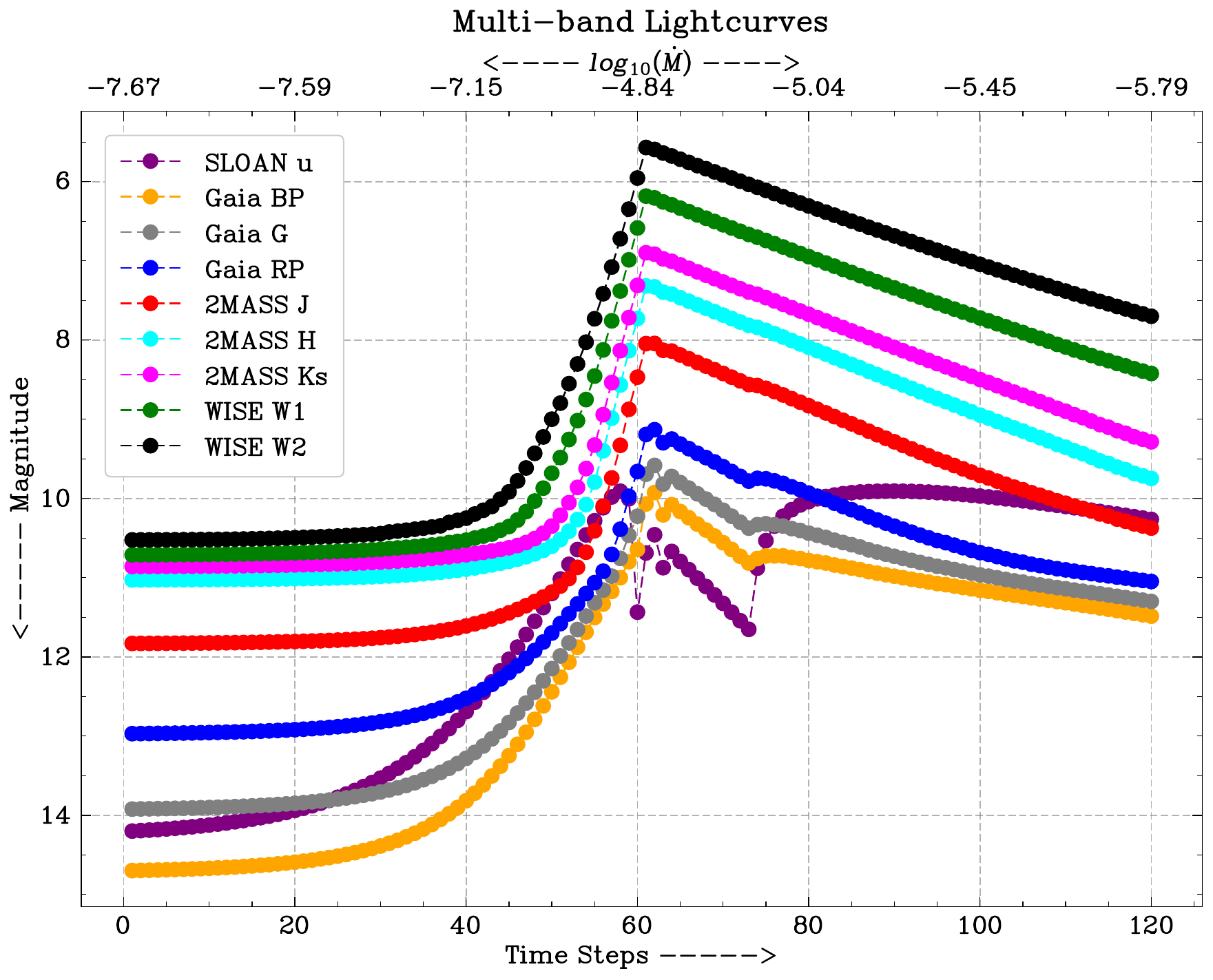}{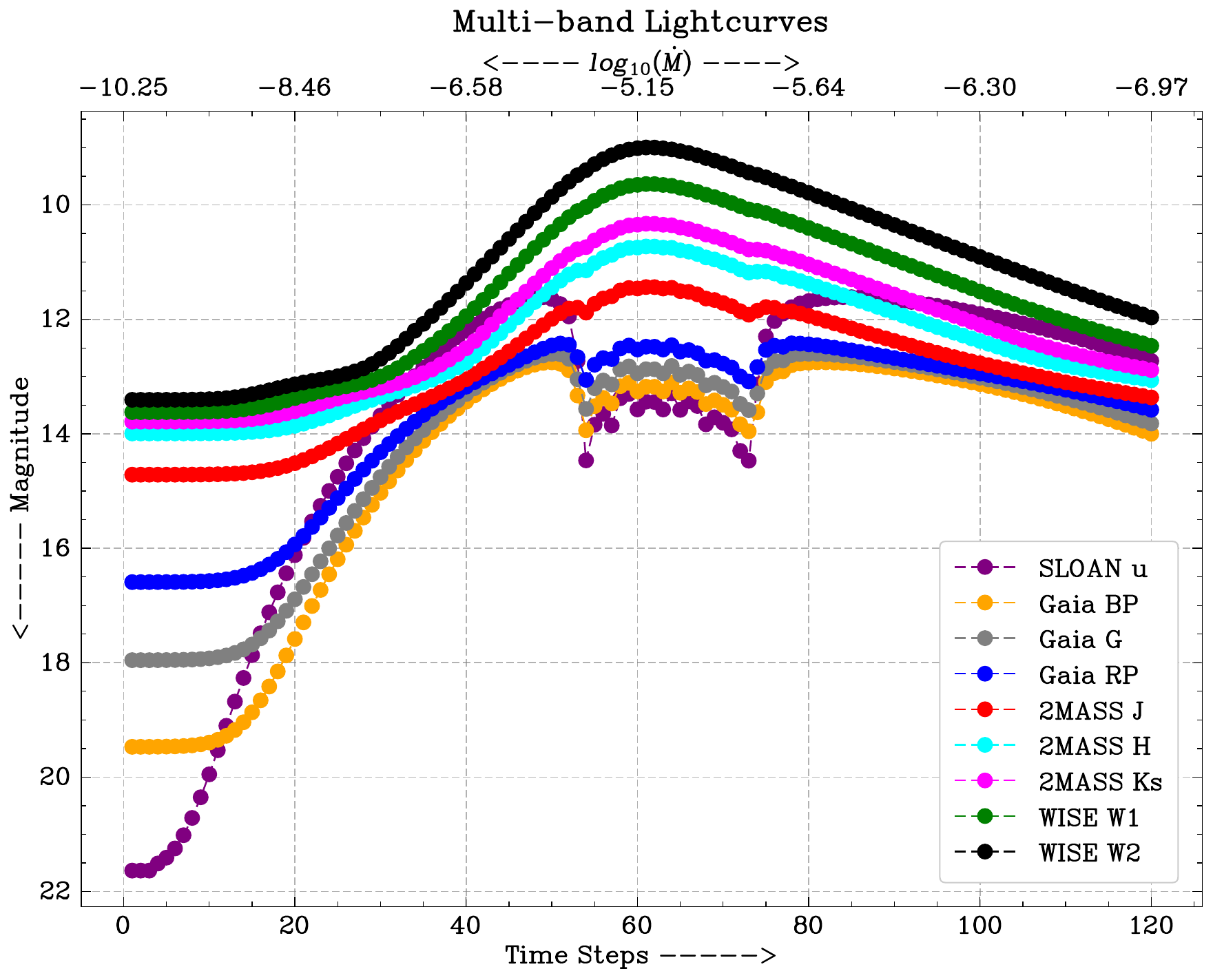}
\plottwo{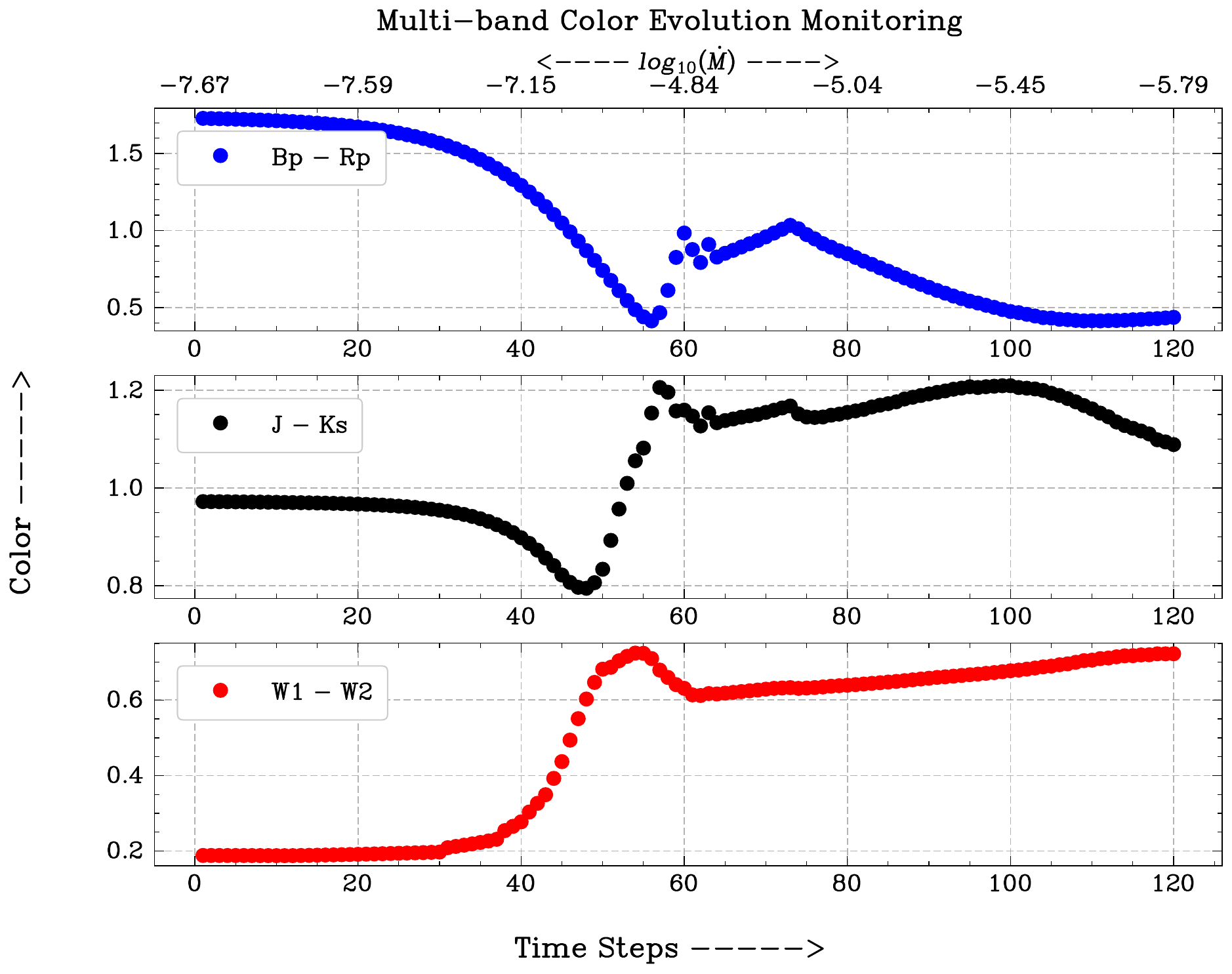}{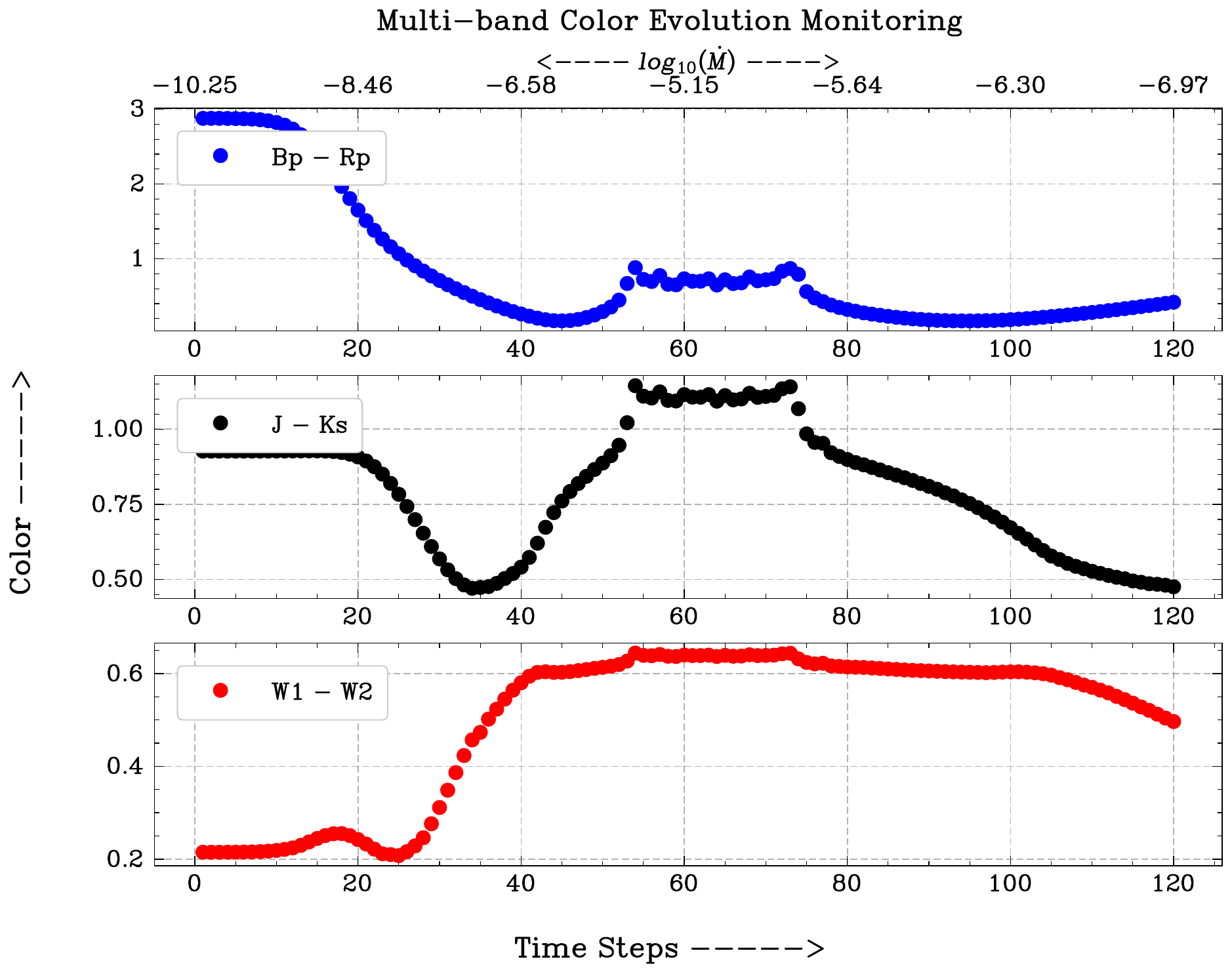}
\caption{{Variation in luminosity (top)}, model lightcurves (middle) and color-curves (bottom) for V960 Mon-Like (left) and Gaia 17bpi-Like (right). 
The small scale jumps are artificial, and due to the temperature grid spacing at higher temperatures; see text. Larger scale variations, however, are real and, e.g. in the \textit{J-Ks} color curve of the Gaia 17bpi-Like case, can be explained by the trade-off between shock emission and gas disk emission in contributing to the total luminosity. 
Until time step $\sim$40, the color evolution is primarily due to the changing shock emission. From time step 50 to about 85, the color evolution is due to the changing brightness of the gas disk. Finally beyond time step of 85, the shock emission again is the major component to the luminosity and thus there is smooth color evolution.
}
  \label{lum_lightcurve_v960_17bpi}
\end{figure}

\subsubsection{Time [0, 40]}
These epochs cover the low-state. The accretion rate is not increasing fast enough for significant change in the other model parameters. Figure \ref{r_in_v960_17bpi} shows that the disk was initially truncated at the co-rotation radius. However, towards the end of this period, the truncation boundary is moving inwards. The gas disk temperature starts rising due to the rise in accretion rate. 
In Figure \ref{hotspot_v960_17bpi}, the hotspot is initially present with $T_{\textrm{hotspot}} = 7500K$ in the low state and reaches a temperature of about $\sim 10000K$ by the end of this period. The luminosity  (Figure \ref{lum_lightcurve_v960_17bpi}) and lightcurve (Figure~\ref{lum_lightcurve_v960_17bpi}) plots are relatively flat during the initial period, though show some features later. By the end of this epoch, the photosphere and the shock emission are contributing equal flux to the total luminosity. In the lightcurve plots, the u-band shows the earliest onset of outburst, followed by optical and then by infrared rise.

\subsubsection{Time [40, 60]}
In this interval, accretion heating makes the disk hot enough to sustain a viscously driven disk atmosphere. 
The gas disk pushes the dust sublimation boundary ($R_{\textrm{sub}}$) farther from the star, and this makes $R_{\textrm{sub}}$ move outwards (Figure \ref{r_in_v960_17bpi}). The temperature of the hotspot continues to increase (Figure \ref{hotspot_v960_17bpi}), but towards the end of this time period it rapidly falls, as the {viscous} gas disk comes very close to the stellar photosphere, shutting down the poloidal accretion activity.

The total luminosity rises very smoothly (Figure \ref{lum_lightcurve_v960_17bpi}). 
For the most part of this epoch, the shock emission dominates the total flux, but in later stages,disk emission contributes significantly. When the higher-temperature gas disk moves inward, its flux increases rapidly with the accretion rate and overwhelms all other components in {contributing to} the total flux. In Figure~\ref{lum_lightcurve_v960_17bpi}, the u-band rise happens {first}, then the WISE bands show the rise. The dust disk luminosity rises earlier than the viscous gas disk luminosity, {This results in the WISE band increase occurring before the} Gaia and 2MASS bands, which show brightening activity {only when} the gas disk {contribution to total flux becomes important}.

\subsubsection{Time [60, 80]}
In this time period, the accretion rate starts to decline from the peak high state. In Figure \ref{r_in_v960_17bpi}, the inner truncation boundary (blue) is held constant for some time at the stellar photosphere boundary, even though the accretion rate has started falling back. The sublimation boundary (black) migrates inward. 
In Figure \ref{hotspot_v960_17bpi}, the {viscous} gas disk touches the stellar surface when the truncation boundary equals the stellar radius. As a result, the poloidal accretion and the flux from the shocks switch \textit{off}. Eventually the disk starts to move outward as the accretion pressure decreases and the magnetospheric accretion again becomes active and the shocks re-appear.

Figure \ref{lum_lightcurve_v960_17bpi} show that the total luminosity is dominated by the gas disk. 
As the disk cools down, the luminosity associated with the gas disk decreases almost log-linearly. Interestingly, this is also the only point in the outburst cycle when the passive dust disk has marginally higher luminosity than the photosphere. 
The discontinuity in the luminosity associated with the shock region is because this is when the poloidal accretion remains switched \textit{off}. 

Figure \ref{lum_lightcurve_v960_17bpi} shows the WISE bands have linearly decreasing luminosity, representing the behavior of the flared dust disk. 
The u-band shows a brief linear decline in the period while the magnetospheric component is still \textit{off}, 
with the luminosity dominated by the high-temperature gas disk, but by time step 80 it begins to probe mainly the hotspot, which is brightening as the viscous gas disk fades. 
During this post-peak epoch the colors all become slightly redder. 

\subsubsection{Time [80, 120]}

{
As the accretion rate decreases, the gas disk inner radius moves outwards while the dust sublimation radius moves inwards (Figure \ref{r_in_v960_17bpi}).
The free-fall velocity increases monotonically.} 
However, although the accretion rate is continuously decreasing, 
the temperature of the hotspot first increases and then decreases (Figure \ref{hotspot_v960_17bpi}),
mirroring the luminosity from the shock region (Figure~\ref{lum_lightcurve_v960_17bpi}). 
By time step of 120, the photosphere and the viscous disk contribute almost equally to the total luminosity, though both of these components are overwhelmed by the shock emission component. 

Figure \ref{lum_lightcurve_v960_17bpi} shows the slope of the lightcurves changing smoothly {with wavelength; longer wavelengths continue their decline from peak 
during this epoch, while shorter wavelengths transition to decline less steeply. 
The behavior in the WISE and 2MASS bands is characteristic of the flared dust disk, which has a linear declining luminosity profile in Figure \ref{lum_lightcurve_v960_17bpi}. 
The Gaia bands show a slope transition because they initially probe inner disk regions with luminosity from the gaseous disk in Figure \ref{lum_lightcurve_v960_17bpi}, 
while later the hotspot switches back \textit{on} due to renewed luminosity associated with magnetospheric shock,
which also causes flattening of the $u$-band lightcurve.
There is continued reddening in} the WISE colors while the 2MASS and Gaia {colors flatten} at earlier stages.

\subsection{Gaia 17bpi-Like Lightcurve}

Gaia 17bpi 
had its increase in brightness recorded at both optical and infrared wavelengths \citep{17bpi_hillenbrand_2018}.
After a brief peak, the source started a decline that is steep enough to indicate a relatively short-lived outburst interval, compared to the other objects discussed here. 
The observed optical amplitude was about -3.5 mag for this source; however,
we have assumed a rise in the accretion rate from $\sim 10^{-10}\ M_{\astrosun} \textrm{yr}^{-1}$ to $\sim 10^{-5.5}\ M_{\astrosun} \textrm{yr}^{-1}$.  This is to fully study in our model the lightcurve behavior from the earliest transition stages between
accretion that is magnetospheric infall and hotspot dominated, and accretion that is FU Ori type and dominated by the viscous inner disk. 
{The adopted accretion profile is shown in Figure~\ref{mdot}.}

The right panels of Figures~ \ref{r_in_v960_17bpi}, \ref{heatmap_v960_17bpi}, \ref{hotspot_v960_17bpi}, and \ref{lum_lightcurve_v960_17bpi}
represent the behavior of the various parameters.

\subsubsection{Time [0, 20]}
Although the accretion rate is rising from its initial very low value, there is not enough heating for other parameters to change yet. Figure \ref{heatmap_v960_17bpi} shows that the temperature is rising in the disk, although it remains below the sublimation temperature. There are no changes to the luminosity or the lightcurve during this epoch.

\subsubsection{Time [20, 40]}
Figure \ref{r_in_v960_17bpi} shows that as the accretion rate becomes high enough ($\sim 10^{-8}\ M_{\astrosun} \textrm{yr}^{-1}$) the gas pressure overcomes the magnetic pressure and thus pushes the inner gas disk boundary towards the star. In Figure \ref{heatmap_v960_17bpi}, the maximum temperature of the gas disk increases very rapidly with the accretion rate. 
The dominant factor irradiating the dust disk thus shifts from photospheric emission to hotspot shock emission, which becomes evident in Figure \ref{lum_lightcurve_v960_17bpi} 
as brightening of dust disk in this epoch. 

In Figure \ref{hotspot_v960_17bpi}, it is clear that the hotspot behavior for Gaia 17bpi is similar to that of V960 Mon.
In Figure \ref{lum_lightcurve_v960_17bpi}, the total luminosity appears to be dominated by the shock emission. The dust disk shows a rise in luminosity due to increasing flux from the heated shock region on the stellar photosphere irradiating the flared disk. Note that the {viscous gas} disk is not responsible for initiating the heating of the flared disk, unlike the other cases discussed above. 

The lightcurve plot of Figure \ref{lum_lightcurve_v960_17bpi} shows the $u$ band rise due to the increased emission from magnetospheric accretion.e The WISE bands begin to rise towards the end of the epoch, due to the higher irradiation.

\subsubsection{Time [40, 60]}
In this epoch, from Figure \ref{r_in_v960_17bpi} it is clear that the gas disk moves towards the stellar surface until time step $\sim 50$, when it touches the stellar surface and shutting \textit{off} the hotspot emission component, which had attained a maximum temperature of $\sim 26,000$ K. In Figure \ref{lum_lightcurve_v960_17bpi}, the shock emission component rises above the photospheric luminosity earlier than the viscous gas disk component, at least for the initial part of this epoch. 

In Figure \ref{lum_lightcurve_v960_17bpi}, there is a noticeable time lag between the rise times of different lightcurves.  
We also note that the $u$-band lightcurve shows some large-scale features pertaining to the relative contributions from hotspot and gas disk emission.

\subsubsection{Time [60, 120]}
This {post-peak} epoch features the slowing down of accretion activity. During the fade, most diagnostic parameters basically mirror their behavior during {the outburst onset} to peak epochs. While the system has not yet returned to its (presumed) pre-outburst low state by the end of 120 time steps, the gas disk has shrunk in size by almost about 8 times compared to the peak-outburst state, from radial thickness of $\sim40R_{*}$ to only about $\sim5R_{*}$. In Figure \ref{lum_lightcurve_v960_17bpi}, the accretion hotspot becomes the major emission component during the later part of the epoch. This reassertion of magnetospheric accretion also shows up in Figure \ref{lum_lightcurve_v960_17bpi}, where the $u$ band captures the restart of accretion hotspot emission.

{As noted above, in Appendix~\ref{sec:parametric} we perform parameter studies around several of the stellar parameters for the Gaia 17bpi-Like case.}

\section{Summary, Discussion{, and Future Directions}}\label{sec:conclusion}

In this paper, we have demonstrated how one can adopt an accretion profile during a YSO accretion outburst, 
and show the consequences of increasing and then decreasing accretion rate on physical parameters
{as well as observables}.  
We illustrate the evolution of several important radial locations, 
temperatures and velocities, of total luminosity {emanating} from different emitting components,
and of emergent lightcurves and color curves.

To simulate the transition from low-state T-Tauri-Like accretion to high-state FU-Ori-Like accretion,
we adopt accretion rate profiles $\dot{M(t)}$ over a fixed interval [0,120] of time steps.
For each accretion profile, the peak {accretion rate} is defined to occur at time step 60
(except for the linearly rising case which does not peak until time the final step, 120).
Peak accretion rates are set to be consistent with detailed accretion disk models of
individual FU Ori stars.  {For the pre-outburst,} low-state accretion rates, {we make an estimation based on the observed outburst} amplitudes of these same FU Ori outbursts.

In the time-dependent model, we additionally assume individualized values for the stellar parameters 
(mass, radius, rotation period, and magnetic field strength), 
At each time step through the changing accretion rate, we calculate 
the resulting flux contributions from magnetospheric accretion, gas disk luminosity, and dust disk luminosity. 

The results from our simulations are:
\begin{itemize}
\item
In the early stages, while the accretion rates remain low, the dust disk which is heated by stellar irradiation,
and the gas disk which is viscously heated due to the accretion, always both contribute to the luminosity.
A contribution from magnetospheric accretion hotspots may or may not be present, depending on their contrast
with the stellar photosphere.  

\item
For the stellar parameters typical of low mass young stars, such as those in Table \ref{tab:fid_parameters}, 
when the values of $\dot{M}$ are $<2-3 \times 10^{-7} ~M_\odot$/yr, 
the effects of the accretion remain in a relatively steady state, and the system lightcurves are essentially constant.

\item
As the accretion rate increases above a few times $10^{-7} ~M_\odot$/yr, both magnetospheric hotspot heating
and inner disk heating increase, causing lightcurves to brighten.
For high enough values of $\dot{M}$, the inner disk boundary moves inward causing lightcurves 
to steepen their brightening, with ultraviolet and blue optical bands showing the earliest effects.

\item
If in the peak outburst state $\dot{M} < 10^{-5} ~M_\odot$/yr, the total flux continues to have contributions from the magnetospheric accretion hotspots.
If the high state $\dot{M}$ exceeds $10^{-5} -10^{-4} ~M_\odot$/yr, the flux becomes dominated by the viscous disk, especially at the shortest wavelengths.

\item 
At such high accretion rates, magnetospheric accretion is suppressed, and the disk material reaches the star through equatorial accretion.  Near the peak of the accretion profile, 
lightcurves peak, and most colors become bluer, at least temporarily.  

\item During the post-peak period, mid-infrared and near-infrared colors generally redden in our modelling,
while optical colors remain bluer for longer, before eventually turning redder as lightcurves fade.

\item
{Throughout an accretion outburst, the general behavior of the multi-wavelength lightcurves we have modelled is that as $\dot{M}$ increases, the u-band brightening happens first, in response to increased heating in magnetospheric accretion hotspots.  The WISE bands brighten next, due to the rising dust disk luminosity.  The viscous gas disk luminosity increases last, and manifests most directly in the Gaia and 2MASS bands.
}
Thus, \textit{during the onset and throughout an accretion outburst, 
red optical and near-infrared lightcurves generally follow the same or very similar form as the accretion profile, 
while mid-infrared lightcurves are more responsive to the geometry and heating of the dust disk.}  

\item
{We highlight that}
the mid-infrared photometric bands can brighten before the visible/near-infrared.
Such an observation is a natural consequence of temperature evolution in the inner disk,
resulting from the {outward} expansion of the dust sublimation front.
{Mid-infrared brightening} is not necessarily related to the accretion outburst triggering event, 
{which when modeled as} propagating outward-in \citep{bae2014,cleaver_2023} 
{has led to some speculation that mid-infrared lightcurves should rise before optical lightcurves}.

\end{itemize}

{We have shown our forward-modeling for several different accretion profiles.
While the results listed above are general to an increasing and then decreasing $\dot{M}$, 
subtleties in the accretion profiles do differentially affect the physical parameters, and thus 
the resulting lightcurves.  For example, some FU Ori outbursts exhibit 
a sharp peak and exponential decline, while others have a more rolling peak and then plateau structure.
Stellar parameters are also important, notably $M/R$ and $B$. 
For example,
V890 Aur had a lower peak accretion rate than HBC 722. However, the HBC 722-Like disk touches the stellar surface when $\Dot{M} \sim 10^{-4}\ M_{\astrosun} \textrm{yr}^{-1}$, whereas for the V890 Aur-Like case, the peak accretion rate of about $\Dot{M} \sim 10^{-5.5}\ M_{\astrosun} \textrm{yr}^{-1}$ could not make the disk touch the stellar surface in an initial model that used our fiducial stellar magnetic field assumption of 1.4 kG. In-order for the disk to reach the stellar surface, the field had to be reduced to 0.8 kG.
}

Our models are meant to be illustrative. Although we have used the observed lightcurves of several actual recently outbursting YSOs as the mock accretion rate profiles for input to our models, our results are not necessarily applicable to the details of particular objects. Nor are they predictive of future lightcurve evolution for any particular object. 

More rigorous theory, such as three-dimensional MHD simulations that show the evolution of an accretion outburst, including the interaction between the stellar magnetic field and the disk, do not yet exist. However, we can compare many of the assumptions in our model with what has been learned from steady-state MHD simulations of FU Ori-Like boundary layer accretion and CTTS-Like magnetospheric accretion. The \citet{Zhu_CTTS_2024MNRAS} radiation-MHD (rMHD) simulation of magnetospheric accretion shows that surface accretion from the disk allows material to enter the accretion funnels from several times $R_\mathrm{in}$ away. These same simulations also show that the disk truncation radius in the simulation is very close to the value calculated from Equation \ref{eq:r_trunc}, even under different assumed stellar rotation rates \citep{Zhu_CTTS_QPOs_2025MNRAS}. 

In our model, we adopt the simplifying assumption that the disk material is loaded into the accretion funnels at the truncation point, which sets a single $v_\mathrm{ff}$ for a given accretion state. \citet{Zhu_CTTS_2024MNRAS} show that the material flowing into the accretion funnel via the surface flow reaches velocities of $\sim 0.1 v_\mathrm{ff}$. This suggests that the ``turning off'' of magnetospheric accretion, which we find as $R_\mathrm{in}$ approaches $R_*$ may not be as rapid as in our model. However, both \citet{Zhu_FUOriSim_2020MNRAS} and \citet{Takasao_boundaryLayer_2025ApJ} demonstrate that during the boundary layer accretion state of the outburst, the disk is highly ionized and is threaded by strong ($\sim$ kG) magnetic fields. These simulations confirm that any previously-existing dipolar field would be crushed during the outburst and likely result in open field oriented along the rotation axis of the disk. 

Our model also does not incorporate possible restructuring of the magnetic field geometry during the rapidly rising stage of the outburst, as was observed in the 2022 outburst of EX Lupi \citep{Singh_EXLup_2024ApJ}. Though the outburst in EX Lup was of much lower amplitude than those we model, the rising state of our lightcurves is physically similar to the high (but still poloidal) accretion state during the peak of an EX Lup-Like outburst. \citet{Singh_EXLup_2024ApJ} report that the hotspot increased in size azimuthally and shifted $-10^\circ$ in latitude (from $75^\circ$ to $65^\circ$) during the outburst. While we do not model these effects in our sources, the azimuthal spreading of the hotspot could offset the rapid decrease in luminosity our models predict. This does, however, affirm our assumption that prior to the outburst, $R_\mathrm{in} \sim R_\mathrm{co-rot}$, but that during onset of the outburst $R_\mathrm{in} = R_\mathrm{trunc} < R_\mathrm{co-rot}$.

Regarding future observations of early-stage FU Ori outbursts, we predict that facilities such as LSST and UVEX should be able to observe ultraviolet and blue-optical lightcurves rising first, when $\dot{M}$ is initially increasing in proto-FU Ori type outbursts, due to increasing hotspot temperatures. Optical through mid-infrared rises should be seen at later times, as the inner disk becomes increasingly viscously heated as $\dot{M}$ increase in outbursts, and approach their peak values.  {Of note in this context is the demonstration by \cite{masley2025} that the predicted infrared ``pre-cursor" to an FU Ori outburst is geometry dependent, and seen only when the viewing angle is close to right down the outflow axis.}

If a lightcurve profile is known from timely monitoring, and the peak high-state accretion rate can be retrieved from spectral energy distribution and/or other disk modelling analysis, physical insight can be gained regarding the underlying accretion profile, as demonstrated in \citep[e.g.][]{liu2022diagnosing, adolfo_v960_2}. 
Our simple models could be applied in such detailed fitting of individual objects, with attention to matching observed lightcurves and multi-wavelength colorcurves in detail. This could result in credible estimates of the low-state accretion rates for known FU Ori objects in their pre-outburst state.

Our infrastructure could also be used for sources currently moving towards an outburst. It would enable monitoring of all the system parameters, like the disk boundaries, temperature profiles, hotspot temperatures, etc., in the system. This means we will have a tool that could tell us roughly at what time and which stellar component will contribute the majority of the flux from the system. 

Since the simulation pipeline is based on the YSOpy framework \citep[][(in-preparation)]{Das2025YSOs,Das2026YSOs}, high-resolution spectra from the system are generated for each time step. Hence, a comprehensive analysis of spectroscopic evolution can be done spanning the optical to the infrared regime.  This will be the topic of a future paper.

This work was supported, in part, by NASA under award \#80NSSC23K0655.
{G.D. acknowledges conversations with Joe P. Ninan, Archis Mukhopadhyay and Koshvendra Singh that contributed to development of the coding infrastructure implemented in this work.
We also thank our referee for comments that helped improve the presentation.}

\newpage

\appendix

\section{Detail of Innovations Beyond the Original YSOpy Code}\label{sec:ysopy}

\begin{itemize}
\item{Simplification to the hydrogen slab model:}
The H-slab model used in \ysopy calculates the transition probability for many states. However, since we are interested in probing only optical and IR wavelengths, we need to calculate only a few transition probabilities by modifying the free-free and bound-free transition probability calculations. This approximation also made the pipeline much faster to execute.

\item{Self-consistency in the $T_{\textrm{hotspot}}$, $n_{e}$ and $\tau$ parameters}: In \ysopy, the temperature of the shock region ($T_{\textrm{slab}}$), electron density associated with the shock region ($n_{e}$), and the optical depth ($\tau$) are chosen arbitrarily by the user. Here, we derived expressions for $T_{\textrm{hotspot}}$ and $n_{e}$ such that they vary with the accretion rate of the system self-consistently (see Section section \ref{subsec:formula}). Currently, $\tau$ is fixed to be $1.0$ at 3000 \AA.

\item{Accounting for cooler gas at larger disk radii:}
Previously, \ysopy did not account for flux from gas at temperatures below 1400 K because we cannot use atmospheric models below $1400$ K. In currently available model grids, spectra with $T_\mathrm{eff} < 1400$ K are computed to reproduce planetary atmospheres of warm Jupiters, which have surface gravity much higher than in accretion disks and often account for the impact of clouds, which are not present in accretion disk atmospheres. Thus, we implemented blackbodies for lower-temperature annuli as an alternative first approximation. 
    
\item{Removal of Doppler convolution:}
Since we are interested in the photometric evolution of the disk, we can neglect the convolution of the model spectra with Keplerian rotation profiles. The filter windows over which we integrate are sufficiently wide that they are not impacted by line broadening. Omitting the typical convolution step does not affect our results and makes the pipeline much faster.

\item{Adopting self-consistent fiducial stellar parameters:}
{In practice, we are adopting stellar parameters consistent with previous accretion disk model 
fits of several FU Ori sources (Table~\ref{tab:fid_parameters}).  However, we can also}
make use of theoretical isochrones in order to connect realistic stellar parameters together.
Adopting the 1 million year isochrone of \cite{Baraffe_2015},  
for a given calculation, we choose a fiducial value for $M_*$ and then interpolate the
corresponding stellar radius $R_{*}$ and stellar temperature $T_{\textrm{photo}}$. 
For the temperature, in practice we approximate from the model atmosphere grid to the nearest 100 K grid point. 
For the stellar luminosity {$L_*$}, we adopt the standard $4\pi R_{*}^2 \int F_\lambda d\lambda = 4\pi R_{*}^2 \sigma T_{\textrm{photo}}^4$.

\item{Addition of filter transmission functions for photometry:}
\ysopy generates only high-resolution spectra. A new feature here is to calculate the photometry from these spectra, as described in \S \ref{sec:filter}.

\end{itemize}

\section{Parametric Study}\label{sec:parametric}

We consider in this section the effects of varying some of the parameters that are generally unknown, but we have assumed in Table~\ref{tab:fid_parameters}.   
{Specifically,} we explore the impact of our choices for the magnetic field strength and for the co-rotation period.

\subsection{Variation in Magnetic Field: (1kG, 2kG and 3kG)}

As the magnetic field strength is increased, the accretion disk now has to overcome a higher accretion pressure in-order to penetrate towards the stellar surface. Since magnetic field strength is one of the parameters that is not very well constrained from observations, we perform a diagnostic analysis changing the magnetic field strengths among $1, 2$ and $3$kG:
\begin{itemize}
    \item {Hotspot Temperature:} The maximum $T_{\textrm{hotspot}}$ attained during an outburst cycle is proportional to the accretion rate $\Dot{M}$ (from Equation \ref{eq:t_hotspot}).
    Thus the maximum value of $T_{\textrm{hotspot}}$ will be larger for disks with stronger magnetic field strengths because with larger $B$, the accretion pressure needs to be higher to overcome the magnetic pressure. The effect of $v_{\textrm{ff}}$ will be insignificant because the change in $\Dot{M}$ is by 1-2 orders of magnitude.
    \item {Luminosity:} As the field strength increases, and the $T_{\textrm{hotspot}}$ correspondingly increases, this increases the overall flux from the shock emission. For lower magnetic field strengths, the shock emission is overwhelmed by the flux from other components for good part of outburst epoch. However for stronger magnetic strengths, for epochs right around peak outburst, the shock emission is the major component of flux.
    \item {Lightcurve:} The mid-infrared WISE bands are insensitive to changes in the magnetic field strengths, and the lightcurves are exactly same for the smaller and larger magnetic field strengths we test. However, the optical and especially $u$ bands are strongly altered because of changes in the disk accretion and magnetospheric accretion interactions. For lower magnetic field strengths, the $u$-band lightcurve is multi-peaked: a first peak corresponds to high-temperature shock emission, a second peak correspond to the peak outburst with the brightest disk accretion emission, and a third peak corresponds to the switching {\it on} again of the magnetospheric accretion flux component. 
    \item {Color Variation with time:} The turning \textit{off} of the magnetospheric accretion flux component just before the peak outbust and subsequent turning back \textit{on} is visible in 
    $Bp-Rp$ color for lower magnetic field strengths. In the case of larger magnetic field strengths, the transition is blended, resulting in a smoother color curve. However in the $J-Ks$ band, not only is the switching \textit{off/on} phenomena more prominent, but also the relative period of time for which the disk was touching the stellar surface is clearly visible. The color becomes maximally red just before the disk touches the stellar surface. It rapidly moves bluewards as the disk moves towards the stellar surface. Finally as the accretion rate falls, the color again reaches a red maximum before finally turning blueward. The time difference between these two red-most points on the color curve basically gives the time period over which the disk was touching the stellar surface.
\end{itemize}

\subsection{Variation in Corotation Period: ($2 - 30$ days)}\label{sec:corotation_period}
The co-rotation radius determines the radius upto which matter orbiting the star is coupled to the stellar rotation period, $P_{\textrm{rot}}$. The co-rotation radius sets an outer boundary within which a disk cannot remain stable solely due to centrifugal force. As the stellar rotation period increases, corresponding centrifugal force decreases and thus the disk remains stable closer to star. 
For very small rotation periods ($T_{*,\textrm{rot}} \leq  1$ day), the large centrifugal force would mean an accretion disk could not be stable, and there would be no accretion activity. For stellar rotation periods ranging from 
$T_{*,\textrm{rot}}$ = $2, 7, 14$ and $30$ days, we perform a diagnostic analysis.

We find that as $T_{*,\textrm{rot}}$ increases, the corresponding $R_{\textrm{co-rot}}$ moves to larger radii. Thus according to Equation \ref{eq:in_disk_boundary}, $R_{\textrm{in}}$ starts from a farther distance in the low state. This means the only component that is affected is the low temperature passively heated dust disk. Again, this is not significant because at such low $\Dot{M}$ rates, the dust disk flux is far below the photospheric flux. This means the lightcurves obtained by assuming $T_{*,\textrm{rot}}$ = 2 days or $T_{*,\textrm{rot}}$ = 30 days is exactly same.

\bibliography{references}{}
\bibliographystyle{aasjournal}
\end{document}